%% file: main_arxiv_submission.tex
\pdfoutput=1
\documentclass[
  reprint,
  aps,
  prx,
  superscriptaddress,
  amsmath,
  amssymb,
  longbibliography
]{revtex4-2}
\usepackage{cmap}
\usepackage{bbm}
\usepackage{float}

\usepackage{braket}
\DeclareMathAlphabet{\mathelt}{OT1}{cmss}{m}{sl}
\DeclareSymbolFont{eltgreek}{OML}{cmssm}{m}{it}
\DeclareMathSymbol{\eltsigma}{\mathord}{eltgreek}{"1B}
\newcommand{\eltfont}[1]{{\let\sigma\eltsigma\mathelt{#1}}}
\newcommand{\grp}[1]{\mathsf{#1}}
\newcommand{\irr}[1]{\text{\normalfont\scshape #1}}

\newcommand{\cls}[1]{\eltfont{#1}}

\newcommand{\elt}[1]{\eltfont{#1}}

\newcommand{\dirprod}{\times}

\newcommand{\Sym}[1]{\grp{Sym_{#1}}}

\newcommand{\iverson}[1]{[#1]}
\newcommand{\restr}[2]{#1\!\downarrow#2}
\newcommand{\Ind}[2]{\operatorname{Ind}_{#1}^{#2}}

\newcommand{\pointr}{\mathbf{r}}

\newcommand{\iunit}{\mathrm{i}}
\newcommand{\expe}{\mathrm{e}}
\newcommand{\diff}{\mathrm{d}}

\newcommand{\rootunity}[1]{\zeta_{#1}}
\newcommand{\irrJ}[1]{D^{(#1)}}
\newcommand{\irrJp}[2]{D^{(#1,#2)}}
\newcommand{\repD}[1]{D^{#1}}
\newcommand{\wignerD}[1]{D^{#1}}

\newcommand{\irchar}[1]{\chi_{#1}}

\newcommand{\diag}{\operatorname{diag}}
\newcommand{\trace}{\operatorname{tr}}
\newcommand{\irrprod}{\otimes}
\newcommand{\mult}{\kappa}

\newcommand{\jm}[2]{\bigl\lvert{}^{#1}_{\,#2}\bigr\rangle}
\newcommand{\cgket}[3]{\bigl\lvert{}^{#1}_{\,#2};#3\bigr\rangle}
\newcommand{\irket}[2]{\bigl\lvert{}^{#1}_{\,#2}\bigr\rangle}

\newcommand{\lensspace}[2]{\mathsf{L}(#1, #2)}

\newcommand{\occ}[1]{\lvert #1\rangle}

\newcommand{\gel}{\elt{g}}             
\newcommand{\hel}{\elt{h}}             
\newcommand{\rotpart}{\elt{R}}         
\newcommand{\rotax}[1]{\elt{R_{#1}}}   
\newcommand{\orient}{\elt{S}}          
\newcommand{\crep}{\elt{w}}           
\newcommand{\gmat}{P(\gel)^{\top}}     
\newcommand{\parity}{p}               
\usepackage{mathtools}
\usepackage{mathrsfs}
\usepackage{tikz}
\usepackage{amsmath,amsthm,amssymb}
\usepackage{booktabs}
\usepackage{multirow}
\newtheorem{theorem}{Theorem}

\newtheorem{corollary}{Corollary}
\newtheorem{lemma}{Lemma}
\newcommand{\ketbra}[2]{|#1\rangle\langle#2|}
\newcommand{\spin}{\mathrel{\upharpoonleft}\kern-2pt\mathrel{\downharpoonright}}

\usetikzlibrary{shapes.geometric, arrows.meta, positioning, fit}

\tikzstyle{block} = [rectangle, draw=teal!80!black, fill=teal!80!black, text=white, 
                      text centered, minimum height=1cm, minimum width=3cm, rounded corners]
\tikzstyle{decision} = [diamond, draw=teal!80!black, fill=teal!80!black, text=white, 
                         aspect=2, text centered, inner sep=0pt, minimum width=3cm, minimum height=1cm]
\tikzstyle{labelbox} = [rectangle, draw, fill=white, text=black, rounded corners, inner sep=1pt]
\tikzstyle{arrow} = [thick, -{Stealth}, draw=teal!80!black]
\usetikzlibrary{shapes,arrows,positioning,decorations.pathreplacing,decorations.markings,calc}

\tikzstyle{block} = [rectangle, draw=teal!80!black, fill=teal!80!black, text=white, 
                      text centered, minimum height=1cm, minimum width=3cm, rounded corners]
\tikzstyle{decision} = [diamond, draw=teal!80!black, fill=teal!80!black, text=white, 
                         aspect=2, text centered, inner sep=0pt, minimum width=3cm, minimum height=1cm]
\tikzstyle{labelbox} = [rectangle, draw, fill=white, text=black, rounded corners, inner sep=1pt]
\tikzstyle{arrow} = [thick, -{Stealth}, draw=teal!80!black]
\usepackage{graphicx}
\usepackage{dcolumn}
\usepackage{bm}
\usepackage[colorlinks=true, 
linkcolor=blue, 
citecolor=blue, 
urlcolor=blue]{hyperref}

\newtheorem{proposition}{Proposition}

\usepackage{array}
\usepackage{longtable}
\providecommand{\cev}[1]{\overleftarrow{#1}}
\providecommand{\spin}{\mathrel{\upharpoonleft}\kern-2pt\mathrel{\downharpoonright}}
\providecommand{\Gmol}{\Gamma_{\mathrm{mol}}}
\providecommand{\Grot}{\Gamma_{\mathrm{r}}}
\providecommand{\Gvib}{\Gamma_{\mathrm{v}}}
\providecommand{\Gspin}{\Gamma_{\mathrm{n}}}
\providecommand{\Real}{\mathbb R}

\providecommand{\Othree}{\grp{O(3)}}
\providecommand{\SOthree}{\grp{SO(3)}}

\newcommand{\benzenemark}{\tikz[baseline=-0.1ex,x=0.36em,y=0.36em,line width=0.06em]{\draw (90:1)--(150:1)--(210:1)--(270:1)--(330:1)--(30:1)--cycle; \draw (0,0) circle (0.52);}}
\DeclareRobustCommand{\seniormark}{\TextOrMath{\benzenemark}{\text{\benzenemark}}}
\makeatletter
\renewcommand\@fnsymbol[1]{\ifcase#1\or \seniormark\or \TextOrMath\textdagger\dagger\or
  \TextOrMath\textdaggerdbl\ddagger\or \TextOrMath\textsection\mathsection\or
  \TextOrMath\textparagraph\mathparagraph\or \TextOrMath\textbardbl\|\or
  \TextOrMath{\textasteriskcentered\textasteriskcentered}{**}\or
  \TextOrMath{\textdagger\textdagger}{\dagger\dagger}\or
  \TextOrMath{\textdaggerdbl\textdaggerdbl}{\ddagger\ddagger}\else\@ctrerr\fi}
\makeatother
\DeclareRobustCommand{\pgatlasfont}[1]{{\fontencoding{OT1}\fontfamily{cmss}\fontshape{n}\selectfont#1}}
\newcommand{\PGAtlasName}{\pgatlasfont{Point Group Atlas}}
\newcommand{\PGAtlas}{\pgatlasfont{PG Atlas}}
\newcommand{\seniorauthornote}{V.V.A. and Y.L. are senior authors. Electronic addresses: \href{mailto:vva@umd.edu}{vva@umd.edu}, \href{mailto:q\string_yuanliu@ncsu.edu}{q\_yuanliu@ncsu.edu}}
\newcommand{\catP}{\mathcal{P}}
\newcommand{\catB}[1]{\mathcal{B}_{#1}}
\newcommand{\catT}[1]{\mathcal{T}_{#1}}

\usepackage{orcidlink}

\begin{document}

\title{
Symmetry classification of hyperfine rovibrational molecular phase space
}

\author{Elin Ranjan Das\orcidlink{0009-0009-8690-1523}}
\affiliation{Department of Electrical and Computer Engineering, North Carolina State University, Raleigh, NC, USA}

\author{Jing-Chen Zhang\orcidlink{0000-0002-0385-8315}}
\affiliation{Department of Electrical and Computer Engineering, North Carolina State University, Raleigh, NC, USA}

\author{Victor V. Albert\orcidlink{0000-0002-0335-9508}}
\thanks{\seniorauthornote}
\affiliation{Joint Center for Quantum Information and Computer Science (QuICS), NIST/University of Maryland, College Park, MD, USA}

\author{Yuan Liu\orcidlink{0000-0003-1468-942X}}
\thanks{\seniorauthornote}
\affiliation{Department of Electrical and Computer Engineering, North Carolina State University, Raleigh, NC, USA}
\affiliation{Department of Computer Science, North Carolina State University, Raleigh, NC, USA}
\affiliation{Department of Physics and Astronomy, North Carolina State University, Raleigh, NC, USA}

\date{\today}

\begin{abstract}

Symmetry-based structures have advanced molecular spectroscopy and quantum information science, yet connections between these fields remain largely unexplored.
We construct a phase space theory of a molecule by jointly treating rotational, vibrational, and nuclear-spin degrees of freedom as the quotient of orientations and normal coordinates by the molecule's point group.
The various combinations of how these three degrees of freedom transform under symmetries organize a molecule's states into molecular species, and we show that each species is defined by a particular absolutely maximally entangled (AME) carrier state, or, for 45 species, by a two-dimensional space of them. 
Up to local unitaries and the order of the sectors, molecular species realize 21 AME states across 75 point groups: a product state, four maximally entangled pairs, and 16 genuinely tripartite states.
A Fourier transform on the quotient relates orientation and angular-momentum descriptions while preserving these carrier states. 
We propose an example experimental protocol for preparing and driving symmetry-adapted bend states of $\mathrm{CaOCH_3}$.

\end{abstract}

\maketitle

\section{Introduction}

Molecules are structured quantum systems readily available in the laboratory and increasingly recognized as promising platforms for quantum information processing.
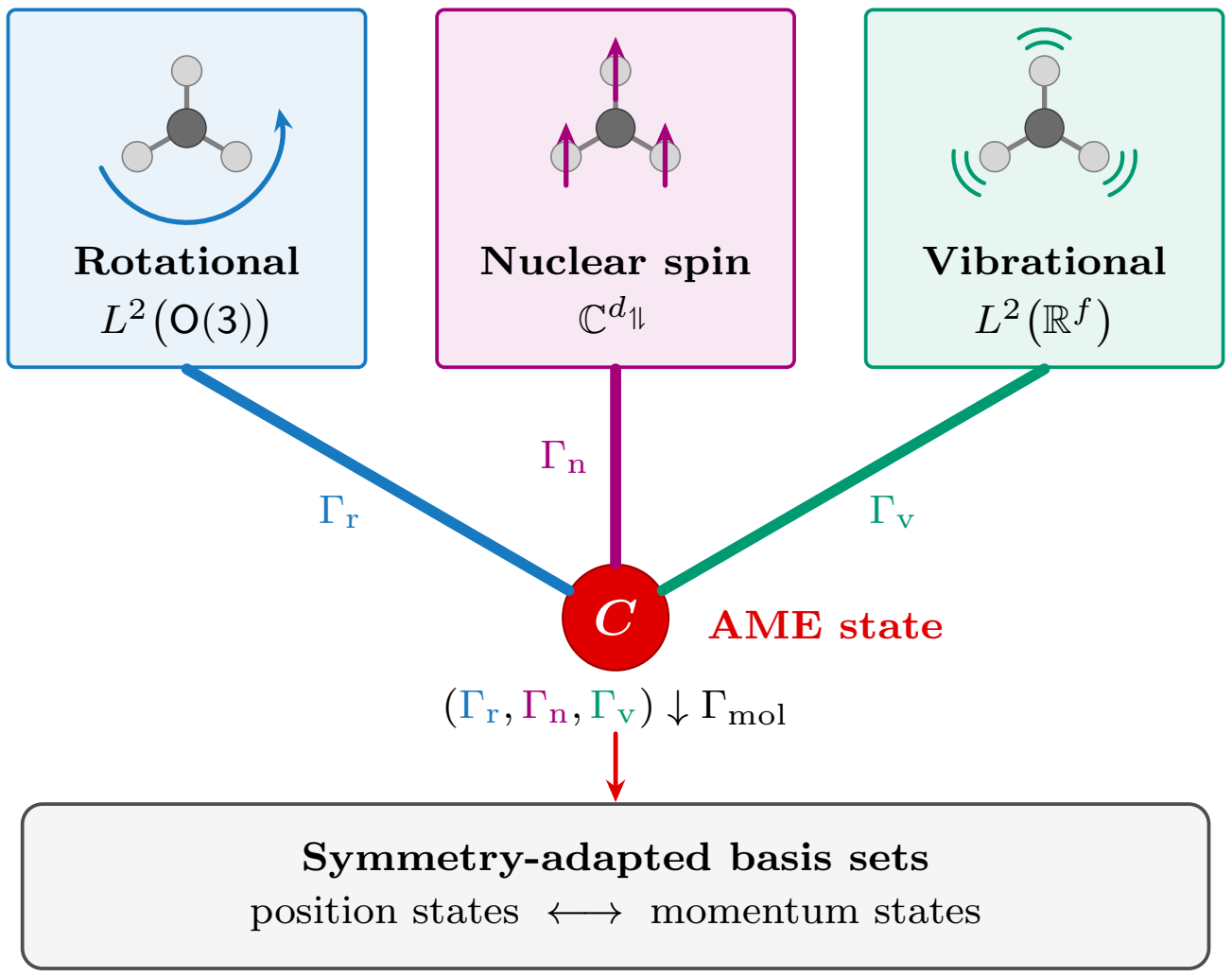
\begin{figure}[!t]
    \centering
    \input{phase_space_overview.tex}
    \caption{Unified representation-theoretic framework for the simultaneous coupling of rotational, nuclear-spin, and vibrational degrees of freedom. The pictures show the three sectors on a pyramidal $\mathrm{XH_3}$ molecule seen down its threefold axis: rotation of the molecule-fixed frame (blue), the proton spins (magenta), and vibration (green). The space of each sector decomposes into irreps of the point group $\grp{G}$, and an admissible triple $(\Grot,\Gspin,\Gvib)\downarrow\Gmol$ defines a species. Symmetry and spin statistics fix the coupling tensor $C$ of the three carriers (physical indices) of the triple (colored legs), or for some species a two-dimensional space of such tensors (namely, $\grp{E}$ irrep subspace). For more general groups, the dimension of $C$ can be higher. Every such tensor is an absolutely maximally entangled (AME) state. The same tensor enters the symmetry-adapted position and momentum states, which are related by the Fourier transform of Sec.~\ref{sec:fourier}.}
    \label{fig:phase_space_formalism}
\end{figure}
Entanglement generation~\cite{Lin2020, Holland2023, Bao2023, Picard2025}, quantum process tomography~\cite{Chuntonov2013, YuenZhou2011}, and the control of vibrational~\cite{Hallas2023, Robichaud2026, Tesch2002} and spin states~\cite{Rugg2019, Ospelkaus2010, Gregory2021} have been demonstrated or analyzed in molecular systems~\cite{10.1063/5.0272970, Cornish2024, Langen2024, Deiss2024, Augenbraun2023}.  However, chemical physics and quantum information science approach these systems from largely disconnected perspectives. In chemical physics, molecular symmetry determines which combinations of rotational, vibrational, and nuclear-spin quantum numbers are physically admissible, governs spectroscopic selection rules, and explains statistical weights and missing spectral lines~\cite{LonguetHiggins1963, BunkerJensen2006, Chan2025}. These symmetry constraints provide a detailed empirical and theoretical map of the molecular states, but do not by themselves accommodate a unified quantum state-space description of the underlying degrees of freedom's abilities to encode and process quantum information. Conversely, quantum information science typically treats only a small portion of molecular degrees of freedom as potential quantum registers~\cite{DeMille2002, Yelin2006, Ni2018}, emphasizing state preparation, entanglement, and coherent control, without leveraging the full molecular Hilbert space nor incorporating the global symmetry constraints that correlate these registers and restrict their accessible states. This disconnect hides a fundamental opportunity: molecular symmetry is not only a constraint on quantum information processing but can itself provide a resource for generating and characterizing structured multipartite entanglement. Moreover, symmetry provides a common framework across the entire molecular phase space to analyze their potential for quantum information processing. Establishing this connection needs a common state-space framework that treats molecular degrees of freedom jointly while retaining their symmetry-imposed correlations.

A phase-space formulation provides such a framework. Here, a phase space is a configuration manifold with two conjugate bases, a position basis of localized configurations and a momentum basis of generator eigenstates, related by the Fourier transform of the underlying group. For a rigid rotor the manifold is the orientation group~\cite{PhysRevX.10.031050, Harter1993}, the momentum basis consists of angular-momentum states $\ket{{}^{J}_{\,m},k}$ measured in rotational spectroscopy, and the position basis consists of oriented configurations prepared by alignment pulses~\cite{Koch2019, Stapelfeldt2003}, which for identical nuclei only approximate localized orientations~\cite{GrohmannManz2018}. 

Reference~\cite{albert2025quantumtheorymolecularorientations} constructed a molecular phase space for the rotational and nuclear-spin sectors, showing that molecular geometry and identical-particle statistics impose Hamiltonian-independent structure on the admissible states, including rotation--spin correlations with no separable representatives. In that construction, however, vibration is treated as a spectator, assumed to transform trivially. The hyperfine theory of Harter and Patterson makes the same assumption~\cite{HarterPatterson1979}. 

The interplay of molecular symmetry and entanglement across internal degrees of freedom has also been explored in several contexts. Pfister, Chardonnet, and Bord\'e already wrote of the entanglement of the rovibrational and nuclear-spin wave functions of phosphine~\cite{Pfister1996}. Nuclear permutation symmetry also entangles rotation with tunneling in CH$_5^+$~\cite{Wodraszka2015} and the inversion mode of ammonia with its other modes~\cite{Bouakline2020}, and it forbids localized states of rotors built from identical nuclei~\cite{GrohmannManz2018}. Entanglement among the vibrational modes of CO$_2$ has been quantified as well~\cite{Rauhut2024}.

Nevertheless, the spectator assumption fails for degenerate fundamentals and overtone manifolds, which carry nontrivial irreps. In these cases, the operations that permute identical nuclei act on the normal coordinates and on the molecular frame at once~\cite{LonguetHiggins1963, Hougen1962}. 

Here we construct the phase space for the three sectors jointly (Fig.~\ref{fig:phase_space_formalism}). The configuration manifold is the quotient of $\grp{O(3)}\times\mathbb R^{f}$ by the diagonal action of the molecular point group $\grp{G}$, with nuclear spin as a finite fiber. Identical-particle statistics selects a one-dimensional molecular species~\cite{BunkerJensen2006}, and the physical state space is the corresponding isotypic component of the unconstrained product. Because $\grp{G}$ acts on orientation and normal coordinates at once, this component is not a product of single-sector subspaces. It factors instead into multiplicity spaces, which symmetry leaves free, and a coupling tensor among the three sectors, which symmetry fixes, uniquely for most point groups. This is the precise sense in which our results are Hamiltonian independent, and it makes the framework complementary to variational methods~\cite{Carrington2017} such as TROVE~\cite{Yurchenko_2017, Chubb2018, Yurchenko2007, Tennyson2016}, DVR3D~\cite{Tennyson2004}, MULTIMODE~\cite{Bowman2003}, and GENIUSH~\cite{Fabri2011}. These programs diagonalize the Hamiltonian on a given potential-energy surface in a numerically built basis, producing energies and eigenvectors. Their hyperfine-resolved extensions add the nuclear spins~\cite{YachmenevKupper2017, OwensYachmenev2018, Coles2019, Yachmenev2022}. In our language each such eigenvector is a superposition, over admissible species, of fixed coupling tensors, and the Hamiltonian determines only the amplitudes in the multiplicity spaces. The framework thus specifies \textit{analytically}, before any calculation, the space in which every variational eigenvector lies and the structure it must carry.

Since the coupling tensor is fixed by the species, so is the entanglement it carries among rotation, nuclear spin, and vibration. We show that every coupling tensor is an absolutely maximally entangled (AME) state~\cite{Helwig2012}, also called a perfect tensor~\cite{Pastawski2015}. It carries the largest entanglement its dimensions allow across every cut between one sector and the other two. The same property makes the tensor invisible to any physical probe of a single sector. Depending on the species, it is a product state, a maximally entangled pair of two sectors, or a genuinely tripartite state. Up to local unitaries and the order of the sectors, the coupling tensors of the 75 point groups of Altmann and Herzig~\cite{AltmannHerzig2011} realize 21 AME states, counting each two-dimensional coupling space by two basis states. Phosphine and calcium monomethoxide host a three-qubit state of GHZ form at definite parity, while methane hosts a totally antisymmetric three-qutrit state. These states are defined by symmetry alone and take the same form in every sector-local basis.

We verify the construction against \(\mathrm{H_2O}\), \(\mathrm{PH_3}\), and \(\mathrm{CH_4}\), whose spectra are established independently, and against the laser-coolable symmetric top \(\mathrm{CaOCH_3}\)~\cite{Mitra2020}, recovering nuclear-spin statistical weights~\cite{Wilson1935, Schmiedt2016, KubischtaTeixeira2026}, species-resolved rotational selection rules, and vibrationally induced species interchange. For each molecule we tabulate the admissible species and write every admissible sector explicitly as tripartite position states, built from orientation cosets, nuclear-spin product states, and normal-coordinate wavefunctions, and as tripartite momentum and spectroscopic states, built from angular-momentum, total-spin, and occupation-number states. 

The rest of the paper is organized as follows. Section~\ref{sec:phasespace} defines the phase space, Sec.~\ref{sec:reduction} reduces each sector by symmetry, and Sec.~\ref{sec:fourier} constructs the Fourier transform on the reduced configuration space. Section~\ref{sec:entang} shows that every coupling tensor is an AME state and classifies these states. Sec.~\ref{sec:examples} works through the molecular examples, and Sec.~\ref{sec:experiment} proposes an experimental preparation of the symmetry-adapted bend states with calcium monomethoxide ($\mathrm{CaOCH_3}$) under $\grp{C_{3v}}$ symmetry as an example. Throughout, we highlight quantities with quantum-information significance, including the dimensions of symmetry-supported subspaces, the multiplicities of repeated irreps, and the transformations generated by molecular symmetry operations. The quantitative treatment of encoding and control is deferred to a companion paper, which is in progress.

\section{Phase Space Description of a molecule}
\label{sec:phasespace}

Following Ref.~\cite{albert2025quantumtheorymolecularorientations}, we take a phase space to be a configuration manifold with a pair of Fourier-conjugate bases on it, a position basis and a momentum basis, together with the Fourier transform relating them.  
This section states what the manifold and the bases are for a molecule with rotational, nuclear-spin, and vibrational degrees of freedom, first on the kinematic space (the entire unreduced Hilbert space from all possible configurations and momentum) and then, in Sec.~\ref{sec:fourier}, after symmetry reduction. Everything later in the paper will point to this space.

First we define the kinematic molecular Hilbert space, which is the threefold product
\begin{equation}
\mathcal H
=
\mathcal H_{\mathrm r}
\otimes
\mathcal H_{\mathrm n}
\otimes
\mathcal H_{\mathrm v},
\label{eq:active_hilbert}
\end{equation}
where $\mathcal H_{\mathrm r}$ describes the rotational orientation of the molecule, $\mathcal H_{\mathrm n}$ the nuclear-spin states of identical nuclei, and $\mathcal H_{\mathrm v}$ the vibrational normal-mode excitations. Identical nuclei are those permuted by the group operations. 

If $\grp{G}\subset\Othree$ denotes the molecular symmetry point group, then each operation $\gel\in\grp{G}$ acts on $\mathcal H$ as 
\begin{align}
\hat{\rho}(\gel)
=
\hat{\rho}_{\mathrm{r}}(\gel)
\otimes
\hat{\rho}_{\mathrm{n}}(\gel)
\otimes
\hat{\rho}_{\mathrm{v}}(\gel).
\label{eq:total_group_action}
\end{align}
Here $\hat\rho_X(\gel)$, $X\in\{\mathrm r,\mathrm n,\mathrm v\}$, are unitary representation operators acting on the rotational, nuclear-spin, and vibrational Hilbert spaces, respectively, in the respective bases.

\subsection{Configuration manifold for the phase space}

Let the molecule have $n$ nuclei and $f = 3n-6$ normal modes, which assumes a nonlinear molecule vibrating about a single equilibrium. An unconstrained molecular configuration is a pair $(\orient,\mathbf q)$: an element $\orient\in\Othree$ fixing the orientation of the molecule-fixed frame relative to the laboratory, and a vector $\mathbf q=(q_1,\ldots,q_f)\in\Real^{f}$ of normal-coordinate displacements. The three Euclidean translation degrees of freedom are ignored. The molecule-fixed frame is the Eckart frame~\cite{Eckart1935, LouckGalbraith1976, LittlejohnReinsch1997}. Linear molecules, with $f=3n-5$, enter through the extended molecular symmetry group~\cite{BunkerJensen2006}, and floppy molecules need the extensions of Refs.~\cite{Schmiedt2015, Fabri2017}. Then,
\begin{equation}
    \gel:\;(\orient,\;\mathbf q)\;\longmapsto\;\left(\orient\gel^{-1},\;D^{\mathrm{v}}(\gel)\,\mathbf q\right),
    \label{eq:diagonal_action}
\end{equation}
where $D^{\mathrm{v}}:\grp{G}\to\grp{O(\mathnormal{f})}$ is the orthogonal representation describing the action of molecular symmetry on the normal coordinates. Configurations related by Eq.~\eqref{eq:diagonal_action} differ only by a relabeling of identical nuclei and are therefore the same physical configuration. The molecular configuration manifold is the quotient
\begin{equation}
\boxed{\;
\mathcal Q
=
\Big(\Othree\times\Real^{f}\Big)\Big/\grp{G}
\;}
\qquad
\dim\mathcal Q = 3n-3
\label{eq:config_manifold}
\end{equation}
The configuration manifold dimension equals the number of independent continuous degrees of freedom: three from molecular orientation ($\Othree$) and $f$ from vibrations. 
In Ref.~\cite{albert2025quantumtheorymolecularorientations} the vibrational factor is absent and the configuration manifold is the orientation manifold $\SOthree/(\grp{G}\cap\grp{SO(3)})$. Without vibrations, our $\mathcal Q$ is $\Othree/\grp{G}$. 
Here the quotient cannot be taken in either factor separately, because $\grp{G}$ acts on both at once.

Nuclear spin carries no coordinate. It enters as a finite-dimensional internal state space $\mathbb C^{d_{\spin}}$, $d_{\spin}=\prod_a(2I_a+1)^{N_a}$, attached to each point of $\mathcal Q$ and permuted by $\grp{G}$ along with the nuclei. In topology, \(\mathcal Q\) is the \emph{base space}, while such internal state space attached to a fixed molecular configuration \(Q\in\mathcal Q\) is called a \emph{fiber}. This is analogous to a spin-\(\tfrac12\) particle, for which each position \(\mathbf r\) carries the two-dimensional space \(\operatorname{span}\{\ket{\uparrow},\ket{\downarrow}\}\).

The kinematic state space is therefore the space of sections of a \emph{finite-rank bundle} over $\mathcal Q$,
\begin{equation}
\mathcal H
=
L^2\!\left(\Othree\right)
\otimes
\left[
\bigotimes_a
\left(\mathbb C^{2I_a+1}\right)^{\otimes N_a}
\right]
\otimes
L^2\!\left(\mathbb R^f\right).
\label{eq:kinematic_space}
\end{equation}
reduced by $\grp{G}$ as described in Sec.~\ref{sec:constraint}. Here, a \emph{finite-rank bundle over \(\mathcal Q\)} is a space built by attaching a fiber to every point $Q\in\mathcal Q$. In the present case, the fiber attached to each configuration is the nuclear-spin Hilbert space $\mathbb C^{d_{\spin}}$. Thus, a point of the total bundle consists of a molecular configuration $Q$ together with a nuclear-spin state in the fiber over $Q$. Over any small region $U\subset\mathcal Q$, the bundle locally looks like the product $U\times\mathbb C^{d_{\spin}}$. A quantum state is a section of this bundle: a map
\[
Q\longmapsto\Psi(Q)
\]
that assigns to each configuration $Q$ a vector $\Psi(Q)$ in the corresponding spin fiber.

\subsection{Kinematic bases}
\label{sec:bases_kin}

The kinematic phase space description~\eqref{eq:kinematic_space} requires three bases summarized in Table~\ref{tab:phase_space}. They are bases of the unreduced space only. The symmetry-reduced and physical bases are obtained from them in Secs.~\ref{sec:reduction} and~\ref{sec:fourier}, respectively. 

The \emph{position} basis consists of localized configurations,
\begin{equation}
\ket{\orient}\otimes\ket{\mathbf m_{I}}\otimes\ket{\mathbf q},
\label{eq:position_basis}
\end{equation}
i.e.\ a definite molecular orientation, a definite set of nuclear-spin projections, and a definite nuclear geometry. The orientation and coordinate labels are continuous, so $\ket{\orient}$ and $\ket{\mathbf q}$ are generalized eigenkets in the usual sense, delta-normalized with respect to Haar measure on $\grp{O(3)}$ and Lebesgue measure on $\mathbb R^f$. We use the unnormalized Haar measure, $\int_{\grp{O(3)}}\diff\gel=16\pi^2$.

The \emph{momentum} basis consists of the Fourier-conjugate states: the Peter--Weyl basis of $\Othree$~\cite[Apps.~1 and~2]{Ezra1982}, which diagonalizes $\mathbf J^2$, the laboratory projection $J_Z$, the molecule-frame projection $J_z$, and parity, and the eigenstates of the normal-mode momenta,
\begin{equation}
\ket{{}^{J,\parity}_{\,m},k}\otimes\ket{\mathbf m_{I}}\otimes\ket{\mathbf p},
\label{eq:momentum_basis}
\end{equation}
where rotational angular momentum $J$ and parity $p=\pm1$ labels an $\Othree$ irrep, and $m$ and $k$ its laboratory- and molecule-frame projections, respectively, and $\mathbf p=(p_1,\ldots,p_f)$ are the normal-mode momenta, the canonical conjugates of $\mathbf q$. We follow the notation of the \PGAtlasName{} (\PGAtlas)~\cite{o3harmonics}, a catalogue of closed subgroups of $\grp{O(3)}$ and their irreps, whose kets $\jm{J,\parity}{m'}$ carry only the projection on which $\grp{G}$ acts, our molecule-frame $k$. We keep both projections \footnote{The \PGAtlas's tables use row vectors, $\pointr\mapsto\pointr P(\gel)$, compose $\gel\hel$ as ``$\gel$ first'', and let $\gel$ act on a function by $(\gel f)(\pointr)=f(\pointr P(\gel))$, that is, by $U(\gel)^{-1}$. We identify states by their wavefunctions, so for the irreps of $\grp{G}$ our labels, partner indices, and coupling coefficients are the \PGAtlas's for the same state. Our explicit symmetry-adapted states agree with its tabulated ones up to the phase of each copy. Our matrices are its matrices at the inverse element, $\repD{\Gamma}(\gel)=\repD{\Gamma}_{\mathrm{atlas}}(\gel^{-1})=\repD{\Gamma}_{\mathrm{atlas}}(\gel)^{\dagger}$, and for a complex-conjugate pair of one-dimensional irreps our character is the conjugate of the tabulated one. The molecule-frame part of our rotor ket $\ket{{}^{J,\parity}_{\,m},k}$ is the \PGAtlas's $(-1)^k\jm{J,\parity}{-k}$, and $\grp{O(3)}/\grp{G}$ is its $\grp{G}\backslash\grp{O(3)}$ read on column vectors. The README of Ref.~\cite{DasRepo2026} states the same dictionary for the irreps of $\grp{G}$.}. Irrep labels are set in small capitals, as in the \PGAtlas.

Position and momentum bases are related by the Fourier transform on the unreduced continuous configuration space $\Othree\times\mathbb R^f$, which factorizes as 
\begin{equation}
\braket{\orient \, | \, {}^{J,\parity}_{\,m},k}
=
\sqrt{\frac{2J+1}{16\pi^{2}}}\;\irrJp{J}{\parity}_{m,k}(\orient),
\qquad
\braket{\mathbf q\,|\,\mathbf p}
=
\frac{\expe^{\iunit\mathbf p\cdot\mathbf q}}{(2\pi)^{f/2}},
\label{eq:fourier_pair}
\end{equation}
with $q_s=Q_s/b_s$ and $p_s=b_sP_s/\hbar$ the dimensionless coordinate and momentum of mode $s$, $b_s=(\hbar/\omega_s)^{1/2}$, and \(Q_s\) is a mass-weighted normal coordinate.

The spectroscopy-resolved third basis is the eigenbasis of the zeroth-order symmetric-top and harmonic Hamiltonian, with the nuclear spins coupled to definite total spin,
\begin{equation}
\ket{{}^{J,\parity}_{\,m},k}\otimes\ket{I_{\mathrm{tot}},M_{I};\mu_{\mathrm n}}\otimes\ket{\mathbf n},
\label{eq:spectroscopic_basis}
\end{equation}
with $(I_{\mathrm{tot}},M_I)$ the total nuclear spin, $\mu_{\mathrm n}$ resolving its multiplicity (e.g., $I_{\mathrm{tot}}=\tfrac12$ with multiplicity $2$ in $\mathrm{PH_3}$ and $I_{\mathrm{tot}}=1,0$ with multiplicities $3,2$ in $\mathrm{CH_4}$), and $\mathbf n=(n_1,\ldots,n_f)$ the vibrational occupation numbers. For an asymmetric top such as water, the rotor Hamiltonian mixes the $k$ states, so the basis is a reference basis.

For $N$ equivalent spin-$\tfrac12$ nuclei, Schur--Weyl duality~\cite{Schmiedt2016, KubischtaTeixeira2026} assigns each total-spin sector $I_{\mathrm{tot}}$ a multiplicity space carrying an irreducible representation of $\Sym{\mathnormal{N}}$. When the permutation action induced by $\grp{G}$ is surjective onto $\Sym{\mathnormal{N}}$, as in all four molecules of Sec.~\ref{sec:examples} ($\Sym{2}$ for water, $\Sym{3}$ for phosphine and $\mathrm{CaOCH_3}$, and $\grp{T_{d}}$, of type $\Sym{4}$, for methane), this representation remains irreducible under $\grp{G}$. Thus, distinct $I_{\mathrm{tot}}$ correspond to distinct nuclear-spin species $\Gspin$, with the multiplicity label $\mu_{\mathrm n}$ serving as the carrier index of $\Gspin$.

For rotation, the spectroscopic labels $(J,\parity,m,k)$ also serve as Fourier momentum labels and symmetric-top quantum numbers. In contrast, vibrational occupations diagonalize the harmonic Hamiltonian $\sum_s\omega_s(\hat n_s+\tfrac12)$ rather than $\hat p_s$, since $\hat n_s$ is conjugate to oscillator phase, not $q_s$. Thus
\(
\braket{\mathbf q|\mathbf n}
=\prod_s\psi_{n_s}(q_s)
\)
defines a harmonic-energy basis transformation, while $\ket{\mathbf m_I}\leftrightarrow \ket{I_{\mathrm{tot}},M_I;\mu_{\mathrm n}}$ is an angular-momentum coupling transformation within the spin fiber. Neither is a Fourier transform. Nevertheless, the occupation basis conveniently represents the Fourier transform on $\mathbb R^f$, whose eigenfunctions are the Hermite functions.

\begin{table}[htbp!]
\caption{The three kinematic bases of the molecular phase space. All are bases of Eq.~\eqref{eq:kinematic_space} before the symmetry constraint of Sec.~\ref{sec:constraint} is imposed.}
\label{tab:phase_space}
\centering
\begin{tabular}{lllll}
\toprule
Sector & Manifold & Position & Momentum & Spectroscopic\\
\midrule
Rotational   & $\grp{O(3)}$      & $\orient$        & $J,\parity,m,k$ & $J,\parity,m,k$\\
Nuclear spin & $\mathbb C^{d_{\spin}}$ & $\mathbf m_I$ & $\mathbf m_I$ & $I_{\mathrm{tot}},M_I;\mu_{\mathrm n}$\\
Vibrational  & $\mathbb R^{f}$     & $\mathbf q$    & $\mathbf p$ & $\mathbf n$\\
\bottomrule
\end{tabular}
\end{table}

This basis choice gives the framework spectroscopic content. Quantum numbers that are exact in the spectroscopic basis, e.g., $J$, the parity $\parity$, the nuclear-spin species, the vibrational occupations, are not independently assignable. The projection correlates them, and the correlations appear directly as statistical weights, missing lines, and forbidden combinations in the spectrum. Section~\ref{sec:examples} shows this basis construction for four molecules, three of whose spectra are textbook cases.

\section{Symmetry reduction of the Phase Space}
\label{sec:reduction}

The kinematic bases in Table~\ref{tab:phase_space} must be converted into symmetry-adapted states carrying the quantum numbers and irrep labels determined by molecular symmetry. Then identical-particle statistics from Lemma~\ref{lemma:unified_admissibility_condition} selects the physically admissible combinations of these labels. Because the point group acts jointly on rotation, nuclear spin, and vibration, the admissible states generally involve coupled symmetry carriers rather than independently selected sector states. The physical Hilbert space is therefore obtained by symmetry reduction of the combined kinematic space.

To implement this projection, we return to the kinematic space constructed in Section~\ref{sec:phasespace}, Eq.~\eqref{eq:kinematic_space}, and decompose each of its three factors into irreducible representations of the same group $\grp{G}$. These sectorwise decompositions provide the ingredients for the joint symmetry selection. They do not impose independent physical constraints. For the vibrational sector, the decomposition is Eq.~\eqref{eq:sfe_map}.

\subsection{The Spin--Statistics Constraint Selection Rule}
\label{sec:constraint}
Let \(D(\gel)\) denote the unitary operator representing the molecular
symmetry operation \(\gel\in\grp{G}\). The physical molecular wavefunction \(\Psi\) must transform according to the one-dimensional irreducible representation \(\Gmol\), i.e.
\begin{equation}
    D(\gel)\Psi
    =
    \chi_{\Gmol}(\gel)\Psi,
    \qquad \forall\,\gel\in\grp{G}.
    \label{eq:psi_transform}
\end{equation}
Equivalently, \(\Psi\) is an eigenfunction of every \(D(\gel)\) with
eigenvalue \(\chi_{\Gmol}(\gel)\). Here
\begin{align}
    \chi_{\Gmol}=\prod_a\operatorname{sign}\!\left[p_a(\gel)\right]^{2I_a}.
    \label{eq:gamma_mol}
\end{align}
where \(p_a(\gel)\) is the permutation induced by $\gel$ among the identical nuclei of species \(a\), each having nuclear spin \(I_a\). For fermionic nuclei ($I_a$ half-integer), odd permutations of identical nuclei contribute a minus sign, while for bosonic nuclei ($I_a$ integer) they contribute a plus sign. Thus eq.~\eqref{eq:gamma_mol} defines a one-dimensional irrep \(\Gmol\in\widehat{\grp{G}}\), which specifies the required symmetry species of $\psi$. The rotational, vibrational, and nuclear-spin factors need not individually transform as $\Gmol$. Only their coupled product must have this overall symmetry. We take the electronic state to be nondegenerate and totally symmetric. A one-dimensional electronic species $\Gamma_{\mathrm{el}}$ would replace $\Gmol$ by $\Gmol\otimes\Gamma_{\mathrm{el}}^{*}$. An electron spin quantized along the laboratory axes, as in Hund's case (b), is invariant under every operation, so it leaves $\Gmol$ unchanged~\cite{Hougen1963, BunkerJensen2006}.

Since $\grp{G}$ acts diagonally on the rotational, nuclear-spin, and vibrational factors through Eq.~\eqref{eq:diagonal_action}, this requirement constrains their combined symmetry rather than the symmetry of each factor separately. 
\begin{lemma}[Symmetry-selection rule.]
A coupled sector can contribute physical states only if
\begin{equation}
    \Grot
    \otimes
    \Gspin
    \otimes
    \Gvib
    \supset
    \Gmol.
\end{equation}
Within each admissible sector, only the component transforming according to $\Gmol$ is retained.
\label{lemma:unified_admissibility_condition}
\end{lemma}

Lemma~\ref{lemma:unified_admissibility_condition} is the spin-statistics rule of the molecular symmetry group~\cite{LonguetHiggins1963, BunkerJensen2006}, with the vibrational species kept explicit. It generalizes the rotational--nuclear-spin rule of Ref.~\cite{albert2025quantumtheorymolecularorientations}, where $\Gvib$ is taken to be the trivial irrep and the condition therefore reduces to $\Grot\otimes\Gspin\supset\Gmol$. 

The symmetry-reduced state space $\mathcal H_{\mathrm{sym}}$ is obtained by projection onto the required molecular species. For an irrep \(\Gamma\in\widehat{\grp{G}}\) of dimension \(d_\Gamma\) and character \(\chi_\Gamma(\gel)\),
\begin{equation}
    \mathcal H_{\mathrm{sym}}
    =
    \left(\frac{1}{|\grp{G}|}
    \sum_{\gel\in\grp{G}}
    \chi_{\Gmol}(\gel)^{*}\,
    \hat{\rho}(\gel)\right)\mathcal H,
    \label{eq:full_group_projector}
\end{equation}
where \(\hat\rho(\gel)\) is the tensor-product action of Eq.~\eqref{eq:total_group_action}.

\subsection{Structure of the three sectors}
\label{sec:isotypic}

Each sector carries a unitary representation of the same finite group $\grp{G}$ and admits an isotypic decomposition
\begin{equation}
\mathcal H_{\alpha}
\cong
\bigoplus_{\Gamma\in\widehat{\grp{G}}}
M^{(\alpha)}_{\Gamma}\otimes V_{\Gamma},
\qquad
\alpha\in\{\mathrm{r},\mathrm{n},\mathrm{v}\},
\label{eq:generic_isotypic}
\end{equation}
where $V_\Gamma\cong\mathbb C^{d_\Gamma}$ is the irrep carrier and $M^{(\alpha)}_\Gamma$ its multiplicity space, with $\mathfrak m^{(\alpha)}_\Gamma
=\dim M^{(\alpha)}_\Gamma$ counting independent copies of $\Gamma$. The group acts as $I_{M^{(\alpha)}_\Gamma}\otimes D^\Gamma(\gel)$, mixing the $d_\Gamma$ carrier components while leaving multiplicity labels unchanged. Consequently, only the carrier factors participate in the spin--statistics projection and intersector symmetry coupling.

Throughout, we fix a reference carrier $V_\Gamma$ with basis $\{\irket{\Gamma}{\mu}\}_{\mu=1}^{d_\Gamma}$ for each irrep and identify its occurrences across all three sectors by unitary intertwiners. The matrices $D^\Gamma(\gel)$ and the partners $\mu$ follow the \PGAtlas{} through the dictionary of Sec.~\ref{sec:bases_kin}. Equivalent realizations $D^\Gamma(\gel)\mapsto U D^\Gamma(\gel)U^\dagger$, related by a unitary change-of-basis operator $U$, are thereby expressed in a common basis, allowing the same coupling tensors to be used across sectors.

The following subsections determine $m^{(\mathrm{r})}_\Gamma$, $m^{(\mathrm{n})}_\Gamma$, and $m^{(\mathrm{v})}_\Gamma$ through character inner products over $\grp{G}$, differing only in the sector representation characters. Table~\ref{tab:sector_summary} summarizes the constructions.

\subsection{Rotational sector and its history}
\label{sec:rot_sector}

The rotational Hilbert space \(L^2(\Othree)\) carries commuting left and right actions associated with laboratory- and molecule-frame transformations~\cite{Harter1978}. For a fixed molecular symmetry species \(\Grot\), the corresponding multiplicity space realizes \(\Ind{\grp{G}}{\Othree}\Grot\) through functions \(f:\Othree\rightarrow V_{\Grot}\) satisfying 
\[
f(\orient\gel)=D^{\Grot}(\gel^{-1})f(\orient),
\qquad \gel\in\grp{G}.
\] 
Here, $\operatorname{Ind}_{\grp G}^{\Othree}\Grot$ denotes the representation of $\Othree$ induced from the irrep $\Grot$ of the subgroup $\grp{G}$. Its Fourier decomposition is obtained from the parity-dressed Wigner functions of Eq.~\eqref{eq:fourier_pair}, labeled by the irreducible representation \(\irrJp{J}{\parity}\) of \(\grp{O(3)}\), with \(J\in\mathbb Z_{\ge0}\) and \(\parity=\pm1\).

Every \(\gel\in\grp{O(3)}\) has unique decomposition as \(\gel=(\det\gel)\,\rotpart\), with \(\rotpart=(\det \gel)\gel \in\grp{SO(3)}\), giving \(\grp{O(3)}=\grp{SO(3)}\dirprod\grp{C_{i}}\). Here, $\grp{C_{i}}=\{\elt{E},\elt{i}\}$ is the inversion group, generated by the parity operation $\elt{i}=-\mathbb I$. Each integer-$J$ irrep therefore has two extensions, distinguished by its inversion parity $\parity=\pm1$:
\begin{equation}
\begin{gathered}
\irrJp{J}{\parity}(\gel)=\parity^{\iverson{\det\gel=-1}}D^J(\rotpart),
\\
\irrJp{J}{+}(\gel)=D^J(\rotpart),\quad \irrJp{J}{-}(\gel)=(\det \gel)\,D^J(\rotpart).
\end{gathered}
\label{eq:O3_wigner}
\end{equation}
Group elements act on states actively, by $U(\gel)$, matrices act on column vectors, and products compose from right to left. Our Wigner matrices $D^J$ are the complex conjugates of the standard ones.

By Frobenius reciprocity~\cite{Serre1977}, the multiplicity of a molecular symmetry species \(\Gamma\) within the rotational manifold \(\irrJp{J}{\parity}\) is
\begin{equation}
m^{(\mathrm r)}_{\Gamma}(J,\parity)
=
\frac{1}{|\grp{G}|}
\sum_{\gel\in\grp{G}}
\parity^{\iverson{\det\gel=-1}}\,
\chi^{J}(\rotpart)\,
\chi_{\Gamma}(\gel)^{*},
\label{eq:frobenius}
\end{equation}
where, if \(\rotpart\) is a proper rotation through angle \(\theta_{\rotpart}\),
\(
\chi^J(\rotpart)
=
\frac{\sin[(J+\tfrac12)\theta_{\rotpart}]}
     {\sin(\theta_{\rotpart}/2)}.
\)

The parity dependence of Eq.~\eqref{eq:frobenius} can be isolated completely. Let
\(
\det_{\grp{G}}
\)
denote the one-dimensional representation obtained by restricting the determinant character of \(\grp{O(3)}\) to \(\grp{G}\). It is trivial when \(\grp{G}\subset\grp{SO(3)}\). In the standard character-table conventions it is \(\irr{a}_{2}\) for \(\grp{C_{2v}}\), \(\grp{C_{3v}}\), and \(\grp{T_{d}}\).

When \(\det_{\grp{G}}\) is nontrivial, the improper elements cancel from the sum over both parities, and Eq.~\eqref{eq:frobenius} gives
\begin{equation}
\begin{split}
&m^{(\mathrm r)}_{\Gamma}(J,+)
+
m^{(\mathrm r)}_{\Gamma}(J,-)
\\
&\quad=
\frac{1}{|\grp{G}\cap\grp{SO(3)}|}
\sum_{\gel\in\grp{G}\cap\grp{SO(3)}}
\chi^{J}(\gel)\,
\chi_{\Gamma}(\gel)^{*}.
\end{split}
\label{eq:parity_counting}
\end{equation}

Tensoring with \(\det_{\grp{G}}=\irr{a}_{2}\) exchanges \(\irr{a}_{1}\leftrightarrow\irr{a}_{2}\) in $\grp{C_{2v}}, \grp{C_{3v}},$ and $\grp{T_{d}},$ where present, \(\irr{b}_{1}\leftrightarrow\irr{b}_{2}\) in $\grp{C_{2v}}$, and \(\irr{t}_{1}\leftrightarrow\irr{t}_{2}\) in $\grp{T_{d}}.$ In symmetric-top descriptions, molecule-frame symmetry operations can exchange \(\ket{{}^{J}_{\,m},k}\) and \(\ket{{}^{J}_{\,m},-k}\), leading to symmetry-adapted linear combinations for $k\neq0.$ This pairing should be distinguished from tunneling parity associated with motion between distinct molecular configurations, such as the umbrella-inversion doublet of ammonia. 

\paragraph{Molecular-symmetry-group labels.}
Spectroscopic compilations (for example the ExoMol line list of methane~\cite{Yurchenko2024}) label rovibrational levels with the molecular symmetry group of feasible permutations and permutation-inversions~\cite{LonguetHiggins1963,BunkerJensen2006}.
In that convention the rotational functions are functions on $\SOthree$, and each operation acts on them through its equivalent rotation, the rotation part $\rotpart$ of the corresponding point-group element.
The laboratory inversion $E^{*}$ enters only through permutation-inversions, and the rotational functions carry no parity of their own.
By Eq.~\eqref{eq:O3_wigner}, the rotational labels of the two conventions coincide at $\parity=+1$ and differ by the factor $\det_{\grp{G}}$ at $\parity=-1$.
The total symmetry of a level is then $\Gmol\otimes\det_{\grp{G}}$ at $\parity=-1$, and the nuclear-spin and vibrational labels agree in the two conventions.
Because $\det_{\grp{G}}\otimes\det_{\grp{G}}$ is trivial, Eq.~\eqref{eq:triple_coupling} assigns the same $N$ to $(\Grot,\Gspin,\Gvib)$ with $\Gmol$ and to $(\Grot\otimes\det_{\grp{G}},\Gspin,\Gvib)$ with $\Gmol\otimes\det_{\grp{G}}$.
The relabeling is therefore a bijection between admissible triples, and it changes no physical state.
For groups of type~I described in Eq.~\eqref{eq:base-space} the factor $\det_{\grp{G}}$ is trivial, and the two conventions give identical labels.

This dictionary also clarifies a disagreement over the symmetry labels of methane levels, which Hougen described in 1974 as a decade-old controversy~\cite{Hougen1974, Hougen1976}.
Jahn assigned rotational species by analogy with multipole expressions~\cite{Jahn1938}, and Hougen introduced the labels now used with the molecular symmetry group~\cite{Hougen1962,Hougen1963}.
Moret-Bailly instead allowed, for each $J$, two rotational functions of parities g and u, our $\parity=+1$ and $\parity=-1$~\cite{MoretBailly1974}.
The schemes disagreed in practice: the $3.39\ \mu\mathrm{m}$ methane line near a helium--neon laser line was labeled $F_1$ by some authors and $F_2$ by others~\cite{MoretBailly1974}.
Berger later classified the levels in the group $\grp{G}\times\Othree$ with rotational functions of both parities~\cite{Berger1977}, which is the convention of this work. Hilico, Berger, and Lo\"ete had introduced this group as the molecular degeneracy group~\cite{HilicoBergerLoete1976}. Refs.~\cite{Rey2003, Michelot2004} discuss how the $\Othree$ representations are oriented in the cubic groups.
Berger's rule for converting to Hougen's labels exchanges the subscripts 1 and 2 at parity u, which is the factor $\det_{\grp{T_{d}}}=\irr{a}_{2}$ at $\parity=-1$ above.
Hougen objected that the u functions are discontinuous, since the $J=0$ one takes opposite signs on two halves of the rotational space~\cite{Hougen1974}.
On $\Othree$ these functions need no cut.
The $J=0$ state with $\parity=-1$ is constant on the proper orientations and takes the opposite value on the improper ones, which form a separate component of $\Othree$.
The state is therefore smooth. Hougen had granted that his scheme, in a sense, ``avoided rather than solved'' the problem of applying improper operations to functions of the Euler angles, because no set of Euler angles describes a left-handed axis system~\cite{Hougen1963}. The improper orientations of $\Othree$ are exactly these left-handed axis systems.

Spectroscopically, Eq.~\eqref{eq:frobenius}, together with Lemma~\ref{lemma:unified_admissibility_condition}, determines which rotational symmetry species are compatible with the spin--statistics constraint for a specified nuclear-spin and vibrational species. Incompatible rotational levels are excluded from that species.

\subsection{Nuclear-spin sector}
\label{sec:spin_sector}

For several sets of equivalent but nonidentical nuclei, with \(N_a\) nuclei of species \(a\) and nuclear spin \(I_a\), the nuclear-spin Hilbert space is
\[
\mathcal H^{(\mathrm n)}_{\spin}
=
\bigotimes_a
\left(\mathbb C^{2I_a+1}\right)^{\otimes N_a}.
\]
The character of this nuclear-spin permutation representation is given by the generalization of the standard cycle-counting formula for identical nuclei~\cite{Oka2011}:
\begin{equation}
\chi^{(\mathrm n)}_{\spin}(\gel)
=
\prod_a
(2I_a+1)^{c[p_a(\gel)]},
\label{eq:nuclear_spin_character}
\end{equation}
where \(c[p_a(\gel)]\) counts all disjoint cycles of \(p_a(\gel)\), including fixed nuclei as one-cycles. 

Character orthogonality then gives the multiplicity of a point-group species \(\Gamma\) in the nuclear-spin sector,
\begin{equation}
m^{(\mathrm n)}_\Gamma
=
\frac{1}{|\grp{G}|}
\sum_{\gel\in\grp{G}}
\chi_\Gamma(\gel)^*\,
\chi^{(\mathrm n)}_{\spin}(\gel).
\label{eq:nuclear_spin_multiplicity}
\end{equation}
Equations~\eqref{eq:gamma_mol} and~\eqref{eq:nuclear_spin_character} are therefore determined by the same induced permutations \(p_a(\gel)\).

Because permutations of equivalent nuclei commute with the collective-spin operators, the nuclear-spin space can be organized simultaneously by total angular momentum and permutation symmetry. For a single set of \(N\) equivalent spin-\(\tfrac12\) nuclei, Schur--Weyl duality~\cite{Schmiedt2016} gives
\[
(\mathbb C^2)^{\otimes N}
\cong
\bigoplus_{I_{\mathrm{tot}}}
\mathcal V_{I_{\mathrm{tot}}}
\otimes
\mathcal M_{I_{\mathrm{tot}}},
\]
where \(\mathcal V_{I_{\mathrm{tot}}}\) is the spin-\(I_{\mathrm{tot}}\) space with basis \(M_I=-I_{\mathrm{tot}},\ldots,I_{\mathrm{tot}}\), while \(\mathcal M_{I_{\mathrm{tot}}}\) carries the permutation symmetry associated with the repeated occurrence of that total spin. When the molecular point group induces the full relevant permutation group, \(\mathcal M_{I_{\mathrm{tot}}}\) may be identified with a point-group carrier \(V_{\Gspin}\), and its basis label \(\mu_{\mathrm n}\) becomes the carrier index of \(V_{\Gspin}\). The spectroscopic spin states may then be written
\(
\cgket{\Gspin}{\mu_{\mathrm n}}{I_{\mathrm{tot}},M_I}.
\)
For several inequivalent nuclear species, each equivalent set may first be coupled separately to a total spin \(I_a^{\mathrm{tot}}\), after which the resulting spin spaces may be coupled further if a total molecular nuclear spin is desired.

\subsection{Vibrational sector}
\label{sec:vib_sector}

The vibrational factor \(L^2(\mathbb R^f)\) is spanned by occupation-number states \(\ket{\mathbf n}=\ket{n_1,\ldots,n_f}\). Since \(\grp{G}\) acts linearly and orthogonally on the normal coordinates through \(D^{\mathrm{v}}(\gel)\), its second-quantized action mixes creation operators only within symmetry-equivalent mode subspaces and preserves the total number of quanta in each such subspace. We therefore group the \(f\) normal coordinates into \(r\) symmetry-degenerate sets, the \(s\)-th carrying a real irrep \(\Gamma_s\) of dimension \(d_s\), and fix an excitation pattern \(\mathbf v=(v_1,\ldots,v_r)\), where
\(
v_s=\sum_{i=1}^{d_s}n_{s,i}
\)
is the total number of quanta in the \(s\)-th set. The corresponding finite-dimensional manifold \(\mathcal H(\mathbf v)\) is invariant under \(\grp{G}\), and we denote the restricted representation by \(U_{\mathbf v}(\gel)\). The blocks are real because $D^{\mathrm v}(\gel)$ is built from the real $3\times3$ matrices of $\grp{G}$. Over $\mathbb C$, a real irrep stays irreducible or splits into a complex-conjugate pair, as the plane of an $\irr{e}$ mode does in $\grp{C_{3}}$, and the decompositions below are over $\mathbb C$.

The creation operators $a^\dagger_{s,i}$, $i=1,\dots,d_s$, of the $s$-th set transform as its coordinates,
\begin{equation}
U(\gel)\,a^\dagger_{s,i}\,U(\gel)^{\dagger}=\sum_{j=1}^{d_s}D^{\Gamma_s}_{j,i}(\gel)\,a^\dagger_{s,j},
\label{eq:creation_transform}
\end{equation}
so $U_{\mathbf v}(\gel)=\bigotimes_s \operatorname{Sym}^{v_s}\!\bigl[D^{\Gamma_s}(\gel)\bigr]$ is a tensor product of symmetric powers.
The \(v\)-excitation subspace of \(d\) bosonic modes is naturally the symmetric tensor power \(\operatorname{Sym}^v\mathbb C^d\): bosonic creation operators commute, so only the completely symmetric part of the \(v\)-fold tensor product occurs. Its dimension is \(\binom{v+d-1}{d-1}\).

For a doubly degenerate mode it is often convenient to pass to the circular combinations $a^\dagger_{\pm}=\tfrac{1}{\sqrt{2}}\,(a^\dagger_{1}\pm\iunit a^\dagger_{2})$, which create opposite circular excitations $\ell=\pm1$ of the degenerate normal-mode doublet. We write the state with $n_-$ quanta of $a^\dagger_-$ and $n_+$ of $a^\dagger_+$ as the \PGAtlas writes an occupation ket, $\occ{n_- n_+}$: the count of each partner without a comma, in the order of the \PGAtlas's partners, $x-\iunit y$ (created by $a^\dagger_-$) before $x+\iunit y$. Its harmonic degree, the \PGAtlas's $\ell$, is $|n_+-n_-|$.
In that basis a rotation $\elt{C_{\mathnormal{n}}}$ about the axis acts diagonally: for a mode of species $\irr{e}_{q}$ ($\irr{e}\equiv\irr{e}_{1}$), on whose plane $\elt{C_{\mathnormal{n}}}$ acts as a rotation by $2\pi q/n$, $U(\elt{C_{\mathnormal{n}}})\occ{n_- n_+}=\expe^{-2\pi\iunit q\ell/n}\occ{n_- n_+}$ with $\ell=n_+-n_-$, and the reflection in the $xz$ plane exchanges $n_+\leftrightarrow n_-$. This basis change requires no physical modification of the molecule. Appendix~\ref{app:circular_vibrations} discusses polarization-controlled preparation and spectroscopic detection of circular excitations, subject to the relevant optical selection rules~\cite{Lin2025}. 

Symmetry-adapted vibrational bases are also constructed in variational rovibrational frameworks such as TROVE \cite{Yurchenko_2017, Chubb2018}, where symmetry adaptation is embedded in the solution of a specified molecular Hamiltonian. The construction used here is different in purpose: the decomposition~\eqref{eq:sfe_map} is determined solely by the representation of the molecular point group on the normal-mode Fock space and therefore requires no Hamiltonian, potential-energy surface, or dynamical model.

By Schur's lemma, each excitation manifold decomposes into irreps of $\grp{G}$,
\begin{equation}
\mathcal H(\mathbf v)\cong\bigoplus_{\Gamma\in\widehat{\grp{G}}}M_\Gamma(\mathbf v)\otimes V_\Gamma,\qquad
m_\Gamma(\mathbf v)=\frac{1}{|\grp{G}|}\sum_{\gel\in\grp{G}}\chi_\Gamma(\gel)^{*}\chi_{\mathbf v}(\gel),
\label{eq:sfe_map}
\end{equation}
with $m_\Gamma(\mathbf v)=\dim M_\Gamma(\mathbf v)$ and $\chi_{\mathbf v}=\trace U_{\mathbf v}$. In this decomposition $\grp{G}$ acts as $\bigoplus_\Gamma I_{M_\Gamma(\mathbf v)}\otimes D^\Gamma(\gel)$, which fixes it up to unitaries on the multiplicity spaces~\cite{BunkerJensen2006}. The projectors $P^\Gamma_{\mu,\nu}=\frac{d_\Gamma}{|\grp{G}|}\sum_{\gel}D^\Gamma_{\mu,\nu}(\gel)^{*}U_{\mathbf v}(\gel)$, applied to occupation states, give the symmetry-adapted states $\cgket{\Gamma}{\mu}{\mult}_{\mathrm v}=\sum_{\mathbf n}C^{\Gamma,\mult,\mu}_{\mathbf n}\ket{\mathbf n}$. These states are also called symmetry-adapted linear combinations (SALCs).
The symmetry-adapted Fock encoding is therefore a kinematic construction rather than a spectroscopic diagonalization procedure. Once the point group, its action on the normal modes, and the excitation pattern \(\mathbf v\) are specified, the carrier--multiplicity decomposition and the coefficients \(C_{\mathbf n}^{\Gamma,\mult,\mu}\) follow without specifying a vibrational Hamiltonian. The symmetry-adapted Fock states of a single degenerate mode, for every point group, are tabulated in the \PGAtlas~\cite{o3harmonics}, together with the species of $\operatorname{Sym}^{v}\Gamma$ for $v\le12$. It lists the states up to $v=6$ for two- and three-dimensional modes and up to $v=3$ for larger ones. When the plane of a mode splits over $\mathbb C$ into a complex-conjugate pair, as in $\grp{C_{3}}$, the \PGAtlas{} gives these states by a rule. Combination bands obey two laws: conservation of the total vibrational angular momentum $\ell_{\mathrm{tot}}\equiv\sum_s k_s\ell_s \pmod N$ for two-dimensional modes, where $N$ is the order of the principal element of the group (a rotation or rotoreflection about its main axis) and $k_s$ is fixed by the phase $\expe^{-2\pi\iunit k_s\ell_s/N}$ with which that element acts on $\occ{n^{(s)}_- n^{(s)}_+}$, and coupling of the harmonic layers, $\ell=|\ell_1-\ell_2|,\ldots,\ell_1+\ell_2$, for vector-like three-dimensional modes.

\begin{table}[t]
\caption{The three sector reductions. Each multiplicity is an inner product of the sector
character with $\chi_\Gamma$ over $\grp{G}$; the sectors differ only in the character.
Together with Lemma~\ref{lemma:unified_admissibility_condition} these determine the physical space.}
\label{tab:sector_summary}
\centering
\footnotesize
\begin{tabular}{llll}
\toprule
Sector & Character & Multiplicity & Observable\\
\midrule
Rotational
& $\parity^{\iverson{\det\gel=-1}}\chi^J(\rotpart)$
& Eq.~\eqref{eq:frobenius}
& allowed rotational levels\\
Nuclear spin
& Eq.~\eqref{eq:nuclear_spin_character}
& Eq.~\eqref{eq:nuclear_spin_multiplicity}
& spin-statistical weights\\
Vibrational
& $\prod_s\chi_{\operatorname{Sym}^{v_s}\Gamma_s}(\gel)$
& Eq.~\eqref{eq:sfe_map}
& combination species\\
\bottomrule
\end{tabular}
\end{table}

\subsection{Symmetry-reduced coupled space}

Throughout we work at fixed rotational manifold $\irrJp{J}{\parity}$ and vibrational excitation pattern $\mathbf v$, on which $m^{(\mathrm r)}_\Gamma$ and $m^{(\mathrm v)}_\Gamma$ respectively depend.

\begin{proposition}[Structure of the physical space]
\label{prop:structure}
With multiplicities from Eqs.~\eqref{eq:frobenius} and~\eqref{eq:nuclear_spin_multiplicity} and Eq.~\eqref{eq:sfe_map},
\begin{align}
\mathcal H_{\mathrm{sym}}
&\cong
\bigoplus_{\Grot,\Gspin,\Gvib}
\underbrace{
M^{(\mathrm r)}_{\Grot}
\otimes
M^{(\mathrm n)}_{\Gspin}
\otimes
M^{(\mathrm v)}_{\Gvib}
}_{\text{symmetry-inert}}
\;\otimes\;
\underbrace{
\mathcal C^{\Gmol}_{\Grot\Gspin\Gvib}
}_{\text{symmetry-constrained}},
\label{eq:coupled_decomposition}
\end{align}
where the coupled space $\mathcal C^{\Gmol}_{\Grot\Gspin\Gvib}$, spanned by point-group coupling coefficients~\cite{Griffith1962, Butler1981, Rey2019}, consists of the vectors $\ket{\psi}=\sum C_{\mu_{\mathrm r}\mu_{\mathrm n}\mu_{\mathrm v}}\irket{\Grot}{\mu_{\mathrm r}}\irket{\Gspin}{\mu_{\mathrm n}}\irket{\Gvib}{\mu_{\mathrm v}}$ satisfying
\begin{equation}
\Bigl[
\repD{\Grot}(\gel)\otimes
\repD{\Gspin}(\gel)\otimes
\repD{\Gvib}(\gel)
\Bigr]\,
\ket{\psi}
=
\irchar{\Gmol}(\gel)\,\ket{\psi}
\label{eq:coupling_space}
\end{equation}
for all $\gel\in\grp{G}$, and has dimension
\begin{equation}
N^{\Gmol}_{\Grot\Gspin\Gvib}
=
\frac{1}{|\grp{G}|}
\sum_{\gel\in \grp{G}}
\irchar{\Grot}(\gel)\,
\irchar{\Gspin}(\gel)\,
\irchar{\Gvib}(\gel)\,
\irchar{\Gmol}(\gel)^{*}.
\label{eq:triple_coupling}
\end{equation}
Hence
\begin{equation}
\dim\mathcal H_{\mathrm{sym}}
=
(2J+1)
\!\!\sum_{\Grot,\Gspin,\Gvib}\!\!
m^{(\mathrm r)}_{\Grot}
m^{(\mathrm n)}_{\Gspin}
m^{(\mathrm v)}_{\Gvib}
N^{\Gmol}_{\Grot\Gspin\Gvib} .
\label{eq:phys_dimension}
\end{equation}
\end{proposition}

\begin{proof}
Insert Eq.~\eqref{eq:generic_isotypic} for each sector into Eq.~\eqref{eq:kinematic_space}. On each summand $\hat\rho(\gel)$ acts as $I_{M_{\mathrm r}}\otimes I_{M_{\mathrm n}}\otimes I_{M_{\mathrm v}}\otimes \repD{\Grot}\otimes \repD{\Gspin}\otimes \repD{\Gvib}$, so the projector $\hat P_{\Gmol}$ of Eq.~\eqref{eq:full_group_projector} acts as the identity on the multiplicity factors and as the $\Gmol$-projector on the carrier product, whose range is Eq.~\eqref{eq:coupling_space}. Its rank is the multiplicity of the one-dimensional $\Gmol$ in the triple product, Eq.~\eqref{eq:triple_coupling}.
\end{proof}
A triple \((\Grot,\Gspin,\Gvib)\) is admissible in the sense of Lemma~\ref{lemma:unified_admissibility_condition} precisely when
\(
N^{\Gmol}_{\Grot\Gspin\Gvib}\neq0,
\)
so Eq.~\eqref{eq:triple_coupling} is the quantitative form of the selection rule. The prefactor \(2J+1\) in Eq.~\eqref{eq:phys_dimension} is the laboratory-frame degeneracy, which is untouched by every molecule-frame symmetry operation.

Equation~\eqref{eq:coupled_decomposition} separates two qualitatively different kinds of degrees of freedom. \(\grp{G}\) acts on the multiplicity factors
\(
M^{(\mathrm r)}_{\Grot}
\otimes
M^{(\mathrm n)}_{\Gspin}
\otimes
M^{(\mathrm v)}_{\Gvib}
\)
as identity, so they are symmetry-inert. Their states may therefore be resolved or mixed only by dynamics acting within the corresponding multiplicity sectors. The coupling factor
\(
\mathcal C^{\Gmol}_{\Grot\Gspin\Gvib},
\)
by contrast, is symmetry-constrained. Its dimension
\(
N^{\Gmol}_{\Grot\Gspin\Gvib}
=
\dim
\mathcal C^{\Gmol}_{\Grot\Gspin\Gvib}
\)
counts the number of independent copies of the required molecular species \(\Gmol\) contained in the product of the three carrier spaces.

For instance, in \(\grp{C_{3v}}\),
\(
\irr{e}\otimes\irr{e}=\irr{a}_{1}\oplus\irr{a}_{2}\oplus\irr{e}.
\)
Hence for \(\Gmol=\irr{a}_{2}\), the fixed triple
\(
(\irr{e},\irr{e},\irr{a}_{1})
\)
contains exactly one copy of \(\irr{a}_{2}\). Once standard bases
\(\{\irket{\irr{e}}{1},\irket{\irr{e}}{2}\}\) are chosen, symmetry fixes the corresponding carrier tensor, up to phase, to the unique antisymmetric combination
\[
\ket{\psi_{\irr{a}_{2}}}
=
\frac{1}{\sqrt{2}}
\left(
\irket{\irr{e}}{1}_{\mathrm r}\irket{\irr{e}}{2}_{\mathrm n}
-
\irket{\irr{e}}{2}_{\mathrm r}\irket{\irr{e}}{1}_{\mathrm n}
\right)
\otimes\ket{\irr{a}_{1}}_{\mathrm v},
\]
up to basis conventions. Here uniqueness has physical content: the relative amplitudes are fixed by symmetry rather than by a Hamiltonian.

Because $\Gmol$ is one-dimensional, $\Gmol\otimes\Gvib^{*}$ is irreducible, and Eq.~\eqref{eq:triple_coupling} reads
\begin{equation}
N^{\Gmol}_{\Grot\Gspin\Gvib}
=
\Bigl\langle
\chi_{\Grot}\otimes\chi_{\Gspin},
\,
\chi_{\Gmol}\otimes\chi_{\Gvib^{*}}
\Bigr\rangle
\equiv
\Bigl\langle
\Grot\otimes\Gspin,
\,
\Gmol\otimes\Gvib^{*}
\Bigr\rangle_{\grp{G}}.
\label{eq:coupling_reduction}
\end{equation}

Equation~\eqref{eq:coupling_reduction} also makes clear when a nontrivial carrier coupling can occur. If all three constituent irreps are one-dimensional, their tensor product is itself one-dimensional and therefore
\(
N^{\Gmol}_{\Grot\Gspin\Gvib}
\in\{0,1\}.
\)

If $N^{\Gmol}_{\Grot\Gspin\Gvib}=1$, the coupling space $\mathcal C^{\Gmol}_{\Grot\Gspin\Gvib}$ is one-dimensional. Once the standard carrier bases are fixed, it determines the coupled carrier state
\begin{equation}
\ket{\psi}
=
\sum_{\mu_{\mathrm r},\mu_{\mathrm n},\mu_{\mathrm v}}
C_{\mu_{\mathrm r}\mu_{\mathrm n}\mu_{\mathrm v}}
\irket{\Grot}{\mu_{\mathrm r}}
\irket{\Gspin}{\mu_{\mathrm n}}
\irket{\Gvib}{\mu_{\mathrm v}}
\label{eq:unique_coupled_state}
\end{equation}
up to an overall phase. For $N>1$, symmetry selects no basis of the coupling space, and any unitary among the copies of $\Gmol$ gives an equally valid basis.

This statement therefore concerns uniqueness \emph{within a fixed species triple}, not uniqueness of the triple itself. When $N=2$, symmetry fixes only a two-dimensional coupling space, treated in Sec.~\ref{sec:entang} and in Sec.~3 of \texttt{species\_tables/species\_tables.pdf} in Ref.~\cite{DasRepo2026}.

We fix the overall phase that Eq.~\eqref{eq:unique_coupled_state} leaves free as the species tables of Ref.~\cite{DasRepo2026} do: each coupling tensor \(C\) is normalized, and its first nonzero component \(C_{\mu_{\mathrm r}\mu_{\mathrm n}\mu_{\mathrm v}}\), with the partner triples \((\mu_{\mathrm r},\mu_{\mathrm n},\mu_{\mathrm v})\) in lexicographic order, is real and positive. All coupling tensors written out in this paper follow this rule.

One consequence of Eq.~\eqref{eq:coupling_reduction} is particularly useful in the examples and makes transparent the spectroscopic effect of activating a nontrivial one-dimensional vibrational species.

\begin{corollary}[Vibrational species shift]
\label{cor:shift}
For one-dimensional \(\Gvib\), the triples
\(
(\Grot,\Gspin,\Gvib)
\quad\text{and}\quad
(\Grot\otimes\Gvib,\Gspin,\Gamma_0)
\),
with \(\Gamma_0\) the trivial irrep of \(\grp{G}\), have naturally isomorphic coupling spaces and are therefore admissible simultaneously. If \(N=1\), their normalized coupling tensors agree up to a phase, once $V_{\Grot}\otimes V_{\Gvib}$ is mapped onto the standard carrier of $\Grot\otimes\Gvib$ by a fixed intertwiner. Thus a nondegenerate vibrational excitation shifts the rotational species by \(\Grot\mapsto\Grot\otimes\Gvib\), leaving the nuclear-spin species and statistical weight unchanged. For Abelian \(\grp{G}\), every admissible triple has \(N=1\), with 
$\Grot
=\Gmol\otimes\Gspin^{*}\otimes\Gvib^{*}.$
For real one-dimensional irreps, \(\Gamma^{*}=\Gamma\).
\end{corollary}
\begin{proof}
See Appendix~\ref{proof:shift}.
\end{proof}
In water, this species shift produces the interchange of the rotational symmetry classes associated with ortho- and para-water in the \(\nu_3\) band (Sec.~\ref{subsec:water}). In phosphine and calcium monomethoxide, an \(\irr{a}_{2}\) vibrational species exchanges the \(\irr{a}_{1}\) and \(\irr{a}_{2}\) rotational species. This exchange of rotational partners also matters for state preparation. Section~\ref{sec:experiment} applies this rule to the symmetric and antisymmetric bend states of calcium monomethoxide.

\section{The Molecular Fourier Transform}
\label{sec:fourier}

Section~\ref{sec:bases_kin} distinguished three useful bases of the unreduced kinematic space: the configuration basis, its Fourier-conjugate momentum basis, and the spectroscopic basis. The preceding symmetry analysis in Sec.~\ref{sec:reduction} has been carried out mainly in the latter, where rotational states are labeled by \((J,\parity,m,k)\), nuclear spins by coupled-spin quantum numbers, and vibrations by occupation numbers. We now ask how the corresponding physical states are represented in orientation space after the molecular symmetry constraint has been imposed.

The multimode oscillator number states are eigenstates of the vibrational Fourier transform of Eq.~\eqref{eq:fourier_pair},
\begin{equation}
\braket{\mathbf p|\mathbf n}=(-\iunit)^{|\mathbf n|}\prod_{s=1}^{f}\psi_{n_s}(p_s),\qquad
\mathscr F_{\mathrm v}=\expe^{-\iunit\pi\hat N/2},
\label{eq:vib_fourier}
\end{equation}
with $|\mathbf n|=\sum_s n_s$, $\hat N$ the total number operator, and $\psi_{n}$ the normalized oscillator eigenfunction. Each manifold of $N$ quanta is thus an eigenspace of $\mathscr F_{\mathrm v}$ with eigenvalue $(-\iunit)^N$, and $\mathscr F_{\mathrm v}$ commutes with the orthogonal action of $\grp{G}$ on $\mathbf q$.

Geometrically, \(\mathscr F_{\mathrm v}\) rotates every dimensionless phase plane \((q_s,p_s)\) by \(\pi/2\). For an individual harmonic mode this is equivalent, up to an overall phase, to evolution through one quarter of its oscillator period. For several modes of unequal frequency, this statement is understood mode by mode rather than as evolution for a single common physical time. Consequently, within any fixed-\(N\) excitation manifold, a symmetry-adapted vibrational state has the same Hermite-function structure in momentum and position representations, with \(\mathbf q\) replaced by \(\mathbf p\) and an overall phase \((-\iunit)^N\). The harmonic approximation therefore renders the vibrational position--momentum duality especially simple within each fixed-total-quanta manifold.

\subsection{Symmetry-reduced Fourier transform on the quotient}
\label{sec:fourier_motivation}

The symmetry-reduced Fourier transform is useful for several related reasons.

\emph{Orientation and angular momentum are complementary descriptions:}
Rotational spectroscopy resolves states labeled by angular momentum and molecular symmetry, whereas alignment and orientation experiments~\cite{Friedrich1991, Loesch1990, Larsen2000}, Coulomb-explosion imaging, and rotational-wavepacket dynamics~\cite{Dooley2003} naturally probe states localized in molecular orientation. For a distinguishable rigid body these descriptions are related by the Peter--Weyl transform on \(\grp{O(3)}\). For a molecule containing identical nuclei, however, the physical orientation space has an additional discrete identification that must be incorporated into the transform.

\emph{The physical orientation base is a quotient:}
A molecule-frame symmetry operation \(\gel\in\grp{G}\) changes the labeled orientation \(\orient\) to \(\orient\gel\) without changing the corresponding unlabeled rigid molecular configuration. The physical orientation base is therefore
\(
\mathcal Q_{\mathrm r}=\grp{O(3)}/\grp{G}.
\)
The covering space \(\grp{O(3)}\) nevertheless remains useful: a rotational state of species \(\Grot\), with partners \(\ket{\psi_\mu}\), is represented there by the equivariant wavefunction with components \(\Psi_\mu(\orient)=\braket{\psi_\mu|\orient}\), satisfying
\begin{equation}
\Psi(\orient\gel)
=
\repD{\Grot}(\gel^{-1})\Psi(\orient),
\qquad
\gel\in\grp{G}.
\label{eq:rot_equivariance}
\end{equation}

Thus molecular symmetry keeps the orientation space and determines how an internal carrier \(V_{\Grot}\) is attached over the quotient.

\emph{The rotational species induces a non-Abelian monodromy:}
A closed path in \(\grp{O(3)}/\grp{G}\) may lift to a path in \(\grp{O(3)}\) whose endpoint differs from its starting point by a molecular symmetry operation \(\gel\). The corresponding fiber then returns transformed by \(\repD{\Grot}(\gel)\). In this sense the rotational species records the monodromy carried by the orientation-space state.

Once a rotational species \(\Grot\) has been selected, the transform between its angular-momentum and orientation-space realizations depends only on the geometry of \(\grp{O(3)}/\grp{G}\) and on the associated carrier \(V_{\Grot}\). Vibrational excitation can nevertheless change which \(\Grot\) is admissible through Lemma~\ref{lemma:unified_admissibility_condition}. Thus different vibrational manifolds may select different associated bundles over the same orientation base space without changing the underlying rotational Fourier transform.

\emph{Orientation-space encodings:}
The quotient-space description is also the natural language for molecular orientation codes~\cite{PhysRevX.10.031050, albert2025quantumtheorymolecularorientations}, whose codewords are superpositions of localized orientations and whose logical operations may be interpreted geometrically as transport on orientation space. Incorporating active vibrational and nuclear-spin sectors therefore requires not a new Fourier kernel, but a symmetry-consistent identification of the rotational bundle selected by the coupled molecular state.

\subsection{Orientational configuration space of the transform}

Subgroups of $\grp{O(3)}$ fall into three types: type I consists of subgroups contained entirely in $\grp{SO(3)}$; type II consists of subgroups containing the inversion $\elt{i}$, necessarily of the form $\grp{G}=(\grp{G}\cap\grp{SO(3)})\dirprod\grp{C_{i}}$ with $\grp{C_{i}}=\{\elt{E},\elt{i}\}$; and type III consists of subgroups containing improper rotations but not the inversion. Accordingly, symmetry reduction yields the orientational base space
\begin{equation}
  \frac{\grp{O(3)}}{\grp{G}}\;\cong\;
  \begin{cases}
    \dfrac{\grp{SO(3)}}{\grp{G}}\times\grp{C_{i}},
      & \grp{G}\ \text{of type I},\\[10pt]
    \dfrac{\grp{SO(3)}}
    {\grp{G}\cap\grp{SO(3)}},
      & \grp{G}\ \text{of type II or III}.
  \end{cases}
  \label{eq:base-space}
\end{equation}
For groups of type I, the two connected components correspond to the two enantiomeric orientations, whereas for types II and III, improper symmetries identify these components, leaving a connected quotient.

For cyclic rotational symmetry, these quotients have a particularly simple topology. A lens space $\lensspace{p}{q}$ is the three-dimensional manifold obtained by quotienting the unit sphere $\mathsf S^{3}\subset\mathbb C^2$ under the cyclic identification
\begin{equation}
  (z_1,z_2)\sim
  \left(\expe^{2\pi\iunit/p}z_1,\expe^{2\pi\iunit q/p}z_2\right),
  \qquad \gcd(p,q)=1.
\end{equation}
Since $\grp{SU(2)}\cong\mathsf S^{3}$ is the double cover of $\grp{SO(3)}$, the preimage of $\grp{C_{\mathnormal{n}}}$ is the cyclic subgroup of order $2n$ of $\grp{SU(2)}$, giving $\grp{SO(3)}/\grp{C_{\mathnormal{n}}}\cong\lensspace{2n}{1}$~\cite{Postnikov1992}. Thus, for water, $\grp{O(3)}/\grp{C_{2v}}\cong\grp{SO(3)}/\grp{C_{2}} \cong \lensspace{4}{1}$; for phosphine and calcium monomethoxide, $\grp{O(3)}/\grp{C_{3v}}\cong\grp{SO(3)}/\grp{C_{3}} \cong \lensspace{6}{1}$; and for methane, $\grp{O(3)}/\grp{T_{d}}\cong\grp{SO(3)}/\grp{T}$, where $\grp{T}$ is the proper tetrahedral group. 

Moreover, position states are no longer localized orientations but \emph{coset} states, superpositions over the symmetry orbit weighted by the inducing irrep, which carry an internal index $\lambda=1,\dots,d_{\Grot}$ in place of a sharp orientation. Here, the symmetry orbit of an orientation $\orient$ is the coset $\orient\grp{G}=\{\orient\gel\mid \gel\in\grp{G}\}$, the set of all orientations related to $\orient$ by molecular symmetry operations. A coset state superposes these orbit elements with irrep-determined amplitudes. The molecular symmetry group acts diagonally on orientation and shape, Eq.~\eqref{eq:diagonal_action}, so the reduced configuration space is not a product but a bundle: the base carries the orientation, and the vibrational carrier, the nuclear-spin carrier, and the coupling channel all live in the fiber above it. Fiber data is carried along by molecule-frame operations and is invisible to laboratory-frame ones. The transform is a statement about the base.

\subsection{Reduced position and momentum bases}
\label{sec:bases}

We now write both bases explicitly, and Sec.~\ref{sec:examples} evaluates them molecule by molecule.

Molecule-frame operations act on the rotational momentum states through the right-regular action, with the phase conventions of Ref.~\cite{albert2025quantumtheorymolecularorientations} extended to improper elements by Eq.~\eqref{eq:O3_wigner}:
\begin{align}
\cev X_{\rotax{z}(\alpha)}\ket{{}^{J,\parity}_{\,m},k}&=\expe^{\iunit k\alpha}\ket{{}^{J,\parity}_{\,m},k},
\nonumber\\
\cev X_{\rotax{y}(\pi)}\ket{{}^{J,\parity}_{\,m},k}&=(-1)^{J+k}\ket{{}^{J,\parity}_{\,m},-k},
\nonumber\\
\cev X_{\elt{i}\,\gel}&=\parity\,\cev X_{\gel},
\label{eq:rot_action}
\end{align}
with $\elt{i}$ the inversion, so that a reflection containing the $z$ axis acts as $\elt{\sigma_{v}}(0)=\elt{i}\,\rotax{y}(\pi)$, the mirror in the $xz$ plane. Here $\elt{\sigma_{v}}(\psi)$ denotes the mirror in the vertical plane that contains the horizontal direction at azimuth $\psi$. On position states $\cev X_{\gel}\ket{\orient}=\ket{\orient\gel^{-1}}$. Nuclear spins are permuted actively, the spin of the nucleus at site $i$ being carried to site $p(i)$, and vibrational quanta transform by Eq.~\eqref{eq:creation_transform}. With these conventions the rovibrational levels of a $\grp{C_{3v}}$ molecule that are of $\irr{a}$ type are those with $k-\ell\equiv0\pmod 3$, as they must be~\cite{Hougen1962, BunkerJensen2006}.

\emph{Momentum and spectroscopic descriptions.} Let $\cgket{\Grot}{\mu_{\mathrm r}}{{}^{J,\parity}_{\,m}\mult_{\mathrm r}}_{\mathrm r}$ be the $\grp{G}$-adapted rotational states, linear combinations of $\ket{{}^{J,\parity}_{\,m},k}$ over $k$ transforming as partner $\mu_{\mathrm r}$ (the column index) of the standard matrices $\repD{\Grot}$, with $\mult_{\mathrm r}=1,\dots,m^{(\mathrm r)}_{\Grot}(J,\parity)$ the multiplicity index; let $\cgket{\Gspin}{\mu_{\mathrm n}}{I_{\mathrm{tot}},M_I}$ be total-spin states of partner $\mu_{\mathrm n}$, with a copy index added to $(I_{\mathrm{tot}},M_I)$ wherever $\Gspin$ occurs more than once at fixed $I_{\mathrm{tot}}$, as surjectivity onto $\Sym{\mathnormal{N}}$ excludes in our examples; and let $\cgket{\Gvib}{\mu_{\mathrm v}}{\mult_{\mathrm v}}_{\mathrm v}$ be the vibrational states of Eq.~\eqref{eq:sfe_map}, lying in a manifold with $N$ total quanta. For an admissible triple with coupling tensor $C$, the \emph{tripartite momentum states} are, in the spectroscopic basis~\eqref{eq:spectroscopic_basis},
\begin{align}
  &\bigl|\Gmol;{}^{J,\parity}_{\,m}\mult_{\mathrm r},I_{\mathrm{tot}}M_I,\mult_{\mathrm v}\bigr\rangle
  \nonumber\\
  &\quad=\!\!\sum_{\mu_{\mathrm r}\mu_{\mathrm n}\mu_{\mathrm v}}\!\!
  C_{\mu_{\mathrm r}\mu_{\mathrm n}\mu_{\mathrm v}}\,
  \cgket{\Grot}{\mu_{\mathrm r}}{{}^{J,\parity}_{\,m}\mult_{\mathrm r}}_{\mathrm r}
  \cgket{\Gspin}{\mu_{\mathrm n}}{I_{\mathrm{tot}}M_I}_{\mathrm n}
  \cgket{\Gvib}{\mu_{\mathrm v}}{\mult_{\mathrm v}}_{\mathrm v},
  \label{eq:spectroscopic_states}
\end{align}
with the vibrational factor expanded in occupation numbers, $\cgket{\Gvib}{\mu_{\mathrm v}}{\mult_{\mathrm v}}_{\mathrm v}=\sum_{\mathbf n}C^{\Gvib,\mult_{\mathrm v},\mu_{\mathrm v}}_{\mathbf n}\ket{\mathbf n}$. In the momentum representation the vibrational factor is $(-\iunit)^{N}\int\diff\mathbf p\,\Psi^{\mu_{\mathrm v}}_{\Gvib,\mult_{\mathrm v}}(\mathbf p)\ket{\mathbf p}$ by Eq.~\eqref{eq:vib_fourier}, with $\Psi$ the wavefunction of the position description below.

\emph{Position basis.} The rotational position states are the coset states
\begin{equation}
  \bigl|\crep;\Grot^{\lambda\mu}\bigr\rangle
  =\sqrt{\frac{d_{\Grot}}{|\grp{G}|}}\sum_{\gel\in \grp{G}}
  \repD{\Grot}_{\lambda,\mu}(\gel)\,
  \bigl|\crep\gel\bigr\rangle,
  \qquad \crep\in \grp{O(3)}/\grp{G},
  \label{eq:coset_states}
\end{equation}
where $\ket{\crep\gel}$ is a sharp orientation in $\grp{O(3)}$. Molecule-frame operations act on the index $\mu$ exactly as $\repD{\Grot}$ acts on a standard basis and leave the fiber index $\lambda$ alone, while laboratory operations do the reverse. The nuclear-spin position states are the product states $\ket{m_1,\ldots,m_N}$ of Table~\ref{tab:phase_space}, and the vibrational position states are the normal-coordinate states $\ket{\mathbf q}$, so a vibrational carrier state becomes a wavefunction $\Psi^{\mu_{\mathrm v}}_{\Gvib,\mult_{\mathrm v}}(\mathbf q)=\bigl\langle\mathbf q\bigm\vert{}^{\Gvib}_{\,\mu_{\mathrm v}};\mult_{\mathrm v}\bigr\rangle$, a polynomial in $\mathbf q$ transforming as $\Gvib$ times the ground-state Gaussian. The \emph{tripartite position states} are
\begin{align}
  &\bigl|\crep,\lambda;I_{\mathrm{tot}}M_I;\mult_{\mathrm v}\bigr\rangle
  =\!\!\sum_{\mu_{\mathrm r}\mu_{\mathrm n}\mu_{\mathrm v}}\!\!
  C_{\mu_{\mathrm r}\mu_{\mathrm n}\mu_{\mathrm v}}\,
  \bigl|\crep;\Grot^{\lambda\mu_{\mathrm r}}\bigr\rangle
  \otimes\bigl|\chi^{\mu_{\mathrm n}}_{I_{\mathrm{tot}}M_I}\bigr\rangle
  \nonumber\\
  &\qquad\otimes\!\int\!\diff\mathbf q\,\Psi^{\mu_{\mathrm v}}_{\Gvib,\mult_{\mathrm v}}(\mathbf q)\ket{\mathbf q},
  \label{eq:position_states}
\end{align}
where $\ket{\chi^{\mu_{\mathrm n}}_{I_{\mathrm{tot}}M_I}}=\sum_{\mathbf m}\bigl\langle\mathbf m\bigm\vert{}^{\Gspin}_{\,\mu_{\mathrm n}};I_{\mathrm{tot}}M_I\bigr\rangle\ket{\mathbf m}$ is the nuclear-spin carrier written in the product basis $\ket{\mathbf m}=\ket{m_1,\ldots,m_N}$.
Applying Eq.~\eqref{eq:coupling_space} shows that the total action $\hat\rho(\hel)$ multiplies both Eq.~\eqref{eq:spectroscopic_states} and Eq.~\eqref{eq:position_states} by $\irchar{\Gmol}(\hel)$ for every $\hel\in\grp{G}$, so both are admissible. The same tensor $C$ appears in both: the coupling is a property of the carriers, not of the basis in which the carriers are written.

Passing from Eq.~\eqref{eq:position_states} to Eq.~\eqref{eq:spectroscopic_states} changes the rotational slot by the transform below. The other two slots only change representation: the nuclear spins pass from the product basis to total spin, and the vibration from $\mathbf q$ to occupation numbers, or to $\mathbf p$ with the phase $(-\iunit)^N$ of Eq.~\eqref{eq:vib_fourier}. Each of these is a $\grp{G}$-equivariant unitary within one sector, so it touches no symmetry label.

\begin{theorem}[Molecular Fourier transform]
\label{thm:fourier}
Fix an admissible triple, its coupling tensor, and the multiplicity labels $(I_{\mathrm{tot}}M_I,\mult_{\mathrm v})$. The tripartite position and momentum states are related by
\begin{align}
  \ket{\crep,\lambda;\cdots}
  &=\!\sum_{J,\parity}\sum_{m,\mult_{\mathrm r}}
   \mathcal Y^{J,\parity}_{m\mult_{\mathrm r}}(\crep,\lambda)\,
  \ket{\Gmol;{}^{J,\parity}_{\,m}\mult_{\mathrm r};\cdots},
  \label{eq:ft_forward}\\
  \ket{\Gmol;{}^{J,\parity}_{\,m}\mult_{\mathrm r};\cdots}
  &=\!\int\!\diff\crep\sum_{\lambda}
   \mathcal Y^{J,\parity}_{m\mult_{\mathrm r}}(\crep,\lambda)^{*}\,
  \ket{\crep,\lambda;\cdots},
  \label{eq:ft_inverse}
\end{align}
with the parity-graded induced harmonic
\begin{equation}
  \boxed{\;
  \mathcal Y^{J,\parity}_{m\mult_{\mathrm r}}(\crep,\lambda)=
  \sqrt{\frac{(2J+1)\,|\grp{G}|}{16\pi^{2}\,d_{\Grot}}}\;
  \irrJp{J}{\parity}_{m;\lambda\mult_{\mathrm r}}(\crep)^{*}\;}
  \label{eq:induced-harmonic}
\end{equation}  
where $\irrJp{J}{\parity}_{m;\lambda\mult_{\mathrm r}}$ is the Wigner matrix element between $\jm{J,\parity}{m}$ and the $\grp{G}$-adapted state of partner $\lambda$ and copy $\mult_{\mathrm r}$, $\int_{\grp{O(3)}/\grp{G}}\diff\crep=16\pi^{2}/|\grp{G}|$, and the sum over $(J,\parity)$ runs over the $\irrJp{J}{\parity}$ with $m^{(\mathrm r)}_{\Grot}(J,\parity)>0$. The harmonics are orthonormal and complete on $L^2(\grp{O(3)}/\grp{G})\otimes\mathbb C^{d_{\Grot}}$, the coupling and multiplicity labels are inert, and the transform is therefore block diagonal.
\end{theorem}
\begin{proof}
    See Appendix~\ref{proof:fourier}.
\end{proof}

The condition $m^{(\mathrm r)}_{\Grot}(J,\parity)>0$ is the selection rule of Eq.~\eqref{eq:frobenius} reappearing as a bandwidth condition on the transform: a position state of a given species contains only those rotational momenta that the species admits.

The discussion above gives us several observations. First, \emph{no vibrational variable is conjugate to orientation.} The vibrational sector carries its own conjugacy, normal-mode coordinate $\mathbf q$ against normal-mode momentum $\mathbf p$, and it acts on a different slot of the fiber from the orientational duality. The occupation numbers and, for a degenerate mode, the vibrational angular momentum $\ell$ belong to neither pair: they are spectroscopic labels, conjugate respectively to oscillator phases and to the angle of the vibrational displacement in the degenerate plane. 

Second, \emph{vibrational excitation selects the rotational monodromy without changing the orientational base space.} Transport around a noncontractible loop acts on the rotational carrier through \(\repD{\Grot}(\gel)\), with \(\gel\) determined by the lifted loop. One-dimensional carriers acquire a phase, whereas an \(\irr{e}\) carrier supports a two-component action. For phosphine and calcium monomethoxide, \(\grp{O(3)}/\grp{C_{3v}}\cong\grp{SO(3)}/\grp{C_{3}}\), so continuous-loop monodromy samples the Abelian restriction \(\restr{\repD{\irr{e}}}{\grp{C_{3}}}\), despite the non-Abelian action of the full point group. Vibrational preparation changes the admissible \(\Grot\) through Lemma~\ref{lemma:unified_admissibility_condition}, thereby selecting the allowed monodromy (Appendix~\ref{app:monodromy}).

\section{Symmetry-Enforced Absolutely Maximally Entangled States}
\label{sec:entang}

The rotational, nuclear-spin, and vibrational carriers give every admissible triple $(\Grot,\Gspin,\Gvib)$ a tripartite structure. Its physical states are the normalized vectors of the coupling space $\mathcal C^{\Gmol}_{\Grot\Gspin\Gvib}$ of Eq.~\eqref{eq:coupling_space},
\begin{equation}
\ket{\psi}
=
\sum_{\mu_{\mathrm r},\mu_{\mathrm n},\mu_{\mathrm v}}
C_{\mu_{\mathrm r}\mu_{\mathrm n}\mu_{\mathrm v}}
\irket{\Grot}{\mu_{\mathrm r}}
\irket{\Gspin}{\mu_{\mathrm n}}
\irket{\Gvib}{\mu_{\mathrm v}},
\label{eq:carrier_state}
\end{equation}
with each $\mu$ running over the partners of its carrier. For $N=1$ this is the state of Eq.~\eqref{eq:unique_coupled_state}. For a sector $X\in\{\mathrm r,\mathrm n,\mathrm v\}$, the reduced state of $\ket{\psi}$ is $\rho_X=\trace_{\overline X}\ketbra{\psi}{\psi}$, with $\overline X$ the other two sectors. We call $\ket{\psi}$ \emph{absolutely maximally entangled} (AME) when $\rho_X=I_{d_{\Gamma_X}}/d_{\Gamma_X}$ for each of the three sectors. With three sectors, every cut separates one sector from the other two. An AME state is thus maximally entangled across every cut, which extends the definition of Ref.~\cite{Helwig2012} to parties of unequal dimension. The cut between $X$ and $\overline X$ then carries the entanglement entropy $\log_2 d_{\Gamma_X}$, the largest value for its dimensions. Equivalently, $\sqrt{d_{\Gamma_X}}\,C$, read as a map from the carrier of $X$ to the other two, is an isometry. The tensor $C$ is thus a perfect tensor~\cite{Pastawski2015}, in a definition that allows legs of different dimensions.

\begin{theorem}[AME coupling spaces]
\label{thm:ame}
Let $(\Grot,\Gspin,\Gvib)$ be admissible for the one-dimensional representation $\Gmol$, and let $\ket{\psi}$ be any normalized vector in its $N$-dimensional coupling space. Then the reduced density matrices satisfy
\begin{equation}
\boxed{
\rho_{\mathrm r}=\frac{I_{d_{\Grot}}}{d_{\Grot}},
\qquad
\rho_{\mathrm n}=\frac{I_{d_{\Gspin}}}{d_{\Gspin}},
\qquad
\rho_{\mathrm v}=\frac{I_{d_{\Gvib}}}{d_{\Gvib}}.
}
\label{eq:marginals}
\end{equation}
Thus $\ket{\psi}$ is AME, for every $N$ and independently of the Hamiltonian that selects a state within the coupling space.
\end{theorem}

\begin{proof}
Every vector in the coupling space obeys Eq.~\eqref{eq:coupling_space}.
Since $\irchar{\Gmol}$ is one-dimensional and unitary, the projector $\rho=\ket{\psi}\!\bra{\psi}$ is invariant under conjugation by the tensor-product representation. Taking the partial trace over the nuclear-spin and vibrational carriers, whose matrices are unitary, gives
\begin{equation}
\repD{\Grot}(\gel)\rho_{\mathrm r}
\repD{\Grot}(\gel)^\dagger=\rho_{\mathrm r}.
\end{equation}
Thus $\rho_{\mathrm r}$ commutes with every matrix of the irreducible complex representation $\Grot$. Schur's lemma implies $\rho_{\mathrm r}=cI$, and $\trace\rho_{\mathrm r}=1$ fixes $c=1/d_{\Grot}$. The nuclear-spin and vibrational reductions follow by the same argument.
\end{proof}
Bryan, Leutheusser, Reichstein, and Van Raamsdonk give this argument for any state that a group fixes up to a phase, with an irreducible representation on each party~\cite{BLRV2019}. Interactions decide which species a level belongs to, and the theorem then fixes its carrier state. Only nuclear-spin terms, which are weak, can change $\Gspin$~\cite{Hougen1963, Chapovsky1999}. First-order Coriolis terms keep the triple of a symmetric top~\cite{Hougen1962} but can mix triples in a spherical top, as in the $J=2$ levels of methane. When interactions mix species with weights $p_\Gamma$, each carrier still has $S(\rho_X)\ge\sum_\Gamma p_\Gamma\log_2 d_{\Gamma_X}$, by concavity of the entropy. The theorem also fixes the kind of entanglement from the carrier dimensions alone.

\begin{corollary}[Kinds of entanglement]
\label{cor:tripartite}
Let $k$ be the number of carriers with dimension above one in an admissible triple. Every normalized vector in its coupling space is a product state if $k=0$, a maximally entangled state of two carriers of equal dimension if $k=2$, and a genuinely tripartite entangled state if $k=3$. No admissible triple has $k=1$.
\end{corollary}

\begin{proof}
By Theorem~\ref{thm:ame}, the cut between a carrier and the other two has Schmidt rank equal to the dimension of that carrier. If $k=1$, every vector would be a product state with a pure marginal on the degenerate carrier, in contradiction with Theorem~\ref{thm:ame}. If $k=2$, the one-dimensional carrier factors out, and the other two carriers share a pure state with flat marginals of equal rank. A pure state is genuinely tripartite entangled exactly when it is entangled across all three cuts, which for $k=3$ follows from the Schmidt ranks.
\end{proof}

For $k=3$, the coupling space is a genuinely entangled subspace~\cite{Cubitt2008,Demianowicz2018}: it contains no biseparable vector. Every genuinely entangled subspace is a completely entangled subspace~\cite{Wallach2002,Parthasarathy2004,Parthasarathy2005,Walgate2008}, one with no fully product vector. Previous work has identified completely rotation--nuclear-spin entangled molecular state spaces arising from symmetry and identical-particle statistics~\cite{albert2025quantumtheorymolecularorientations}. Entangled rovibrational and nuclear-spin wave functions appear already in Refs.~\cite{Pfister1996, BordeChardonnet2000}. Here we extend this principle to the coupled rotational, nuclear-spin, and vibrational carriers. A nondegenerate vibrational carrier has $d_{\Gvib}=1$ and necessarily factors out. The two-sector construction of Ref.~\cite{albert2025quantumtheorymolecularorientations}, which fixes the vibrational carrier to the trivial irrep, therefore cannot support genuine tripartite entanglement among the three carriers.

Each AME state has the smallest norm on its orbit under stochastic local operations and classical communication (SLOCC)~\cite{PhysRevA.65.052112}, here the invertible local operators of unit determinant. By the Kempf--Ness theorem, two AME states on the same SLOCC orbit are related by local unitaries (LU)~\cite{BLRV2019}. We thus classify coupling tensors up to LU and the order of the carriers. We call the carrier dimensions of a triple, sorted in increasing order, its \emph{format}. Bryan, Reichstein, and Van Raamsdonk, who call AME states locally maximally entangled, give the dimension of the space of these states up to LU for each format~\cite{BRV2018}. In the formats $(1,d,d)$, $(2,2,2)$, and $(2,3,3)$, this space is a single point, so the AME property alone fixes the state. These states are the product state, the maximally entangled pairs, the three-qubit GHZ state~\cite{DurVidalCirac2000}, and the trine-GHZ state of a qubit and two qutrits. In the other formats of Table~\ref{tab:ame_catalogue}, the AME states form continuous families, from which the point groups pick particular members.

Table~\ref{tab:ame_catalogue} lists the states of the coupling tensors of the 75 point groups of Altmann and Herzig~\cite{AltmannHerzig2011}, as tabulated in the \PGAtlas~\cite{o3harmonics}. For an $N=2$ space, the table lists its two members of definite $K$, the total angular momentum of a pair of equal carriers. Up to LU and the order of the carriers, there are exactly 21: the product state $\catP$, four maximally entangled pairs $\catB{1}$ to $\catB{4}$, and 16 tripartite states $\catT{1}$ to $\catT{16}$. Symmetry thus enforces an AME state of each kind of entanglement in Corollary~\ref{cor:tripartite}. Only $\catP$, $\catB{1}$, $\catB{2}$, and $\catT{1}$ to $\catT{4}$ occur outside the icosahedral groups $\grp{I}$ and $\grp{I_{h}}$. For the linear groups $\grp{C_{\infty v}}$ and $\grp{D_{\infty h}}$, the rotation--bend pair $\catB{1}$ depends on the factorization: in the frame of Hougen, Bunker, and Johns, the parity states of a $\Pi$ level are product states~\cite{HougenBunkerJohns1970}. Reference~\cite{DasRepo2026} gives every coupling tensor, the triples of each state in each group, and the checks of this classification.

\begin{table}[t]
\centering
\footnotesize
\setlength{\tabcolsep}{4pt}
\begin{tabular}{lccll}
\toprule
id & format & $N$ & state & groups\\
\midrule
$\catP$ & $(1,1,1)$ & 1 & product & all\\
\midrule
$\catB{1}$ & $(1,2,2)$ & 1 & pair & axial, $\grp{T_{d}}$, $\grp{O}$, $\grp{O_{h}}$\\
$\catB{2}$ & $(1,3,3)$ & 1 & pair & cubic, icos.\\
$\catB{3}$ & $(1,4,4)$ & 1 & pair & icos.\\
$\catB{4}$ & $(1,5,5)$ & 1 & pair & icos.\\
\midrule
$\catT{1}$ & $(2,2,2)$ & 1 & GHZ & axial, $\grp{T_{d}}$, $\grp{O}$, $\grp{O_{h}}$\\
$\catT{2}$ & $(2,3,3)$ & 1 & trine-GHZ & $\grp{T_{d}}$, $\grp{O}$, $\grp{O_{h}}$\\
$\catT{3}$ & $(3,3,3)$ & 1, 2 & $\epsilon_{ijk}$ & cubic, icos.\\
$\catT{4}$ & $(3,3,3)$ & 1, 2 & $|\epsilon_{ijk}|$ & cubic\\
$\catT{5}$ & $(3,3,4)$ & 1 & $\irr{t}_{1}\irrprod\irr{t}_{2}\supset\irr{g}$ & icos.\\
$\catT{6}$ & $(3,3,5)$ & 1 & $3j(1,1,2)$ & icos.\\
$\catT{7}$ & $(3,3,5)$ & 1 & $\irr{t}_{1}\irrprod\irr{t}_{2}\supset\irr{h}$ & icos.\\
$\catT{8}$ & $(3,4,4)$ & 1 & 't~Hooft symbol & icos.\\
$\catT{9}$ & $(3,4,5)$ & 1 & $\irr{t}_{1}\irrprod\irr{g}\supset\irr{h}$ & icos.\\
$\catT{10}$ & $(3,5,5)$ & 1 & $3j(1,2,2)$ & icos.\\
$\catT{11}$ & $(4,4,4)$ & 1 & simplex cubic & icos.\\
$\catT{12}$ & $(4,4,5)$ & 1 & $\irr{g}\irrprod\irr{g}\supset\irr{h}$ & icos.\\
$\catT{13}$ & $(4,5,5)$ & 2 & $K=3$ & icos.\\
$\catT{14}$ & $(4,5,5)$ & 2 & $K=4$ & icos.\\
$\catT{15}$ & $(5,5,5)$ & 2 & $3j(2,2,2)$, $K=2$ & icos.\\
$\catT{16}$ & $(5,5,5)$ & 2 & $K=4$ & icos.\\
\bottomrule
\end{tabular}

\caption{The 21 AME states of the coupling tensors of the 75 point groups, up to LU and the order of the carriers. The format lists the carrier dimensions in increasing order, and $N$ is the dimension of the coupling space. In the last column, axial means the non-Abelian axial groups together with $\grp{C_{\infty v}}$ and $\grp{D_{\infty h}}$. Cubic means $\grp{T}$, $\grp{T_{h}}$, $\grp{T_{d}}$, $\grp{O}$, and $\grp{O_{h}}$, and icos.\ means $\grp{I}$ and $\grp{I_{h}}$. The state $\catT{1}$ does not occur in $\grp{C_{4v}}$, $\grp{D_{4}}$, $\grp{D_{2d}}$, or $\grp{D_{4h}}$. The states $\catT{3}$ and $\catT{4}$ have $N=2$ in $\grp{T}$ and $\grp{T_{h}}$ and $N=1$ elsewhere. For $N=2$ the states listed have definite $K$, the total angular momentum of a pair of equal carriers. Trine-GHZ is $\tfrac{1}{\sqrt3}\sum_{i}\ket{a_i}\ket{i}\ket{i}$, with $\ket{a_i}$ the three trine states of a qubit. Reference~\cite{DasRepo2026} gives every tensor and the triples of each state.}
\label{tab:ame_catalogue}

\end{table}

Several of these states are Wigner $3j$ symbols. This happens when each carrier is a whole irrep of $\grp{O(3)}$ restricted to $\grp{G}$ and $\Gmol$ is the character of the $3j$ symbol. That character is $1$ on the rotations and the product of the three parities $\parity$ on the improper elements. In methane, $\irr{t}_{2}$ is the vector irrep $\irrJp{1}{-}$. The triple $(\irr{t}_{2},\irr{t}_{2},\irr{t}_{2})$ with $\Gmol=\irr{a}_{2}$ thus carries $3j(1,1,1)$, the Levi-Civita state $\catT{3}$~\cite{FitziGisinMaurer2001, Cabello2002}. In $\grp{I}$, whose irreps $\irr{t}_{1}$ and $\irr{h}$ are $\irrJ{1}$ and $\irrJ{2}$, the symbols $3j(1,1,2)$, $3j(1,2,2)$, and $3j(2,2,2)$ give $\catT{6}$, $\catT{10}$, and $\catT{15}$. More generally, each tripartite state of Table~\ref{tab:ame_catalogue} comes from a $3j$ symbol for one of its triples~\cite{DasRepo2026}. There the state is the restriction of that symbol to three carriers, each contained once in an irrep of $\grp{O(3)}$.

The molecules of Sec.~\ref{sec:examples} realize seven of these states. In $\grp{C_{3v}}$, the triple $(\irr{e},\irr{e},\irr{e})$ has format $(2,2,2)$, so its coupling tensor is the GHZ state $\catT{1}$. Phosphine and calcium monomethoxide host this state at definite parity, along with $\catP$ and $\catB{1}$, whose $\mathrm{CaOCH_3}$ form Yu et al.\ wrote as the symmetry-allowed states~\cite{Yu2019, YuHutzler2021}. Reference~\cite{albert2025quantumtheorymolecularorientations}, which uses proper rotations only, treats such molecules in $\grp{C_{3}}$, where every nuclear-spin isomer is separable. Here $\catT{1}$ enters through the reflections, so a superposition of its two degenerate parity partners can be a product state. Methane, in $\grp{T_{d}}$, hosts these three states together with $\catB{2}$, the trine-GHZ state $\catT{2}$, and the $(3,3,3)$ states $\catT{3}$ and $\catT{4}$. The Levi-Civita state $\catT{3}$ also follows from a determinant identity. If $\Grot\cong\Gspin\cong\Gvib\cong\Gamma$ with $\repD{\Gamma}(\gel)\in\grp{O(3)}$ and $\irchar{\Gmol}(\gel)=\det\repD{\Gamma}(\gel)$, then $\sum_{ijk}D_{ai}D_{bj}D_{ck}\epsilon_{ijk}=\det(D)\,\epsilon_{abc}$ puts $\ket{\epsilon}=\tfrac{1}{\sqrt{6}}\sum_{ijk}\epsilon_{ijk}\irket{\Gamma}{i}_{\mathrm r}\irket{\Gamma}{j}_{\mathrm n}\irket{\Gamma}{k}_{\mathrm v}$ in the coupling space.

Theorem~\ref{thm:ame} covers $N=2$ as well, so every member of such a coupling space is AME. Among the 75 groups, $N=2$ occurs only for $(\irr{t},\irr{t},\irr{t})$ in $\grp{T}$ and $\grp{T_{h}}$ and for $(\irr{g},\irr{h},\irr{h})$ and $(\irr{h},\irr{h},\irr{h})$ in $\grp{I}$ and $\grp{I_{h}}$, up to order and parity labels. Symmetry then fixes a two-dimensional subspace of AME states. 
Each such space is spanned by two states of Table~\ref{tab:ame_catalogue}, its members of definite $K$. The pairs are $\catT{3}$ and $\catT{4}$ for $(\irr{t},\irr{t},\irr{t})$, $\catT{13}$ and $\catT{14}$ for $(\irr{g},\irr{h},\irr{h})$, and $\catT{15}$ and $\catT{16}$ for $(\irr{h},\irr{h},\irr{h})$. Any superposition of these two states is allowed, and the molecular Hamiltonian picks the physical one. The proof of Theorem~\ref{thm:ame} extends to pairs of members: every operator on a single carrier has the same expectation value in all members and zero matrix elements between orthogonal ones~\cite{Knill1997}. Each such space is thus a pure quantum code of distance two on the three carriers~\cite{DasRepo2026}, in the sense of Refs.~\cite{Scott2004, HuberGrassl2020, WangFeng2012}. Gauging a symmetry yields codes of this kind as well~\cite{BLRV2019}.

\begin{corollary}[Representation independence]
\label{cor:rep_independence}
The position and momentum descriptions of an admissible carrier state, Eqs.~\eqref{eq:position_states} and~\eqref{eq:spectroscopic_states}, are related by unitary transformations local to the rotational, nuclear-spin, and vibrational sectors. Consequently, the AME property and the state of Table~\ref{tab:ame_catalogue} are the same in these descriptions.
\end{corollary}

\begin{proof}
By Theorem~\ref{thm:fourier} and the sector-local changes of basis to the spectroscopic states, the two descriptions differ by a local unitary $U_{\mathrm r}\otimes U_{\mathrm n}\otimes U_{\mathrm v}$. Such a unitary preserves the reduced spectra and the LU class of a state.
\end{proof}

These statements concern the internal sectors of a single molecule, not spatially separated laboratories, and they assume a fixed decomposition into rotation, nuclear spin, and vibration. This decomposition is the one on which the zeroth-order Hamiltonian is local, with each sector addressed by its own interactions~\cite{ZanardiLidarLloyd2004}. Only its cut between the nuclear spins and the rest is free of a frame choice, since the Eckart frame fixes the split between rotation and vibration. Relabeling plays the role of a gauge group, and in lattice gauge theory the analogous $\log d$ term of a cut is attributed to the embedding of the physical Hilbert space~\cite{VanAcoleyen2016}. Vaccaro et al.\ identify the same term as a correlation forced by superselection~\cite{Vaccaro2008}. The carrier entanglement still has consequences: it is why no probe of one sector sees the tensor, and it makes each $N=2$ space a code of distance two. We treat free rotors, since for deeply hindered rotors a description by localized pocket states~\cite{Huller1977} can remove this term. Symmetry does not imply that a thermal ensemble occupies a pure state of one coupling space. Probes of one sector act on each carrier as scalars, so they cannot witness this entanglement. Certifying it in a prepared molecule therefore needs a mixed-state analysis and a measurement of the species triple, which fixes the marginals through Theorem~\ref{thm:ame}.

\section{Examples}
\label{sec:examples}

We work through three molecules with increasing point-group size, $\mathrm{H_2O}$ ($\grp{C_{2v}}$), $\mathrm{PH_3}$ ($\grp{C_{3v}}$), and $\mathrm{CH_4}$ ($\grp{T_{d}}$). In addition, in Sec.~\ref{sec:experiment}, concrete pulses for state preparation will be discussed for the laser-coolable symmetric top $\mathrm{CaOCH_3}$ ($\grp{C_{3v}}$), which has the same point group and proton-spin structure as $\mathrm{PH_3}$. Each of the first three examples follows four steps below:
\begin{enumerate}
\item \emph{Species table.} The admissible triples $(\Grot,\Gspin,\Gvib)$ with $N\neq0$ from Eq.~\eqref{eq:triple_coupling}, their carrier dimensions and AME state (Table~\ref{tab:ame_catalogue}), and the lowest rotational manifold $\irrJp{J}{\parity}$, nuclear-spin multiplet, and vibrational manifold realizing each factor.
\item \emph{Tripartite position states}, Eq.~\eqref{eq:position_states}: coset states on $\grp{O(3)}/\grp{G}$, nuclear spins in the product basis, and vibrational wavefunctions in normal coordinates.
\item \emph{Tripartite momentum states}, Eq.~\eqref{eq:spectroscopic_states}: $\grp{G}$-adapted $\ket{{}^{J,\parity}_{\,m},k}$ combinations, total-spin states, and occupation-number states.
\item \emph{A worked instance} with the quantum numbers, harmonic frequencies, and transform coefficients inserted.
\end{enumerate}
Table~\ref{tab:molecular_examples} collects the input data. Vibrational wavefunctions are written in dimensionless normal coordinates $q_s=Q_s/b_s$, with $Q_s$ the mass-weighted coordinate and $b_s=(\hbar/\omega_s)^{1/2}$ the zero-point length, and $\psi_0(\mathbf q)=\pi^{-f/4}\expe^{-|\mathbf q|^2/2}$ is the vibrational ground state. Every coupling tensor below was obtained by projection onto Eq.~\eqref{eq:coupling_space} with the standard matrices stated in each subsection, and every explicit state was checked numerically to transform with $\chi_{\Gmol}$ under the full action of Eq.~\eqref{eq:total_group_action}.

\begin{table*}[htbp!]
\caption{Symmetry data for the four molecules treated here. Fundamentals are observed band origins in $\mathrm{cm^{-1}}$~\cite{Shimanouchi1972,NISTCaOCH3}; for $\mathrm{CaOCH_3}$, $\nu_7$, used in Appendix~\ref{app:caoch3}, has no band origin in Ref.~\cite{NISTCaOCH3}. $\mathrm{CaOCH_3}$ is placed like $\mathrm{PH_3}$ (Sec.~\ref{subsec:ammonia}), with the $z$ axis along the Ca--O--C axis, proton $j$ at azimuth $2\pi(j-1)/3$, and the same $\elt{C_{3}}$, $\elt{\sigma_{v}}(0)$, and $\irr{e}$ matrices. $\mathcal H^{\mathrm{(n)}}_{\spin}$ is the decomposition of Eq.~\eqref{eq:nuclear_spin_multiplicity} for the permuted protons; $\Gmol$ is the species required by Eq.~\eqref{eq:gamma_mol}. Statistical weights are the dimensions $d_\Gamma m^{\mathrm{(n)}}_\Gamma$ of the spin sectors.}
\label{tab:molecular_examples}
\centering
\small
\begin{tabular}{llllll}
\toprule
Molecule & $\grp{G}$ & fundamentals $\nu_s(\Gamma_s)$ & $\mathcal H^{\mathrm{(n)}}_{\spin}$
& $\Gmol$ & spin species (weight)\\
\midrule
$\mathrm{H_2O}$ & $\grp{C_{2v}}$ & $\nu_1(\irr{a}_{1})\,3657$, $\nu_2(\irr{a}_{1})\,1595$, $\nu_3(\irr{b}_{2})\,3756$ & $3\irr{a}_{1}\oplus \irr{b}_{2}$ & $\irr{b}_{2}$
& ortho $I{=}1$ (3), para $I{=}0$ (1)\\
$\mathrm{PH_3}$ & $\grp{C_{3v}}$ & $\nu_1(\irr{a}_{1})\,2323$, $\nu_2(\irr{a}_{1})\,992$, $\nu_3(\irr{e})\,2328$, $\nu_4(\irr{e})\,1118$ & $4\irr{a}_{1}\oplus 2\irr{e}$ & $\irr{a}_{2}$
& ortho $I{=}\tfrac32$ (4), para $I{=}\tfrac12$ (4)\\
$\mathrm{CH_4}$ & $\grp{T_{d}}$ & $\nu_1(\irr{a}_{1})\,2917$, $\nu_2(\irr{e})\,1534$, $\nu_3(\irr{t}_{2})\,3019$, $\nu_4(\irr{t}_{2})\,1306$ & $5\irr{a}_{1}\oplus \irr{e}\oplus 3\irr{t}_{2}$ & $\irr{a}_{2}$
& meta $I{=}2$ (5), para $I{=}0$ (2), ortho $I{=}1$ (9)\\
$\mathrm{CaOCH_3}$ & $\grp{C_{3v}}$ & $\nu_3(\irr{a}_{1})\,1156$, $\nu_4(\irr{a}_{1})\,488$, $\nu_7(\irr{e})$, $\nu_8(\irr{e})\,144$ & $4\irr{a}_{1}\oplus 2\irr{e}$ & $\irr{a}_{2}$
& ortho $I{=}\tfrac32$ (4), para $I{=}\tfrac12$ (4)\\
\bottomrule
\end{tabular}
\end{table*}
For a detailed construction, in Appendix \ref{subsec:tour} we follow a single symmetry operation through vibration alone, rotation and vibration, rotation and nuclear spin, and all three sectors together, so that each ingredient of the construction can be seen acting on explicit states.

\subsection{Abelian group example: Water}
\label{subsec:water}

Water has $\grp{G}=\grp{C_{2v}}$, two identical protons, and three normal modes $\nu_1(\irr{a}_{1})$, $\nu_2(\irr{a}_{1})$, $\nu_3(\irr{b}_{2})$, with the molecule in the $yz$ plane and the $\elt{C_{2}}$ axis along $z$. The $\elt{C_{2}}$ operation and the reflection $\elt{\sigma_{v}}$ in the $xz$ plane each exchange the protons, so Eq.~\eqref{eq:gamma_mol} gives $\chi_{\Gmol}=(1,-1,-1,1)$ on $\{\cls{E},\cls{C_{2}},\cls{\sigma_{v}},\cls{\sigma_{d}}\}$, with $\elt{\sigma_{d}}$ the reflection in the molecular ($yz$) plane, i.e.\ $\Gmol=\irr{b}_{2}$. The cycle counts $c[p(\gel)]=(2,1,1,2)$ give $\chi^{\mathrm{(n)}}_{\spin}=(4,2,2,4)$ and
\begin{equation}
\mathcal H^{\mathrm{(n)}}_{\spin}(\mathrm{H_2O})\cong 3\irr{a}_{1}\oplus \irr{b}_{2} ,
\end{equation}
the $3\irr{a}_{1}$ sector carrying the $I_{\mathrm{tot}}=1$ ortho triplet, with $\mult_{\mathrm n}=M_I$, and the $\irr{b}_{2}$ sector the $I_{\mathrm{tot}}=0$ para singlet. Every vibrational state is a product of nondegenerate oscillators, of species $\irr{a}_{1}$ when $n_3$ is even and $\irr{b}_{2}$ when $n_3$ is odd, so no vibrational state of water has species $\irr{a}_{2}$ or $\irr{b}_{1}$.

\emph{Species table.} Because $\grp{C_{2v}}$ is Abelian, Corollary~\ref{cor:shift} fixes $\Grot=\irr{b}_{2}\otimes\Gspin\otimes\Gvib$ uniquely, and with $\Gspin,\Gvib\in\{\irr{a}_{1},\irr{b}_{2}\}$ four triples survive (Table~\ref{tab:water_species}). A corollary is that rotational states of species $\irr{a}_{2}$ and $\irr{b}_{1}$, for instance the $J=0$, $\parity=-1$ state, are absent from water in every vibrational state.

\begin{table}[h]
\caption{Admissible species of $\mathrm{H_2O}$, $\Gmol=\irr{b}_{2}$. All carriers are one-dimensional, so every coupling tensor is the product state $\catP$. The last three columns give the lowest realization of each factor: rotational manifold $\irrJp{J}{\parity}$ with its $k$ parity (along the $\elt{C_{2}}$ axis), nuclear-spin species, and vibrational manifold.}
\label{tab:water_species}
\centering
\begin{tabular}{cccccc}
\toprule
$\Grot$ & $\Gspin$ & $\Gvib$ & $\irrJp{J}{\parity}$ ($k$) & spin & vib.\\
\midrule
$\irr{b}_{2}$ & $\irr{a}_{1}$ & $\irr{a}_{1}$ & $\irrJp{1}{+},\irrJp{1}{-}$ ($k$ odd) & ortho & ground\\
$\irr{a}_{1}$ & $\irr{b}_{2}$ & $\irr{a}_{1}$ & $\irrJp{0}{+}$ ($k$ even) & para & ground\\
$\irr{a}_{1}$ & $\irr{a}_{1}$ & $\irr{b}_{2}$ & $\irrJp{0}{+}$ ($k$ even) & ortho & $\nu_3$\\
$\irr{b}_{2}$ & $\irr{b}_{2}$ & $\irr{b}_{2}$ & $\irrJp{1}{+},\irrJp{1}{-}$ ($k$ odd) & para & $\nu_3$\\
\bottomrule
\end{tabular}
\end{table}

In the vibrational ground state ortho requires $\Grot=\irr{b}_{2}$ and para requires $\Grot=\irr{a}_{1}$. With one excitation in $\nu_3$, $\Gvib=\irr{b}_{2}$ and the allowed rotational species \emph{interchange}: ortho now pairs with $\Grot=\irr{a}_{1}$ and para with $\Grot=\irr{b}_{2}$. The rotational level pattern of the $\nu_3$ band is the complement of the ground-state pattern, level by level. With only $\Grot\otimes\Gspin$ constrained, the two-sector treatment predicts the same pattern in both bands, and is wrong for any band whose upper state carries a non-totally-symmetric species, as seen from Corollary~\ref{cor:shift}.

\emph{Position states.} For one-dimensional $\Grot$ the coset states~\eqref{eq:coset_states} have no fiber index,
\begin{align}
\ket{\crep;\Grot}=\tfrac12\bigl[&\ket{\crep}+\chi_{\Grot}(\elt{C_{2}})\ket{\crep\elt{C_{2}}}\nonumber\\
&+\chi_{\Grot}(\elt{\sigma_{v}})\ket{\crep\elt{\sigma_{v}}}+\chi_{\Grot}(\elt{\sigma_{d}})\ket{\crep\elt{\sigma_{d}}}\bigr],
\label{eq:water_coset}
\end{align}
so $\ket{\crep;\irr{b}_{2}}$ carries the sign pattern $(+,-,-,+)$ over the four orbit points and $\ket{\crep;\irr{a}_{1}}$ the pattern $(+,+,+,+)$. The nuclear-spin position states are $\ket{\uparrow\uparrow}$, $\tfrac{1}{\sqrt{2}}\,(\ket{\uparrow\downarrow}+\ket{\downarrow\uparrow})$, $\ket{\downarrow\downarrow}$ (ortho) and $\tfrac{1}{\sqrt{2}}\,(\ket{\uparrow\downarrow}-\ket{\downarrow\uparrow})$ (para), and the vibrational wavefunctions are $\Psi_{\mathbf n}(\mathbf q)=\prod_s\psi_{n_s}(q_s)$ with $\psi_n$ the Hermite functions. The four species of Table~\ref{tab:water_species} are single products. With $M_I=1$ for ortho,
\begin{align}
\text{ortho, ground:}&\quad \ket{\crep;\irr{b}_{2}}\otimes\ket{\uparrow\uparrow}\otimes\psi_0(\mathbf q),\nonumber\\
\text{para, ground:}&\quad \ket{\crep;\irr{a}_{1}}\otimes\tfrac{1}{\sqrt{2}}\,(\ket{\uparrow\downarrow}-\ket{\downarrow\uparrow})\otimes\psi_0(\mathbf q),\nonumber\\
\text{ortho, }\nu_3:&\quad \ket{\crep;\irr{a}_{1}}\otimes\ket{\uparrow\uparrow}\otimes\sqrt2\,q_3\,\psi_0(\mathbf q),\nonumber\\
\text{para, }\nu_3:&\quad \ket{\crep;\irr{b}_{2}}\otimes\tfrac{1}{\sqrt{2}}\,(\ket{\uparrow\downarrow}-\ket{\downarrow\uparrow})\otimes\sqrt2\,q_3\,\psi_0(\mathbf q).
\label{eq:water_position}
\end{align}
In the position picture the species interchange is a transfer of sign: the proton-exchanging operations $\elt{C_{2}}$ and $\elt{\sigma_{v}}$ reverse $q_3$, so an odd vibrational wavefunction removes the sign they would otherwise require of the orbit weights, and the rotational coset switches from the $\irr{b}_{2}$ to the $\irr{a}_{1}$ pattern.

\emph{Momentum states.} Quantizing $k$ along the $\elt{C_{2}}$ axis, Eq.~\eqref{eq:rot_action} gives $\elt{C_{2}}\ket{k}=(-1)^k\ket{k}$ and $\elt{\sigma_{d}}\ket{{}^{J,\parity}_{\,m},k}=(-1)^J\parity\ket{{}^{J,\parity}_{\,m},-k}$, so the adapted rotational states are
\begin{align}
\ket{\irr{a}_{1};{}^{J,\parity}_{\,m} k}&=\tfrac{1}{\sqrt{2}}\,\bigl(\ket{{}^{J,\parity}_{\,m},k}+(-1)^J\parity\ket{{}^{J,\parity}_{\,m},-k}\bigr),\quad k>0\text{ even},\nonumber\\
\ket{\irr{b}_{2};{}^{J,\parity}_{\,m} k}&=\tfrac{1}{\sqrt{2}}\,\bigl(\ket{{}^{J,\parity}_{\,m},k}+(-1)^J\parity\ket{{}^{J,\parity}_{\,m},-k}\bigr),\quad k\text{ odd},
\label{eq:water_rot}
\end{align}
together with $\ket{{}^{J,\parity}_{\,m},0}$, which is $\irr{a}_{1}$ for $(-1)^J\parity=+1$. The opposite relative signs give the absent $\irr{a}_{2}$ and $\irr{b}_{1}$ states. The tripartite momentum states are the products
\begin{equation}
\ket{\irr{b}_{2};{}^{J,\parity}_{\,m}k;I_{\mathrm{tot}}M_I;\mathbf n}
=\ket{\Grot;{}^{J,\parity}_{\,m} k}\otimes\ket{I_{\mathrm{tot}},M_I}\otimes\ket{n_1,n_2,n_3},
\label{eq:water_momentum}
\end{equation}
with $(\Grot,I_{\mathrm{tot}},n_3\bmod 2)$ taken from a row of Table~\ref{tab:water_species}.

\emph{Worked instance.} At $J=1$ the ortho species of the ground state has the two states, 
\begin{align*}
    \ket{\irr{b}_{2};{}^{1,\pm}_{\,m}1}=\tfrac{1}{\sqrt{2}}\,(\ket{k{=}1} \mp \ket{k{=}{-1}}),
\end{align*}
which transform under molecule-frame operations as the out-of-plane ($x$) and in-plane ($y$) components of a vector. They are the two ortho levels $1_{10}$ and $1_{01}$, with zero angular-momentum projection on the out-of-plane $c$ axis and the in-plane $a$ axis respectively, while the single para level $1_{11}$ is $\ket{{}^{1,-}_{\,m},0}$ of species $\irr{a}_{1}$. By Theorem~\ref{thm:fourier}, with $|\grp{G}|=4$ and $d_{\Grot}=1$, these enter the ortho position state with
\begin{equation}
\mathcal Y^{1,\parity}_{m}(\crep)=\frac{\sqrt3}{2\pi}\cdot\frac{1}{\sqrt2}\,\bigl[\irrJp{1}{\parity}_{m,1}(\crep)^{*}-\parity\,\irrJp{1}{\parity}_{m,-1}(\crep)^{*}\bigr],
\end{equation}
which at the reference orientation $\crep=\elt{E}$ equals $0.195\,[\delta_{m,1}-\parity\,\delta_{m,-1}]$. After a $\nu_3$ excitation ($b_3=0.095\,\mathrm{amu^{1/2}\,\text{\AA}}$), the same ortho spin state pairs instead with $\irr{a}_{1}$ rotational states, beginning with $0_{00}$: the lowest ortho level of the $\nu_3$ band has $J=0$.

\subsection{Non-Abelian pyramidal group: Phosphine}
\label{subsec:ammonia}

Phosphine has $\grp{G}=\grp{C_{3v}}$, three identical protons, and six normal modes $\nu_1(\irr{a}_{1})$, $\nu_2(\irr{a}_{1})$, $\nu_3(\irr{e})$, $\nu_4(\irr{e})$. As in Sec.~\ref{subsec:tour}, we take $\elt{C_{3}}=\rotax{z}(2\pi/3)$, the reflection $\elt{\sigma_{v}}(0)=\elt{i}\,\rotax{y}(\pi)$ in the $xz$ plane, proton $j$ at azimuth $2\pi(j-1)/3$ (the \PGAtlas's orbit point $j-1$), and the standard $\irr{e}$ matrices $D^{\irr{e}}(\elt{C_{3}})=\diag(\rootunity{3},\rootunity{3}^{2})$, $D^{\irr{e}}(\elt{\sigma_{v}}(0))=\bigl(\begin{smallmatrix}0&1\\1&0\end{smallmatrix}\bigr)$, $\rootunity{3}=\expe^{2\pi\iunit/3}$ (by the \PGAtlas{} conventions of Sec.~\ref{sec:bases_kin}, the \PGAtlas tabulates the first of these matrices at $\elt{C_{3}^{2}}$), so that $\elt{C_{3}}\irket{\irr{e}}{c}=\rootunity{3}^{c}\irket{\irr{e}}{c}$ and $\elt{\sigma_{v}}(0)$ exchanges $\irket{\irr{e}}{1}\leftrightarrow\irket{\irr{e}}{2}$. Each $\elt{C_{3}}$ is an even permutation and each $\elt{\sigma_{v}}$ an odd one, so $\Gmol=\irr{a}_{2}$. The cycle counts $c=(3,1,2)$ give $\chi^{\mathrm{(n)}}_{\spin}=(8,2,4)$ on $\{\cls{E},\cls{2C_{3}},\cls{3\sigma_{v}}\}$ and
\begin{equation}
\mathcal H^{\mathrm{(n)}}_{\spin}(\mathrm{PH_3})\cong 4\irr{a}_{1}\oplus 2\irr{e} .
\end{equation}
Explicitly, with $\ket{\downarrow_j}$ ($\ket{\uparrow_j}$) the product state with proton $j$ down (up) and the other two up (down),
\begin{align}
\irr{a}_{1}:&\quad \ket{\tfrac32,\tfrac32}=\ket{\uparrow\uparrow\uparrow},\quad \ket{\tfrac32,\tfrac12}=\tfrac{1}{\sqrt{3}}\,\textstyle\sum_j\ket{\downarrow_j},\ \nonumber\\
&\quad\ket{\tfrac32,-\tfrac12}=\tfrac{1}{\sqrt3}\textstyle\sum_j\ket{\uparrow_j},\quad\ket{\tfrac32, -\tfrac32}=\ket{\downarrow\downarrow\downarrow}
\nonumber\\
\irr{e}:&\quad \cgket{\irr{e}}{c}{\tfrac12,+\tfrac12}=\tfrac{1}{\sqrt{3}}\,\textstyle\sum_{j=1}^{3}\rootunity{3}^{-c(j-1)}\ket{\downarrow_j},\nonumber\\
&\quad \cgket{\irr{e}}{c}{\tfrac12,-\tfrac12}=\tfrac{1}{\sqrt{3}}\,\textstyle\sum_{j=1}^{3}\rootunity{3}^{-c(j-1)}\ket{\uparrow_j}.
\label{eq:nh3_spin}
\end{align}
The four-dimensional $\irr{a}_{1}$ multiplicity space is the $I_{\mathrm{tot}}=\tfrac32$ ortho quartet, and the $\irr{e}$ sector is the pair of $I_{\mathrm{tot}}=\tfrac12$ para doublets, the doublet label being the carrier index $c$ and $M_I$ the multiplicity index. The $^{31}$P spin is an unpermuted spectator factor $\mathbb C^2$. The ortho and para weights are equal, as observed. $\irr{a}_{2}$ is absent from the nuclear-spin sector: admissibility can be met only by correlating sectors.

Phosphine is a rigid pyramid. Its inversion splittings should become observable only from the umbrella overtone $4\nu_2$ upward~\cite{SousaSilva2016}, and a sub-Doppler search in $3\nu_2$ found none~\cite{OkudaSasada2018}. The $\parity=\pm1$ partners of Eq.~\eqref{eq:O3_wigner}, distinct points of the momentum basis of Table~\ref{tab:phase_space}, are therefore degenerate, as in the rigid $\grp{C_{3v}}$ description used here. They pair with the same nuclear-spin species and, for $k\neq0$, carry the same $\Grot$ (Table~\ref{tab:ammonia_species}). Hyperfine terms lift this degeneracy at the kilohertz scale, so that the two parity members of a $\nu_4$ level show different hyperfine structures~\cite{Butcher1993, Pfister1996}. A $k=0$ level of the ground state has a single allowed parity (Sec.~\ref{subsec:tour}).

\emph{Species table.} Exhaustive enumeration gives eight triples, all with $N=1$ (Table~\ref{tab:ammonia_species}). Every $\Gvib$ occurs: $\irr{a}_{1}$ in the ground state, $\irr{e}$ in the fundamentals of $\nu_3$ and $\nu_4$, and $\irr{a}_{2}$ first in $3\nu_4$ (harmonic estimate $3354\ \mathrm{cm^{-1}}$), where $\operatorname{Sym}^{3}\irr{e}=\irr{a}_{1}\oplus \irr{a}_{2}\oplus \irr{e}$~\cite{o3harmonics}.

\begin{table}[h]
\caption{Admissible species of $\mathrm{PH_3}$, $\Gmol=\irr{a}_{2}$, with $j\in\{1,2\}$. The state column gives the AME state of Table~\ref{tab:ame_catalogue}, with the two entangled sectors of $\catB{1}$ in parentheses. The last two columns give the lowest rotational and vibrational manifolds realizing each factor ($\Gspin=\irr{a}_{1}$ is ortho, $\irr{e}$ para).}
\label{tab:ammonia_species}
\centering
\begin{tabular}{cccclll}
\toprule
$\Grot$ & $\Gspin$ & $\Gvib$ & $(d_{\mathrm{r}},d_{\mathrm{n}},d_{\mathrm{v}})$ & state & $\irrJp{J}{\parity}$ & vib.\\
\midrule
$\irr{a}_{2}$ & $\irr{a}_{1}$ & $\irr{a}_{1}$ & $(1,1,1)$ & $\catP$ & $\irrJp{0}{-}$ & ground\\
$\irr{a}_{1}$ & $\irr{a}_{1}$ & $\irr{a}_{2}$ & $(1,1,1)$ & $\catP$ & $\irrJp{0}{+}$ & $3\nu_4$\\
$\irr{a}_{j}$ & $\irr{e}$   & $\irr{e}$   & $(1,2,2)$ & $\catB{1}$ ($\mathrm{n\!:\!v}$) & $\irrJp{0}{+},\irrJp{0}{-}$ & $\nu_4$\\
$\irr{e}$   & $\irr{a}_{1}$ & $\irr{e}$   & $(2,1,2)$ & $\catB{1}$ ($\mathrm{r\!:\!v}$) & $\irrJp{1}{+},\irrJp{1}{-}$ & $\nu_4$\\
$\irr{e}$   & $\irr{e}$   & $\irr{a}_{j}$ & $(2,2,1)$ & $\catB{1}$ ($\mathrm{r\!:\!n}$) & $\irrJp{1}{+},\irrJp{1}{-}$ & ground, $3\nu_4$\\
$\irr{e}$   & $\irr{e}$   & $\irr{e}$   & $(2,2,2)$ & $\catT{1}$ (GHZ) & $\irrJp{1}{+},\irrJp{1}{-}$ & $\nu_4$\\
\bottomrule
\end{tabular}
\end{table}

\emph{Coupling tensors.} Invariance under $\elt{C_{3}}$ requires $c_{\mathrm{r}}+c_{\mathrm{n}}+c_{\mathrm{v}}\equiv0 \pmod 3$ with $c=0$ for $\irr{a}_{1}$ and $\irr{a}_{2}$ carriers, so every coupling has at most two terms, and $\elt{\sigma_{v}}(0)$, which sends $c\to-c$ in all factors at once, fixes their relative sign through $\chi_{\irr{a}_{2}}(\elt{\sigma_{v}})=-1$. The resulting tensors, unique up to phase because $N=1$, are
\begin{align}
(\irr{a}_{i},\irr{a}_{1},\irr{a}_{i'}):\;&\ket{\irr{a}_{i}}_{\mathrm{r}}\ket{\irr{a}_{1}}_{\mathrm{n}}\ket{\irr{a}_{i'}}_{\mathrm{v}},\ \ i\neq i',
\label{eq:nh3_s1}\\[2pt]
(\irr{a}_{j},\irr{e},\irr{e}):\;&
\tfrac{1}{\sqrt{2}}\,\ket{\irr{a}_{j}}_{\mathrm{r}}
\bigl(
\irket{\irr{e}}{1}_{\mathrm{n}}\irket{\irr{e}}{2}_{\mathrm{v}}
+(-1)^{j}
\irket{\irr{e}}{2}_{\mathrm{n}}\irket{\irr{e}}{1}_{\mathrm{v}}
\bigr),
\label{eq:nh3_s3}\\[2pt]
(\irr{e},\irr{a}_{1},\irr{e}):\;&
\tfrac{1}{\sqrt{2}}\,\ket{\irr{a}_{1}}_{\mathrm{n}}
\bigl(
\irket{\irr{e}}{1}_{\mathrm{r}}\irket{\irr{e}}{2}_{\mathrm{v}}
-
\irket{\irr{e}}{2}_{\mathrm{r}}\irket{\irr{e}}{1}_{\mathrm{v}}
\bigr),
\label{eq:nh3_s4}\\[2pt]
(\irr{e},\irr{e},\irr{a}_{j}):\;&
\tfrac{1}{\sqrt{2}}\,\ket{\irr{a}_{j}}_{\mathrm{v}}
\bigl(
\irket{\irr{e}}{1}_{\mathrm{r}}\irket{\irr{e}}{2}_{\mathrm{n}}
+(-1)^{j}
\irket{\irr{e}}{2}_{\mathrm{r}}\irket{\irr{e}}{1}_{\mathrm{n}}
\bigr),
\label{eq:nh3_s5}\\[2pt]
(\irr{e},\irr{e},\irr{e}):\;&
\tfrac{1}{\sqrt{2}}\,
\bigl(
\irket{\irr{e}}{1}_{\mathrm{r}}\irket{\irr{e}}{1}_{\mathrm{n}}\irket{\irr{e}}{1}_{\mathrm{v}}
-
\irket{\irr{e}}{2}_{\mathrm{r}}\irket{\irr{e}}{2}_{\mathrm{n}}\irket{\irr{e}}{2}_{\mathrm{v}}
\bigr).
\label{eq:nh3_ghz}
\end{align}
In Eq.~\eqref{eq:nh3_ghz} all three carriers are two-dimensional, so the tensor is the GHZ state $\catT{1}$ of Table~\ref{tab:ame_catalogue}. 

\emph{Position states.} With $|\grp{G}|=6$ and the standard $\irr{e}$ matrices, the four coset states of Eq.~\eqref{eq:coset_states} are
\begin{align}
\ket{\crep;\irr{e}^{11}}&=\tfrac{1}{\sqrt{3}}\,\bigl(\ket{\crep}+\rootunity{3}\ket{\crep\elt{C_{3}}}+\rootunity{3}^{2}\ket{\crep\elt{C_{3}^{2}}}\bigr),\nonumber\\
\ket{\crep;\irr{e}^{12}}&=\tfrac{1}{\sqrt{3}}\,\bigl(\ket{\crep\elt{\sigma_{v}}(0)}+\rootunity{3}\ket{\crep\elt{C_{3}}\elt{\sigma_{v}}(0)}\nonumber\\
&\qquad+\rootunity{3}^{2}\ket{\crep\elt{C_{3}^{2}}\elt{\sigma_{v}}(0)}\bigr),\nonumber\\
\ket{\crep;\irr{e}^{21}}&=\tfrac{1}{\sqrt{3}}\,\bigl(\ket{\crep\elt{\sigma_{v}}(0)}+\rootunity{3}^{2}\ket{\crep\elt{C_{3}}\elt{\sigma_{v}}(0)}\nonumber\\
&\qquad+\rootunity{3}\ket{\crep\elt{C_{3}^{2}}\elt{\sigma_{v}}(0)}\bigr),\nonumber\\
\ket{\crep;\irr{e}^{22}}&=\tfrac{1}{\sqrt{3}}\,\bigl(\ket{\crep}+\rootunity{3}^{2}\ket{\crep\elt{C_{3}}}+\rootunity{3}\ket{\crep\elt{C_{3}^{2}}}\bigr),
\label{eq:c3v_coset}
\end{align}
while $\ket{\crep;\Gamma}=\tfrac{1}{\sqrt{6}}\,\sum_{\gel}\chi_{\Gamma}(\gel)\ket{\crep\gel}$ for $\Gamma=\irr{a}_{1},\irr{a}_{2}$. The diagonal states $\irr{e}^{11}$, $\irr{e}^{22}$ superpose the three proper orbit points, the off-diagonal states $\irr{e}^{12}$, $\irr{e}^{21}$ the three improper ones. The vibrational carrier of a single $\nu_4$ excitation is
\begin{equation}
\braket{\mathbf q|{}^{\irr{e}}_{\,1}}_{\mathrm{v}}=\sqrt2\,q_{4,-}\,\psi_0(\mathbf q),\qquad
\braket{\mathbf q|{}^{\irr{e}}_{\,2}}_{\mathrm{v}}=\sqrt2\,q_{4,+}\,\psi_0(\mathbf q),
\label{eq:nh3_vib_position}
\end{equation}
with $q_{4,\pm}=\tfrac{1}{\sqrt{2}}\,(q_{4,x}\pm\iunit q_{4,y})$, and the nuclear-spin carriers are Eq.~\eqref{eq:nh3_spin}. Inserting these into Eq.~\eqref{eq:position_states} with the tensor~\eqref{eq:nh3_ghz} gives the GHZ position state. For fiber index $\lambda=1$ and $M_I=+\tfrac12$,
\begin{align}
&\ket{\crep,1;\tfrac12\tfrac12;\nu_4}=\frac{1}{\sqrt{6}}\,\Bigl[\bigl(\ket{\crep}+\rootunity{3}\ket{\crep\elt{C_{3}}}+\rootunity{3}^{2}\ket{\crep\elt{C_{3}^{2}}}\bigr)\nonumber\\
&\otimes\cgket{\irr{e}}{1}{\tfrac12\tfrac12}\otimes\sqrt2\,q_{4,-}\psi_0(\mathbf q)\nonumber\\
&-\bigl(\ket{\crep\elt{\sigma_{v}}(0)}+\rootunity{3}\ket{\crep\elt{C_{3}}\elt{\sigma_{v}}(0)}+\rootunity{3}^{2}\ket{\crep\elt{C_{3}^{2}}\elt{\sigma_{v}}(0)}\bigr)\nonumber\\
&\otimes\cgket{\irr{e}}{2}{\tfrac12\tfrac12}\otimes\sqrt2\,q_{4,+}\psi_0(\mathbf q)\Bigr],
\label{eq:nh3_ghz_position}
\end{align}
with the spin kets written in the product basis through Eq.~\eqref{eq:nh3_spin}. The proper orbit points are correlated with one para doublet and one sense of vibrational circulation, the improper orbit points with the other. The fiber index $\lambda=2$ gives the state with the roles of the two cosets exchanged.

\emph{Momentum states.} From Eq.~\eqref{eq:rot_action}, $\elt{C_{3}}\ket{{}^{J,\parity}_{\,m},k}=\rootunity{3}^{k}\ket{{}^{J,\parity}_{\,m},k}$ and $\elt{\sigma_{v}}(0)\ket{{}^{J,\parity}_{\,m},k}=(-1)^{J+k}\parity\ket{{}^{J,\parity}_{\,m},-k}$, so the adapted rotational states are
\begin{align}
\irr{e}:&\quad \cgket{\irr{e}}{1}{{}^{J,\parity}_{\,m} k}=\ket{{}^{J,\parity}_{\,m},k},\nonumber\\
&\quad \cgket{\irr{e}}{2}{{}^{J,\parity}_{\,m} k}=(-1)^{J+k}\parity\ket{{}^{J,\parity}_{\,m},-k},\nonumber\\
\irr{a}_{1},\irr{a}_{2}:&\quad \tfrac{1}{\sqrt{2}}\,\bigl(\ket{{}^{J,\parity}_{\,m},k'}\pm(-1)^{J+k'}\parity\ket{{}^{J,\parity}_{\,m},-k'}\bigr),
\label{eq:c3v_rot}
\end{align}
with $k\equiv1\pmod 3$ and $k'\equiv0\pmod 3$, $k'>0$, together with $\ket{{}^{J,\parity}_{\,m},0}$, of species $\irr{a}_{1}$ ($\irr{a}_{2}$) for $(-1)^J\parity=+1$ ($-1$). For vibration, $U(\elt{C_{3}})=\rootunity{3}^{-\ell}$ on circular occupation states $\occ{n_- n_+}$ with $\ell=n_+-n_-$, and $\elt{\sigma_{v}}(0)$ exchanges $n_\pm$, so the $\irr{e}$ carriers of one degenerate mode are $\irket{\irr{e}}{1}=\ket{\ell}$ and $\irket{\irr{e}}{2}=\ket{-\ell}$ with $\ell\equiv2\pmod3$, and $\ell\equiv0$ gives $\irr{a}_{1}$ and $\irr{a}_{2}$ combinations. For a single $\nu_4$ excitation, $\irket{\irr{e}}{1}_{\mathrm{v}}=\ket{\ell_4{=}{-1}}$ and $\irket{\irr{e}}{2}_{\mathrm{v}}=\ket{\ell_4{=}{+1}}$. The GHZ momentum state at $J=1$, $\parity=+1$, $M_I=+\tfrac12$ is
\begin{align}
&\ket{\irr{a}_{2};{}^{1,+}_{\,m};\tfrac12\tfrac12;\nu_4}
=\frac{1}{\sqrt{2}}\,\Bigl[\ket{{}^{1,+}_{\,m},k{=}1}\otimes\cgket{\irr{e}}{1}{\tfrac12\tfrac12}\nonumber\\
&\otimes\ket{\ell_4{=}{-1}}
-\ket{{}^{1,+}_{\,m},k{=}{-1}}\otimes\cgket{\irr{e}}{2}{\tfrac12\tfrac12}\otimes\ket{\ell_4{=}{+1}}\Bigr].
\label{eq:nh3_ghz_momentum}
\end{align}
The molecule-frame projection $k$, the para doublet index, and the vibrational angular momentum are perfectly correlated, and each is maximally uncertain on its own. The tensors~\eqref{eq:nh3_s3}--\eqref{eq:nh3_ghz} also reproduce the familiar symmetric-top rule~\cite{Hougen1962}: rovibrational factors with $k-\ell\equiv0\pmod3$ (the $\irr{a}$ type) pair only with ortho spin, and those with $k-\ell\not\equiv0$ (the $\irr{e}$ type) only with para spin.

\emph{Worked instance.} For $\Grot=\irr{e}$, $|\grp{G}|=6$, and $J=1$, Theorem~\ref{thm:fourier} gives $\mathcal Y^{1,\parity}_m(\crep,\lambda)=\tfrac{3}{4\pi}\irrJp{1}{\parity}_{m;\lambda}(\crep)^{*}$, and at the reference orientation $\mathcal Y^{1,\parity}_m(\elt{E},1)=0.239\,\delta_{m,1}$ and $\mathcal Y^{1,\parity}_m(\elt{E},2)=0.239\,\parity\,\delta_{m,-1}$: at $\crep=\elt{E}$ the fiber index of the $J=1$ component is the sign of the laboratory projection. The state~\eqref{eq:nh3_ghz_momentum} lies one $\nu_4$ excitation ($1118\ \mathrm{cm^{-1}}$, $b_4=0.174\ \mathrm{amu^{1/2}\,\text{\AA}}$) above the ground-state $J=1$, $|k|=1$ para level. Phosphine therefore hosts a symmetry-enforced GHZ state at definite parity across rotation, nuclear spin, and vibration, reachable with a single degenerate-mode excitation and para nuclear spin, and by Corollary~\ref{cor:rep_independence} it is a GHZ state in both Eqs.~\eqref{eq:nh3_ghz_position} and~\eqref{eq:nh3_ghz_momentum}.

\subsection{Non-Abelian Tetrahedral Group: Methane}
\label{subsec:methane}

Methane has $\grp{G}=\grp{T_{d}}$, four identical protons, and nine vibrational degrees of freedom,
\begin{equation}
\Gamma_{\mathrm v}^{\mathrm{fund}}=\irr{a}_{1}\oplus \irr{e}\oplus 2\irr{t}_{2} .
\label{eq:methane_modes}
\end{equation}
We place proton $j$ at $\mathbf r_j\in\{(1,1,1),(1,-1,-1),(-1,1,-1),(-1,-1,1)\}$ (the \PGAtlas's orbit points $1,2,3,0$, in this order) and use the standard matrices $D^{\irr{t}_{2}}(\gel)=\gmat$ (the Cartesian $3\times3$ matrix, with $P(\gel)$ the \PGAtlas's matrix of $\gel$, which acts on row vectors), $D^{\irr{t}_{1}}(\gel)=\det(\gel)\,\gmat$, $D^{\irr{a}_{2}}(\gel)=\det \gel$, and for $\irr{e}$ the complex basis in which the rotation $\elt{C_{3}}$ about $(1,1,1)$ acts as $\diag(\rootunity{3},\rootunity{3}^{2})$ and the reflection $\elt{\sigma_{d}}$ in the plane $x=y$ exchanges the two partners. Cartesian indices $i=x,y,z$ are identified with $1,2,3$, the partners of $\irr{t}_{1}$ and $\irr{t}_{2}$, also in powers of $\rootunity{3}$. Both infrared-active fundamentals $\nu_3$ and $\nu_4$ carry $\irr{t}_{2}$, so Eq.~\eqref{eq:methane_modes} already contains a repeated irrep, and overtones raise the multiplicity further. By the \PGAtlas's tables~\cite{o3harmonics}, $\operatorname{Sym}^{2}\irr{t}_{2}=\irr{a}_{1}\oplus \irr{e}\oplus \irr{t}_{2}$ and $\operatorname{Sym}^{3}\irr{t}_{2}=\irr{a}_{1}\oplus \irr{t}_{1}\oplus 2\irr{t}_{2}$. The four protons give $\chi^{\mathrm{(n)}}_{\spin}=(16,4,4,2,8)$ on $\{\cls{E},\cls{8C_{3}},\cls{3C_{2}},\cls{6S_{4}},\cls{6\sigma_{d}}\}$ and
\begin{equation}
\mathcal H^{\mathrm{(n)}}_{\spin}(\mathrm{CH_4})\cong 5\irr{a}_{1}\oplus \irr{e}\oplus 3\irr{t}_{2} ,
\label{eq:methane_spin_decomposition}
\end{equation}
the meta ($I_{\mathrm{tot}}=2$), para ($I_{\mathrm{tot}}=0$), and ortho ($I_{\mathrm{tot}}=1$) species, with $\Gmol=\irr{a}_{2}$ from the odd permutations $\elt{S_{4}}$ and $\elt{\sigma_{d}}$. The carriers are, with $\ket{\downarrow_j}$ the state with proton $j$ down and the rest up,
\begin{align}
\irr{t}_{2}\ (\text{ortho}):&\quad \cgket{\irr{t}_{2}}{i}{1,1}=\tfrac12\textstyle\sum_{j=1}^{4}(\mathbf r_j)_i\ket{\downarrow_j},\nonumber\\
\irr{e}\ (\text{para}):&\quad \cgket{\irr{e}}{1}{0,0}=\tfrac{1}{\sqrt{6}}\,\textstyle\sum_i\rootunity{3}^{-i}P_i,\nonumber\\
&\quad \cgket{\irr{e}}{2}{0,0}=\tfrac{1}{\sqrt{6}}\,\textstyle\sum_i\rootunity{3}^{i}P_i,
\label{eq:ch4_spin}
\end{align}
where $P_x=\ket{\uparrow\uparrow\downarrow\downarrow}+\ket{\downarrow\downarrow\uparrow\uparrow}$, $P_y=\ket{\uparrow\downarrow\uparrow\downarrow}+\ket{\downarrow\uparrow\downarrow\uparrow}$, $P_z=\ket{\uparrow\downarrow\downarrow\uparrow}+\ket{\downarrow\uparrow\uparrow\downarrow}$ pair the protons exchanged by $\elt{C_{2x}}$, $\elt{C_{2y}}$, $\elt{C_{2z}}$. The other ortho projections follow by lowering, and the meta quintet is the symmetric $I_{\mathrm{tot}}=2$ multiplet. 

\emph{Species table.} Enumeration over $\Grot\in\grp{\widehat{T}_{d}}$, $\Gspin\in\{\irr{a}_{1},\irr{e},\irr{t}_{2}\}$, $\Gvib\in\grp{\widehat{T}_{d}}$ gives $26$ triples, every one with $N=1$ (Table~\ref{tab:methane_species}). All five vibrational species occur: $\irr{t}_{2}$ at $\nu_4$, $\irr{e}$ at $\nu_2$, $\irr{t}_{1}$ first at $\nu_2+\nu_4$ ($2840\ \mathrm{cm^{-1}}$ harmonic), and $\irr{a}_{2}$ first at $\nu_2+2\nu_4$ ($4146\ \mathrm{cm^{-1}}$). In the vibrational ground state the table leaves exactly $(\irr{a}_{2},\irr{a}_{1},\irr{a}_{1})$, $(\irr{t}_{1},\irr{t}_{2},\irr{a}_{1})$, and $(\irr{e},\irr{e},\irr{a}_{1})$, i.e.\ a meta level at $J=0$, an ortho level at $J=1$, and a para level at $J=2$ accompanied by an ortho level of species $\irr{t}_{1}$ ($J=2$, $\parity=-1$), which is the known species assignment of the lowest rotational levels of methane.

\begin{table*}[t]
\caption{All $26$ admissible species of $\mathrm{CH_4}$, $\Gmol=\irr{a}_{2}$, each with $N=1$. The state column gives the AME state of Table~\ref{tab:ame_catalogue}. The lowest realizations are: $\Grot=\irr{a}_{1},\irr{a}_{2},\irr{t}_{1},\irr{t}_{2},\irr{e}$ at $\irrJp{0}{+},\irrJp{0}{-},\irrJp{1}{+},\irrJp{1}{-},\irrJp{2}{\pm}$; $\Gspin=\irr{a}_{1},\irr{e},\irr{t}_{2}$ are meta, para, ortho; $\Gvib=\irr{a}_{1},\irr{t}_{2},\irr{e},\irr{t}_{1},\irr{a}_{2}$ first at the ground state, $\nu_4$, $\nu_2$, $\nu_2{+}\nu_4$, $\nu_2{+}2\nu_4$. The last column refers to the tensor families (a)--(g) of Eq.~\eqref{eq:td_tensors}.}
\label{tab:methane_species}
\centering
\small
\begin{tabular}{ccccll@{\qquad}ccccll}
\toprule
$\Grot$ & $\Gspin$ & $\Gvib$ & $(d_{\mathrm{r}},d_{\mathrm{n}},d_{\mathrm{v}})$ & state & $C$ &
$\Grot$ & $\Gspin$ & $\Gvib$ & $(d_{\mathrm{r}},d_{\mathrm{n}},d_{\mathrm{v}})$ & state & $C$\\
\midrule
$\irr{a}_{1}$ & $\irr{a}_{1}$ & $\irr{a}_{2}$ & $(1,1,1)$ & $\catP$ & (a) & $\irr{t}_{1}$ & $\irr{e}$ & $\irr{t}_{1}$ & $(3,2,3)$ & $\catT{2}$ & (e$_-$)\\
$\irr{a}_{2}$ & $\irr{a}_{1}$ & $\irr{a}_{1}$ & $(1,1,1)$ & $\catP$ & (a) & $\irr{t}_{1}$ & $\irr{e}$ & $\irr{t}_{2}$ & $(3,2,3)$ & $\catT{2}$ & (e$_+$)\\
$\irr{a}_{1}$ & $\irr{e}$ & $\irr{e}$ & $(1,2,2)$ & $\catB{1}$ & (b) & $\irr{t}_{2}$ & $\irr{e}$ & $\irr{t}_{1}$ & $(3,2,3)$ & $\catT{2}$ & (e$_+$)\\
$\irr{a}_{2}$ & $\irr{e}$ & $\irr{e}$ & $(1,2,2)$ & $\catB{1}$ & (b) & $\irr{t}_{2}$ & $\irr{e}$ & $\irr{t}_{2}$ & $(3,2,3)$ & $\catT{2}$ & (e$_-$)\\
$\irr{e}$ & $\irr{a}_{1}$ & $\irr{e}$ & $(2,1,2)$ & $\catB{1}$ & (b) & $\irr{t}_{1}$ & $\irr{t}_{2}$ & $\irr{e}$ & $(3,3,2)$ & $\catT{2}$ & (e$_+$)\\
$\irr{e}$ & $\irr{e}$ & $\irr{a}_{1}$ & $(2,2,1)$ & $\catB{1}$ & (b) & $\irr{t}_{2}$ & $\irr{t}_{2}$ & $\irr{e}$ & $(3,3,2)$ & $\catT{2}$ & (e$_-$)\\
$\irr{e}$ & $\irr{e}$ & $\irr{a}_{2}$ & $(2,2,1)$ & $\catB{1}$ & (b) & $\irr{e}$ & $\irr{t}_{2}$ & $\irr{t}_{1}$ & $(2,3,3)$ & $\catT{2}$ & (e$_+$)\\
$\irr{e}$ & $\irr{e}$ & $\irr{e}$ & $(2,2,2)$ & $\catT{1}$ & (c) & $\irr{e}$ & $\irr{t}_{2}$ & $\irr{t}_{2}$ & $(2,3,3)$ & $\catT{2}$ & (e$_-$)\\
$\irr{a}_{1}$ & $\irr{t}_{2}$ & $\irr{t}_{1}$ & $(1,3,3)$ & $\catB{2}$ & (d) & $\irr{t}_{2}$ & $\irr{t}_{2}$ & $\irr{t}_{2}$ & $(3,3,3)$ & $\catT{3}$ & (f)\\
$\irr{a}_{2}$ & $\irr{t}_{2}$ & $\irr{t}_{2}$ & $(1,3,3)$ & $\catB{2}$ & (d) & $\irr{t}_{1}$ & $\irr{t}_{2}$ & $\irr{t}_{1}$ & $(3,3,3)$ & $\catT{3}$ & (f)\\
$\irr{t}_{1}$ & $\irr{a}_{1}$ & $\irr{t}_{2}$ & $(3,1,3)$ & $\catB{2}$ & (d) & $\irr{t}_{1}$ & $\irr{t}_{2}$ & $\irr{t}_{2}$ & $(3,3,3)$ & $\catT{4}$ & (g)\\
$\irr{t}_{2}$ & $\irr{a}_{1}$ & $\irr{t}_{1}$ & $(3,1,3)$ & $\catB{2}$ & (d) & $\irr{t}_{2}$ & $\irr{t}_{2}$ & $\irr{t}_{1}$ & $(3,3,3)$ & $\catT{4}$ & (g)\\
$\irr{t}_{1}$ & $\irr{t}_{2}$ & $\irr{a}_{1}$ & $(3,3,1)$ & $\catB{2}$ & (d) & & & & & & \\
$\irr{t}_{2}$ & $\irr{t}_{2}$ & $\irr{a}_{2}$ & $(3,3,1)$ & $\catB{2}$ & (d) & & & & & & \\
\bottomrule
\end{tabular}
\end{table*}

\emph{Coupling tensors.} The $26$ triples fall into seven closed-form families, written here for the carrier factors in the order of the triple (for the families with a one-dimensional factor, the two degenerate factors are listed in their order in the triple):
\begin{align}
&\text{(a)}\ \ket{\Grot}\ket{\Gspin}\ket{\Gvib};\nonumber\\
&\text{(b)}\ \tfrac{1}{\sqrt{2}}\,\bigl(\irket{\irr{e}}{1}\irket{\irr{e}}{2}\pm\irket{\irr{e}}{2}\irket{\irr{e}}{1}\bigr),\nonumber\\
&\qquad\text{with }+\text{ for }(\irr{a}_{2},\irr{e},\irr{e}),(\irr{e},\irr{e},\irr{a}_{2})\text{ and }-\text{ otherwise};\nonumber\\
&\text{(c)}\ \tfrac{1}{\sqrt{2}}\,\bigl(\irket{\irr{e}}{1}\irket{\irr{e}}{1}\irket{\irr{e}}{1}-\irket{\irr{e}}{2}\irket{\irr{e}}{2}\irket{\irr{e}}{2}\bigr);\nonumber\\
&\text{(d)}\ \tfrac{1}{\sqrt{3}}\,\textstyle\sum_i\ket{i}\ket{i};\nonumber\\
&\text{(e}_\pm\text{)}\ \tfrac{1}{\sqrt{6}}\,\rootunity{3}^{2}\textstyle\sum_i\bigl(\rootunity{3}^{i}\irket{\irr{e}}{1}\pm\rootunity{3}^{-i}\irket{\irr{e}}{2}\bigr)\ket{i}\ket{i};\nonumber\\
&\text{(f)}\ \tfrac{1}{\sqrt{6}}\,\textstyle\sum_{ijk}\epsilon_{ijk}\ket{i}\ket{j}\ket{k};\nonumber\\
&\text{(g)}\ \tfrac{1}{\sqrt{6}}\,\textstyle\sum_{ijk}|\epsilon_{ijk}|\ket{i}\ket{j}\ket{k}.
\label{eq:td_tensors}
\end{align}
Families (a)--(g) are the states $\catP$, $\catB{1}$, $\catT{1}$, $\catB{2}$, $\catT{2}$, $\catT{3}$, and $\catT{4}$ of Table~\ref{tab:ame_catalogue}, given by the $\grp{T_{d}}$ coupling coefficients of Refs.~\cite{MoretBailly1965, Champion1977} up to basis conventions. In family (e) the $\irr{e}$ carrier may sit in any of the three slots, and the factor $\rootunity{3}^{2}$ makes the first component, at $i=1$, real and positive (the phase rule stated after Eq.~\eqref{eq:unique_coupled_state}). The sign is $(+)$ when the two three-dimensional carriers are different ($\irr{t}_{1}$ and $\irr{t}_{2}$) and $(-)$ when they are equal. The pattern is set by the determinant: $\irr{t}_{1}={\det_{\grp{G}}}\otimes \irr{t}_{2}$, so each $\irr{t}_{1}$ carrier contributes a factor $\det \gel$, and the product of carriers must supply exactly one such factor to produce $\Gmol=\irr{a}_{2}$. For three vector carriers this selects the antisymmetric invariant $\epsilon_{ijk}$ when the number of $\irr{t}_{1}$ factors is even, and the symmetric invariant $|\epsilon_{ijk}|$, the $\grp{T_{d}}$-invariant cubic $xyz$, when it is odd.

\emph{Position states.} Because $D^{\irr{t}_{2}}(\gel)=\gmat$, the $\irr{t}_{2}$ coset states take a particularly transparent form,
\begin{equation}
\ket{\crep;\irr{t}_{2}^{\lambda\mu}}=\frac{1}{2\sqrt{2}}\,\sum_{\gel\in \grp{T_{d}}}\gmat_{\lambda,\mu}\,\ket{\crep\gel},
\label{eq:td_coset}
\end{equation}
and $\irr{t}_{1}$ differs by the factor $\det \gel$. The vibrational carrier of one $\nu_4$ excitation is $\braket{\mathbf q|{}^{\irr{t}_{2}}_{\,k}}_{\mathrm{v}}=\sqrt2\,q_{4,k}\psi_0(\mathbf q)$. Inserting Eq.~\eqref{eq:td_coset}, the ortho spin carriers~\eqref{eq:ch4_spin}, and tensor (f) into Eq.~\eqref{eq:position_states}, the contraction $\sum_{ijk}\epsilon_{ijk}\gmat_{\lambda,i}S_jQ_k$ is the $\lambda$-th component of the rotated cross product, and the Levi-Civita position state is
\begin{equation}
\ket{\crep,\lambda;1,1;\nu_4}=\frac{1}{4\sqrt{3}}\,\sum_{\gel\in \grp{T_{d}}}\ket{\crep\gel}\otimes\bigl[\gmat\,(\vec S\times\vec Q)\bigr]_\lambda ,
\label{eq:ch4_position}
\end{equation}
where $\vec S=(\cgket{\irr{t}_{2}}{1}{1,1},\cgket{\irr{t}_{2}}{2}{1,1},\cgket{\irr{t}_{2}}{3}{1,1})$ is the triplet of spin kets of Eq.~\eqref{eq:ch4_spin} in the product basis and $\vec Q=\sqrt2\,(q_{4,x},q_{4,y},q_{4,z})\,\psi_0(\mathbf q)$ the triplet of vibrational wavefunctions. The molecule's orientation, the direction encoded in the nuclear spins, and the direction of the vibrational displacement form a right-handed frame, averaged over the orbit.

\emph{Momentum states.} The adapted rotational states needed for the lowest realizations are, from Eq.~\eqref{eq:rot_action},
\begin{align}
\irr{t}_{1},\irr{t}_{2}:&\quad \irket{\Gamma}{1}=\tfrac{1}{\sqrt{2}}\,(\ket{k{=}{-1}}-\ket{k{=}1}),\nonumber\\
&\quad \irket{\Gamma}{2}=-\tfrac{\iunit}{\sqrt2}\,(\ket{k{=}{-1}}+\ket{k{=}1}),\quad \irket{\Gamma}{3}=\ket{k{=}0},\nonumber\\
\irr{e}:&\quad \irket{\irr{e}}{1},\irket{\irr{e}}{2}=\tfrac{1}{\sqrt{2}}\,\ket{k{=}0}\mp\tfrac{\iunit}{2}(\ket{k{=}2}+\ket{k{=}{-2}}),
\label{eq:td_rot}
\end{align}
where $\Gamma$ is $\irr{t}_{1}$ at $J=1$, $\parity=+1$ and $\irr{t}_{2}$ at $J=1$, $\parity=-1$, and the $\irr{e}$ states are at $J=2$, $\parity=+1$. The states $\irr{a}_{1}$ and $\irr{a}_{2}$ are given by $\ket{k{=}0}$ at $J=0$ with $\parity=+1$ and $-1$. The vibrational carriers are $\irket{\irr{t}_{2}}{i}=a^\dagger_{s,i}\ket0$ for $s=3,4$; $\irket{\irr{e}}{1},\irket{\irr{e}}{2}=\tfrac{1}{\sqrt{2}}\,(a^\dagger_{2u}\mp \iunit a^\dagger_{2v})\ket0$ for $\nu_2$, whose real coordinates $q_{2u}$, $q_{2v}$ transform as $2z^2-x^2-y^2$ and $\sqrt3(x^2-y^2)$; $\irket{\irr{t}_{1}}{i}=\tfrac{1}{\sqrt{2}}\,(\rootunity{3}^{i}\irket{\irr{e}}{1}_{\nu_2}-\rootunity{3}^{-i}\irket{\irr{e}}{2}_{\nu_2})\irket{\irr{t}_{2}}{i}_{\nu_4}$ in $\nu_2+\nu_4$; and $\ket{\irr{a}_{2}}=\tfrac{1}{\sqrt{2}}\,(\irket{\irr{e}}{1}_{\nu_2}\irket{\irr{e}}{2}_{2\nu_4}-\irket{\irr{e}}{2}_{\nu_2}\irket{\irr{e}}{1}_{2\nu_4})$ in $\nu_2+2\nu_4$, with $\irket{\irr{e}}{1}_{2\nu_4},\irket{\irr{e}}{2}_{2\nu_4}=\tfrac{1}{\sqrt{3}}\,\sum_i\rootunity{3}^{\mp i}\ket{2_i}$ and $\ket{2_i}=\tfrac{1}{\sqrt{2}}\,(a^\dagger_{4,i})^2\ket0$. The Levi-Civita momentum state at $J=1$, $\parity=-1$, $M_I=1$, one $\nu_4$ excitation is
\begin{equation}
\ket{\irr{a}_{2};{}^{1,-}_{\,m};1,1;\nu_4}=\frac{1}{\sqrt{6}}\,\sum_{ijk}\epsilon_{ijk}\,\cgket{\irr{t}_{2}}{i}{{}^{1,-}_{\,m}}\otimes\cgket{\irr{t}_{2}}{j}{1,1}\otimes a^\dagger_{4,k}\ket0 .
\label{eq:methane_levicivita}
\end{equation}

\emph{Worked instance.} This is the totally antisymmetric three-qutrit state~$\catT{3}$. Its marginals are $\rho=I_3/3$ in each sector, and the Schmidt rank is $3$ across every cut, so methane supports symmetry-enforced tripartite entanglement one dimension higher than phosphine, one $\nu_4$ excitation ($1306\ \mathrm{cm^{-1}}$, $b_4=0.161\ \mathrm{amu^{1/2}\,\text{\AA}}$) above the ground-state $J=1$ ortho level. For $\Grot=\irr{t}_{2}$, $|\grp{G}|=24$, $J=1$, Theorem~\ref{thm:fourier} gives $\mathcal Y^{1,-}_m(\crep,\lambda)=\tfrac{\sqrt6}{2\pi}\irrJp{1}{-}_{m;\lambda}(\crep)^{*}$. At $\crep=\elt{E}$ its values $0.390\times\{\tfrac{1}{\sqrt{2}}\,(\delta_{m,-1}-\delta_{m,1}),\,\tfrac{\iunit}{\sqrt2}\,(\delta_{m,-1}+\delta_{m,1}),\,\delta_{m,0}\}$ for $\lambda=x,y,z$ show that the fiber of the ortho species at the reference orientation is a laboratory-frame vector. The triples $(\irr{t}_{2},\irr{t}_{2},\irr{t}_{1})$ and $(\irr{t}_{1},\irr{t}_{2},\irr{t}_{2})$, also tripartite with ranks $(3,3,3)$, carry the symmetric state $\catT{4}$ of family (g) rather than $\epsilon_{ijk}$. The all-$\irr{e}$ triple reproduces the phosphine GHZ state $\catT{1}$ inside $\grp{T_{d}}$. In the ExoMol line list~\cite{Yurchenko2024}, whose labels are those of the molecular symmetry group, this Levi-Civita level is the $J=1$ level of species $F_2$ at $1316.297\ \mathrm{cm^{-1}}$. The $(\irr{t}_{1},\irr{t}_{2},\irr{t}_{2})$ level at $\parity=+1$, which carries $|\epsilon_{ijk}|$, is the $F_1$ level at $1325.822\ \mathrm{cm^{-1}}$. These $J=1$ tensors are Jahn's Coriolis sublevels of definite $R=J-l$~\cite{Jahn1938, Hecht1961}: $R=0$ is the rotation--vibration pair, $R=1$ is $\epsilon_{ijk}$, and $R=2$ holds $|\epsilon_{ijk}|$ and the trine-GHZ level. The fundamental conserves $R$, so R(0) and P(1) end on the two pairs, Q(1) on $\epsilon_{ijk}$, and the two P(2) lines on $|\epsilon_{ijk}|$ and the trine-GHZ level. In $\nu_3$, which has the same tensors, Juurlink, Smith, and Utz prepared the pair, $\epsilon_{ijk}$, and the trine-GHZ level in a molecular beam~\cite{Juurlink2000}.

\section{Preparing symmetry-adapted bend states of calcium monomethoxide}
\label{sec:experiment}

We consider the polyatomic molecule calcium monomethoxide, $^{40}\mathrm{Ca}^{16}\mathrm{O}^{12}\mathrm{CH}_3$, as a candidate platform for preparing symmetry-adapted states with $\grp{C_{3v}}$ symmetry. This prolate symmetric top has a $\widetilde X\,{}^2A_1$ electronic ground state and has been extensively characterized spectroscopically~\cite{Whitham1998,Namiki1998,Crozet2002}. Both of its proton-spin isomers have been laser-cooled, to a transverse beam temperature of $1.8\pm0.7$~mK~\cite{Mitra2020}, and arrays of such molecules have been proposed for quantum computing~\cite{Yu2019, Wei2011}. Rotational states of the trapped symmetric top CH$_3$F have been prepared by optical pumping~\cite{Glockner2015}. Like phosphine, it has three equivalent protons, so the results from Sec.~\ref{subsec:ammonia} carry over (Appendix~\ref{app:caoch3}).

\begin{figure}[tb]
\centering
\begin{minipage}[b]{0.345\columnwidth}\raggedright
{\small (a)}\\[-1pt]
\input{caoch3_molecule_axes.tex}
\end{minipage}\hfill
\begin{minipage}[b]{0.645\columnwidth}\raggedright
{\small (b)}\\[-1pt]
\input{ladder_tree_compact.tex}
\end{minipage}
\caption{(a) $\mathrm{CaOCH_3}$ and its axes. The molecule-fixed $z$ axis is the $\elt{C_{3}}$ symmetry axis, pointing from the methyl group to Ca~\cite{Mitra2020}; $k$ is the projection of the rotational angular momentum $\mathbf N$ on $z$, and $Z$ is the laboratory axis along the field polarization, at an angle $\beta$ to $z$ that is not fixed in any level. The $\nu_8$ bend quanta circulate counterclockwise ($n_+$, blue) or clockwise ($n_-$, orange) when viewed from $+z$, so $\ell=n_+-n_-$. (b) Pulse sequences that prepare the four target levels with $n_+ + n_-=3$. Gray labels give the bend species of each target, and $L_{12}$ and $L_{21}$ are two rotational copies of the $\irr{e}$ species. Solid (dashed) arrows are the driven (undriven) links. P1--P7: blue (orange) ones exchange a $q=+1$ ($q=-1$) bend quantum.}
\label{fig:caoch3_tree}
\end{figure}
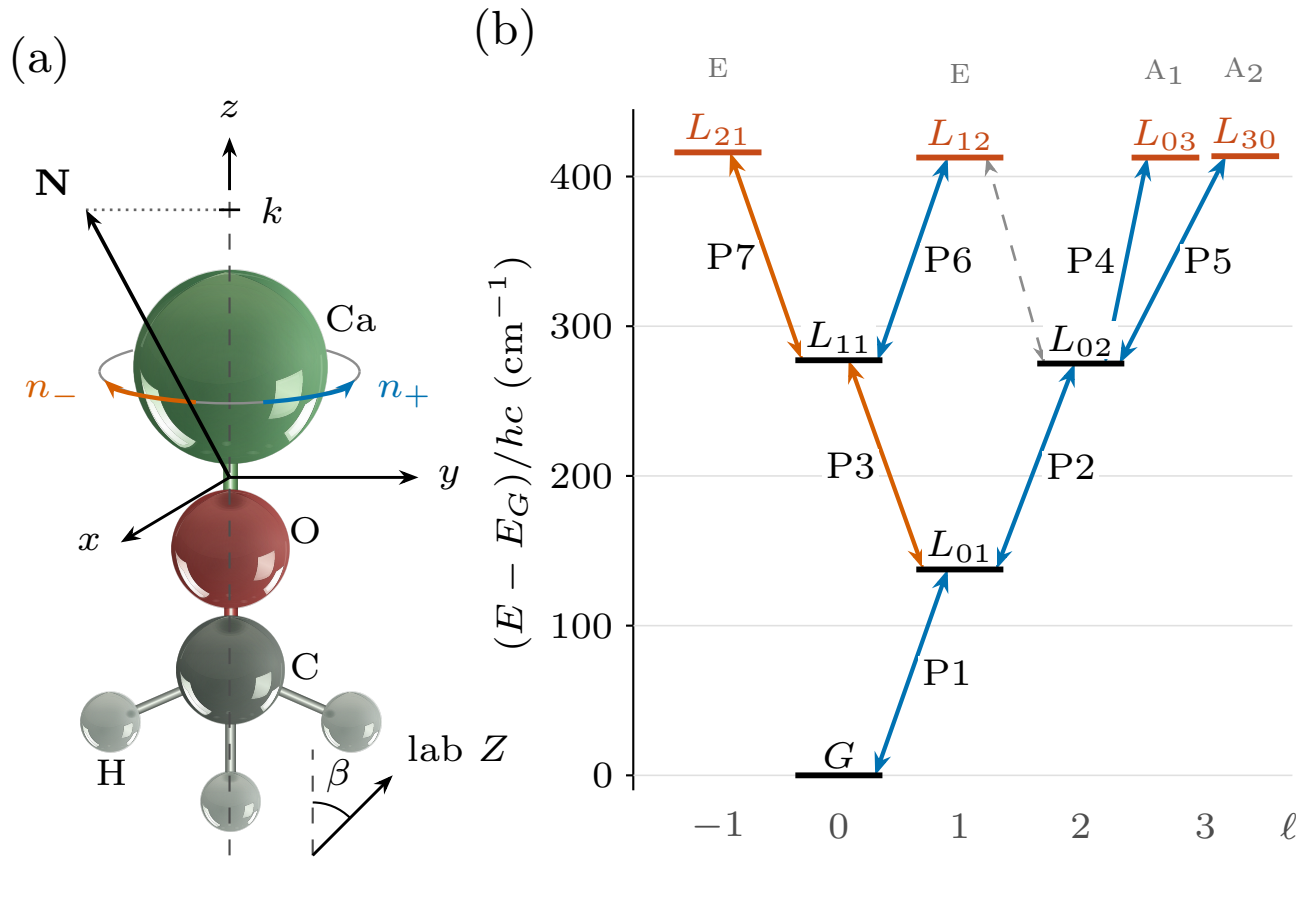

The molecule has twelve normal coordinates: four of them transform as $\irr{a}_{1}$, and the other eight form four degenerate pairs that transform as $\irr{e}$. We focus on the doubly degenerate Ca--O--C bending mode $\nu_8$. Its two coordinates describe bending in two perpendicular planes. Its fundamental has been reported at $144\pm5\ \mathrm{cm^{-1}}$~\cite{Kozyryev2019}.  We label bend states $\occ{n_-n_+}$ by the numbers $n_+$ and $n_-$ of quanta circulating counterclockwise and clockwise about the symmetry axis (Fig.~\ref{fig:caoch3_tree}(a) and Appendix~\ref{app:circular_vibrations}), so that $v=n_++n_-$ and the vibrational angular momentum is $\ell\hbar$, with $\ell=n_+-n_-$. Using this mode, we show how to prepare an arbitrary superposition of four molecular levels whose bend parts are built from the SALCs of Sec.~\ref{sec:vib_sector} with $n_-+n_+=3$ quanta (Eq.~\eqref{eq:caoch3_target}). These SALCs are $\ket{\irr{a}_{1}}=\tfrac{1}{\sqrt{2}}\,(\occ{03}+\occ{30})$, $\ket{\irr{a}_{2}} =\tfrac{1}{\sqrt{2}}\,(\occ{03}-\occ{30})$, $\irket{\irr{e}}{1}=\occ{21}$, and $\irket{\irr{e}}{2}=\occ{12}$. The levels also differ in their rotational states, so tracing out rotation and spin leaves the bend in a mixture of these SALCs (Appendix~\ref{app:caoch3_ham}).

\emph{Hamiltonian and selection rules.} $\mathrm{CaOCH_3}$ has one unpaired electron, so we write the rotational angular momentum of Sec.~\ref{sec:rot_sector}, called $J$ there, as $\mathbf N$, with projection $k$ on the symmetry axis. This axis, the molecule-fixed $z$ axis of Fig.~\ref{fig:caoch3_tree}(a), is the $\elt{C_{3}}$ axis: it passes through the Ca, O, and C nuclei and through the center of the three protons, and we take it to point from the methyl group to Ca, as in Ref.~\cite{Mitra2020}. Both $k$ and $\ell$ are measured along this $z$ axis, so they have the same sign when the molecule rotates and the bend quanta circulate in the same sense about $z$. We couple $\mathbf N$ to the electron spin $S=\tfrac12$ and the total proton spin $I$ as $\mathbf J=\mathbf N+\mathbf S$ and $\mathbf F=\mathbf J+\mathbf I$, with $M_F$ the projection of $\mathbf F$ on the laboratory axis $Z$. The ground state follows Hund's case (b)~\cite{Whitham1998,Crozet2002}, so the electron spin is not tied to the molecular axis $z$, and we quantize it along the laboratory axis $Z$. Such a spin is invariant under every symmetry operation and stays outside the triples (Sec.~\ref{sec:constraint}). With only $\nu_8$ excited, the energy levels of this symmetric top are~\cite{JohnstonDennison1935,Herzberg1945,Nielsen1951,BunkerJensen2006}
\begin{align}
\frac{E}{hc}={}&\tilde\omega_8v+x_{8,8}v(v-1)+g_{8,8}\ell^2+BN(N+1)\nonumber\\
&+(A-B)k^2-2A\zeta_8k\ell ,
\label{eq:caoch3_hamiltonian}
\end{align}
where $\tilde\omega_8$ is the bend wavenumber, $x_{8,8}$ and $g_{8,8}$ are anharmonic constants, $A$ and $B$ are rotational constants, and $\zeta_8$ is the Coriolis constant. The constant $g_{8,8}$ includes the term $A\zeta_8^2$ from the vibrational angular momentum~\cite{Watson1968}. Every term depends only on $N$ and on the projections $k$ and $\ell$ on the molecular axis $z$, so none depends on the angle $\beta$ between $z$ and the laboratory axis $Z$ or on $M_F$. This angle therefore has no fixed value in any level: in the rotational state $\ket{NkM_N}$, where $M_N$ is the projection of $\mathbf N$ on $Z$, its probability density is proportional to $|d^N_{M_Nk}(\beta)|^2\sin\beta$, with $d^N$ the Wigner small-$d$ matrix~\cite{Zare1988}. The axis $Z$ enters only through the field polarization in the transition dipoles below. We keep in Eq.~\eqref{eq:caoch3_hamiltonian} only the terms needed for the symmetry analysis. Centrifugal distortion~\cite{Herzberg1945,Crozet2002} and $\ell$-type doubling~\cite{Nielsen1951} shift levels by much less than the rotational spacings, and spin--rotation~\cite{VanVleck1951,Raynes1964,Hirota1985} and hyperfine interactions shift them only at the megahertz scale~\cite{Namiki1998,Mitra2020,Yu2019}. We neglect these terms here and estimate their values in Appendix~\ref{app:caoch3_ham}. The upper block of Table~\ref{tab:caoch3_constants} lists the constants of Eq.~\eqref{eq:caoch3_hamiltonian}: $\zeta_8$, $x_{8,8}$, and $g_{8,8}$ are estimated through ab initio calculations (Appendix~\ref{app:caoch3_ham}), and the others come from experiment~\cite{Crozet2002,Namiki1998,Kozyryev2019}.

Within the approximation of Eq.~\eqref{eq:caoch3_hamiltonian}, the eigenstates of the molecular Hamiltonian are labeled by $(v,\ell,k)$. An electric field of amplitude $\mathcal E$ polarized along the laboratory axis $Z$ drives a transition $\ket{\mathrm{lo}}\to\ket{u}$ with Rabi frequency $\Omega=|d|\mathcal E/\hbar$, where $d=\bra{u}\mu_Z\ket{\mathrm{lo}}$ is the transition dipole matrix element. Each absorbed photon adds one bend quantum to the molecule with vibrational angular momentum $q\hbar$, $q=\pm1$. Such an absorption obeys $\Delta v=1$, $\Delta\ell=\Delta k=q$~\cite{Herzberg1945}, $\Delta N,\Delta J,\Delta F=0,\pm1$, and $\Delta M_F=0$. Appendix~\ref{app:caoch3_dipoles} derives these selection rules from the transition dipole matrix elements~\cite{Edmonds1957,Zare1988,BrownCarrington2003}. Because $d$ depends on $M_F$, a pulse has a well-defined area only within one $M_F$ sublevel, so all tones share the polarization axis $Z$. Because of anharmonicity, rotation, and Coriolis coupling, consecutive transitions with the same $q$ between levels of Eq.~\eqref{eq:caoch3_hamiltonian} generally have distinct frequencies.  

\emph{Target and intermediate states.} Before preparing superpositions, we write the target levels that carry the SALCs with all their quantum numbers, $\ket{L_{n_-n_+}}=\tfrac{1}{\sqrt{2}}\,(\ket{k}\occ{n_-n_+}\pm\ket{-k}\occ{n_+n_-})\ket{\irr{a}_{1}}_{\mathrm{n}}$ as in Eq.~\eqref{eq:caoch3_levels}, where $\ket{k}$ is the rotational state with projection $k$, and $\ket{\irr{a}_{1}}_{\mathrm{n}}$ is the ortho nuclear spin state. The reflection $\elt{\sigma_{v}}(0)$ reverses $k$ and $\ell$, so it swaps the first term in $\ket{L_{n_-n_+}}$, defined as the branch term, with the second, its mirror. Spin statistics fixes their relative sign so that the level has species $\irr{a}_{2}$. Both branch and mirror terms have equal energies, so one tone drives both.

In $L_{03}$ and $L_{30}$, the bend part is the $\irr{a}$ SALC $\tfrac{1}{\sqrt{2}}\,(\occ{03}\pm\occ{30})$ only if $\ket{k}=\ket{-k}$, that is, $k=0$ and we have for instance:
\begin{equation}
\ket{L_{30}}=\ket{0}\,\tfrac{1}{\sqrt{2}}\,(\occ{03}-\occ{30})\ket{\irr{a}_{1}}_{\mathrm{n}} .
\label{eq:caoch3_a_levels}
\end{equation}
Pairing $\ket{k}\occ{03}$ with $\ket{k}\occ{30}$ fails for $k\neq0$, because the Coriolis term $-2A\zeta_8k\ell$ of Eq.~\eqref{eq:caoch3_hamiltonian} splits the branch and mirror terms in energy, so their relative phase changes in time and no single tone is resonant with both. In both terms of Eq.~\eqref{eq:caoch3_a_levels}, $k-\ell\equiv0\pmod3$, so $\elt{C_{3}}$ leaves the rotation--bend part of the level unchanged. Spin statistics requires the whole level to have species $\Gmol=\irr{a}_{2}$, which $\elt{C_{3}}$ also leaves unchanged. The proton spin state must therefore be unchanged by $\elt{C_{3}}$ as well, and the only such state is the ortho state $\ket{\irr{a}_{1}}_{\mathrm{n}}$, with $I=\tfrac32$, as in phosphine (Sec.~\ref{subsec:ammonia} and Appendix~\ref{app:caoch3}). Bend transitions conserve $k-\ell$, which modulo 3 is Hougen's rigorous selection rule for electric-dipole transitions~\cite{Hougen1962}, so we start from a single $M_F$ sublevel of an ortho ground-state level $G$ with the same $k-\ell$, which can be prepared from the thermally populated levels after laser cooling. The target is
\begin{equation}
\ket{\Psi_T}=c^{\irr{a}_{1}}\ket{L_{03}}+c^{\irr{a}_{2}}\ket{L_{30}}+c^{\irr{e}}_{1}\ket{L_{21}}+c^{\irr{e}}_{2}\ket{L_{12}} ,
\label{eq:caoch3_target}
\end{equation}
with each $c=|c|\,\expe^{\iunit\gamma}$. Figure~\ref{fig:caoch3_tree} sketches the pulse sequences from $G$ to the four target levels, and Table~\ref{tab:caoch3_levels} lists the quantum numbers of every level.

\emph{Protocol.} In the frame rotating with each tone, a pulse of area $\Theta=\int\Omega\,\diff t$ and phase $\phi$ on the link $\mathrm{lo}\leftrightarrow u$ is the rotation gate~\cite{AllenEberly1975}
\begin{equation}
R_{u,\mathrm{lo}}(\Theta,\phi)=\exp\Bigl[-\iunit\tfrac{\Theta}{2}\bigl(\expe^{\iunit\phi}\ket{u}\bra{\mathrm{lo}}+\expe^{-\iunit\phi}\ket{\mathrm{lo}}\bra{u}\bigr)\Bigr],
\label{eq:caoch3_gate}
\end{equation}
which rotates between $\{\ket{\mathrm{lo}},\ket{u}\}$ for that link only. The gate requires its tone to be resonant with $\omega_{u,\mathrm{lo}}=(E_u-E_{\mathrm{lo}})/\hbar$ and detuned from every other transition by much more than the Rabi frequency $\Omega$ (Appendix~\ref{app:caoch3_pulses}). Writing $R_n\equiv R_{u_n,\mathrm{lo}_n}(\Theta_n,\phi_n)$ for pulse P$n$, the sequence must satisfy
\begin{equation}
R_7R_6\cdots R_1\ket{G}=\iunit\ket{\Psi_T}.
\label{eq:caoch3_sequence}
\end{equation}
The factor $\iunit$ is a global phase convention, so we can set $\phi=0$ for P1--P3 and let P4--P7 set the phases of the coefficients in Eq.~\eqref{eq:caoch3_target}.

\emph{Solutions.} We find the pulse areas and phases by solving Eq.~\eqref{eq:caoch3_sequence} algebraically. (a) An arbitrary superposition of the four levels uses all seven pulses. Their areas satisfy:
\begin{align}
\tan\tfrac{\Theta_2}{2}&=\biggl(\frac{|c^{\irr{a}_{1}}|^2+|c^{\irr{a}_{2}}|^2}{|c^{\irr{e}}_{1}|^2+|c^{\irr{e}}_{2}|^2}\biggr)^{1/2},\nonumber\\
\tan\tfrac{\Theta_4}{2}&=\frac{|c^{\irr{a}_{1}}|}{|c^{\irr{a}_{2}}|},\qquad\tan\tfrac{\Theta_6}{2}=\frac{|c^{\irr{e}}_{2}|}{|c^{\irr{e}}_{1}|}.
\label{eq:caoch3_areas}
\end{align}
The other four are $\pi$ pulses. Their phases satisfy: P1--P3 have $\phi=0$, and P4, P5, P6, and P7 have $\phi=\gamma^{\irr{a}_{1}}$, $\gamma^{\irr{a}_{2}}$, $\gamma^{\irr{e}}_{2}$, and $\gamma^{\irr{e}}_{1}$. For example, the equal superposition with every $c=\tfrac12$ needs $\pi/2$ pulses for P2, P4, and P6, $\pi$ pulses for P1, P3, P5, and P7, and $\phi=0$ throughout. (b) A single level needs only the three $\pi$ pulses on its route, all with $\phi=0$: P1, P2, and P5 prepare $L_{30}$, whose bend part is $\ket{\irr{a}_{2}}$, and P1, P2, and P4 prepare $L_{03}$, whose bend part is $\ket{\irr{a}_{1}}$. (c) Likewise, P1, P3, and P7 prepare $L_{21}$, and P1, P3, and P6 prepare $L_{12}$. These two are rotational copies of one $\irr{e}$-bend species, and each leaves the bend in $I_{\irr{e}}/2$. With all tones at an assumed intensity of $1.5\ \mathrm{mW/cm^2}$, low enough for each pulse to resolve the hyperfine structure, a single pulse lasts at most $51~\mu\mathrm{s}$, a single-level sequence $91$ to $112~\mu\mathrm{s}$, and any superposition at most $217~\mu\mathrm{s}$. Appendix~\ref{app:caoch3_pulses} and Table~\ref{tab:caoch3_pulses} derive the parameters of the pulse sequences.

\section{Discussion and Conclusion}
\label{sec:discussion}

This work establishes a symmetry-reduced quantum classification of the various state spaces for molecular rotation, nuclear spin, and vibration. 
Admissible combinations of molecular quantum numbers, conventionally described through spectroscopic selection rules and nuclear-spin statistical weights, follow directly from molecular symmetry and identical-particle statistics~\cite{LonguetHiggins1963, BunkerJensen2006}, independently of the rovibrational Hamiltonian. 
The admissible spin-statistics-respecting states of a molecule can be grouped into species, with each species defined by the triple of irreps describing how each of the three degrees of freedom transforms under symmetries. 

While this formalism has been known for nearly a century~\cite{Wilson1935, Wilson1935b} and applied, mostly numerically, on a per-molecule basis, no classification across all possible molecular species has been done to our knowledge.
This delay and the mathematical techniques used echo the classification of band insulators~\cite{Bradlyn2017, Po2017, Kruthoff2017}, which is also recent relative to the existence of band theory.

We show that the physical states of each species are characterized by invariant coupling tensors. Every such tensor is an absolutely maximally entangled state of its carriers, independently of the Hamiltonian. Interactions can mix species, but each species keeps its AME state. The number of degenerate carriers fixes its kind, from product states through maximally entangled pairs to genuinely tripartite states. Up to local unitaries and the order of the carriers, these tensors realize 21 AME states across the 75 point groups, with at least one of each kind, counting each two-dimensional coupling space by two basis states. The seven states that occur outside the icosahedral groups all appear in methane.

The phase-space formulation expresses this structure in complementary position and momentum representations, connected by a symmetry-adapted Fourier transformation that preserves admissibility and coupling structure. Beyond spectroscopy, carrier dimensions and coupling multiplicities provide a natural basis for organizing molecular qudit encodings~\cite{Sawant2020}, much like the noiseless subsystems of Refs.~\cite{ZanardiRasetti1997, LidarChuangWhaley1998, KnillLaflammeViola2000, LidarWhaley2003}. 

The molecular examples demonstrate the generality of this framework. In water, antisymmetric stretching interchanges the rotational symmetry species compatible with a given nuclear-spin species, while phosphine and methane illustrate more intricate coupling involving multidimensional irreducible representations. The symmetric-top molecule $\mathrm{CaOCH_3}$ further extends the construction to systems relevant for quantum-state preparation and control. In each case, the established statistical weights and rotational restrictions are recovered from a unified representation-theoretic prescription.

Our results inspire ample future directions. The resulting symmetry-enforced correlations motivate further investigation of their preparation, measurement, and use in quantum information processing~\cite{PRXQuantum.5.040336, PRXQuantum.5.030310, SciPostPhys.19.1.020, Zeppenfeld2023, Ding2023, Ding2023Corrigendum, Low2025}. In particular, determining whether carrier or multiplicity spaces support protected logical encodings requires an explicit analysis of the relevant noise operators~\cite{Xu_2024, johnson2026symmetryprotectedquantumcomputingusing, Jain2024, Zhong2016, Stickler2018}, while identifying non-Clifford operations requires specifying both the encoding and the available controls~\cite{luo2026hybridqubitrotorquantumsystems}. The $N=2$ coupling spaces are candidates for protected encodings: each is a pure code of distance two on the three carriers~\cite{DasRepo2026}. Extensions to strongly anharmonic or large-amplitude molecular motion may also require more general configuration spaces, such as those that treat a bending or torsional coordinate together with the orientation~\cite{HougenBunkerJohns1970, BunkerJensen2006}.

\section*{Acknowledgments}

The authors thank Sergei Yurchenko, Jonathan Tennyson, Eric R. Hudson, Wesley C. Campbell, Hao Wu, Lee R. Liu, and Stephen R. Leone for helpful discussions. E.R.D., J.C.Z., and Y.L. acknowledge support from the U.S. Department of Energy, Office of Science, Basic Energy Sciences, under Award No. DE-SC0026636. E.R.D, V.V.A., and Y.L. also acknowledge the support from the National Science Foundation under Grant No. 2120757, QLCI-CI: Quantum Leap Challenge Institute for Robust Quantum Simulation (via a seed award).

Parts of this manuscript were prepared with assistance from ChatGPT (GPT-5.6), developed by OpenAI, and Claude Code (using Claude Opus 4.6 \& 5), developed by Anthropic. These tools were used to derive explicit basis elements and to edit portions of the manuscript, following the authors’ instructions. 
All content, claims, and conclusions have been reviewed and verified by the authors to ensure accuracy and originality.
Certain commercial products are identified in this publication. Such identification is not intended to imply recommendation or endorsement by NIST.

\section*{Data and code availability}
The species tables, the AME states of Table~\ref{tab:ame_catalogue}, and the code that generates and verifies them, as well as the scripts, inputs, and outputs of the ab initio calculations for calcium monomethoxide (Appendix~\ref{app:caoch3_ham}) are available at Ref.~\cite{DasRepo2026}. The symmetry-adapted vibrational Fock states of single degenerate modes are tabulated in the \PGAtlas~\cite{o3harmonics}. 

\bibliography{references}

\clearpage
\onecolumngrid
\appendix
\setcounter{section}{0}
\setcounter{equation}{0}
\setcounter{table}{0}
\setcounter{figure}{0}

\renewcommand{\thesection}{\Alph{section}}
\renewcommand{\thesubsection}{\arabic{subsection}}

\renewcommand{\theequation}{\thesection.\thesubsection.\arabic{equation}}
\renewcommand{\thetable}{\thesection.\arabic{table}}
\renewcommand{\thefigure}{\thesection.\arabic{figure}}

\counterwithin*{equation}{subsection}
\counterwithin{table}{section}
\counterwithin{figure}{section}

\input{appendix}

\end{document}

%% file: phase_space_overview.tex
\definecolor{secr}{HTML}{1779BF}
\definecolor{secn}{HTML}{A3007A}
\definecolor{secv}{HTML}{009A72}
\definecolor{entg}{HTML}{E00000}
\tikzset{
  fig1/.style={x=1cm,y=1cm,font=\footnotesize,>={Stealth[length=4.5pt,width=3.8pt]},line cap=round,line join=round},
  sector/.style={rectangle,rounded corners=1.5pt,draw=#1,fill=#1!9,line width=0.7pt,
                 minimum width=2.5cm,minimum height=2.5cm,inner sep=2pt},
  basisbox/.style={rectangle,rounded corners=4pt,draw=black!70,fill=black!4,line width=0.7pt,align=center,
                   minimum width=8.3cm,minimum height=1.15cm,inner sep=2pt},
  atomX/.style={circle,fill=black!58,draw=black!75,line width=0.3pt,inner sep=0pt,minimum size=0.27cm},
  atomH/.style={circle,fill=black!16,draw=black!50,line width=0.3pt,inner sep=0pt,minimum size=0.21cm},
  pics/xh3/.style={code={
    \foreach \a in {90,210,330} {\draw[line width=1.0pt,black!50] (0,0)--(\a:0.40);}
    \foreach \a in {90,210,330} {\node[atomH] at (\a:0.40) {};}
    \node[atomX] at (0,0) {};
  }},
  pics/xh3rot/.style={code={
    \pic {xh3};
    \draw[secr,line width=1.0pt,->] (205:0.66) arc[start angle=205,end angle=372,radius=0.66];
  }},
  pics/xh3nuc/.style={code={
    \pic {xh3};
    \foreach \a in {90,210,330} {\draw[secn,line width=1.0pt,->] ($(\a:0.40)+(0,-0.20)$)--($(\a:0.40)+(0,0.24)$);}
  }},
  pics/xh3vib/.style={code={
    \pic {xh3};
    \foreach \a in {90,210,330} {
      \draw[secv,line width=0.8pt] ($(\a:0.40)+(\a-38:0.20)$) arc[start angle=\a-38,end angle=\a+38,radius=0.20];
      \draw[secv,line width=0.8pt] ($(\a:0.40)+(\a-38:0.29)$) arc[start angle=\a-38,end angle=\a+38,radius=0.29];
    }
  }},
}
\begin{tikzpicture}[fig1]
\foreach \nm/\x/\col/\ttl/\spc/\pc in {
    r/-3.0/secr/Rotational/{$L^2\bigl(\grp{O(3)}\bigr)$}/xh3rot,
    n/0.0/secn/Nuclear spin/{$\mathbb C^{d_{\spin}}$}/xh3nuc,
    v/3.0/secv/Vibrational/{$L^2\bigl(\mathbb R^{f}\bigr)$}/xh3vib} {
  \node[sector=\col] (\nm) at (\x,0) {};
  \pic at ($(\nm.north)+(0,-0.84)$) {\pc};
  \node[anchor=base,inner sep=0pt] at ($(\nm.south)+(0,0.66)$) {\textbf{\ttl}};
  \node[anchor=base,inner sep=0pt] at ($(\nm.south)+(0,0.25)$) {\spc};
}
\node[circle,fill=entg,draw=entg!70!black,line width=0.4pt,inner sep=0pt,minimum size=0.74cm,text=white]
  (C) at (0,-3.0) {\normalsize$\boldsymbol{C}$};
\node[anchor=west,text=entg,inner sep=1pt] at ($(C.east)+(0.22,-0.05)$) {\textbf{AME state}};
\draw[secr,line width=2.2pt] (r.south) -- (C.150) node[pos=0.5,below left=-1.5pt,text=secr] {$\Grot$};
\draw[secn,line width=2.2pt] (n.south) -- (C.north) node[pos=0.45,left=1pt,text=secn] {$\Gspin$};
\draw[secv,line width=2.2pt] (v.south) -- (C.30) node[pos=0.5,below right=-1.5pt,text=secv] {$\Gvib$};
\node[anchor=north,inner sep=1pt] (tri) at ($(C.south)+(0,-0.07)$)
  {$({\color{secr}\Grot},{\color{secn}\Gspin},{\color{secv}\Gvib})\downarrow\Gmol$};
\node[basisbox,anchor=north] (b) at ($(tri.south)+(0,-0.48)$) {\textbf{Symmetry-adapted basis sets}\\[1pt]
  position states $\;\longleftrightarrow\;$ momentum states};
\draw[->,line width=0.9pt,draw=entg] (tri.south) -- (b.north);
\end{tikzpicture}

%% file: caoch3_molecule_axes.tex
\definecolor{caoplus}{HTML}{0072B2}
\definecolor{caominus}{HTML}{D55E00}
\begin{tikzpicture}[font=\scriptsize,
  ax/.style={-{Stealth[length=1.3mm]},line width=0.55pt},
  atm/.style={inner sep=0.3pt}]
\node[anchor=south west,inner sep=0] at (0,0) {\includegraphics[width=2cm]{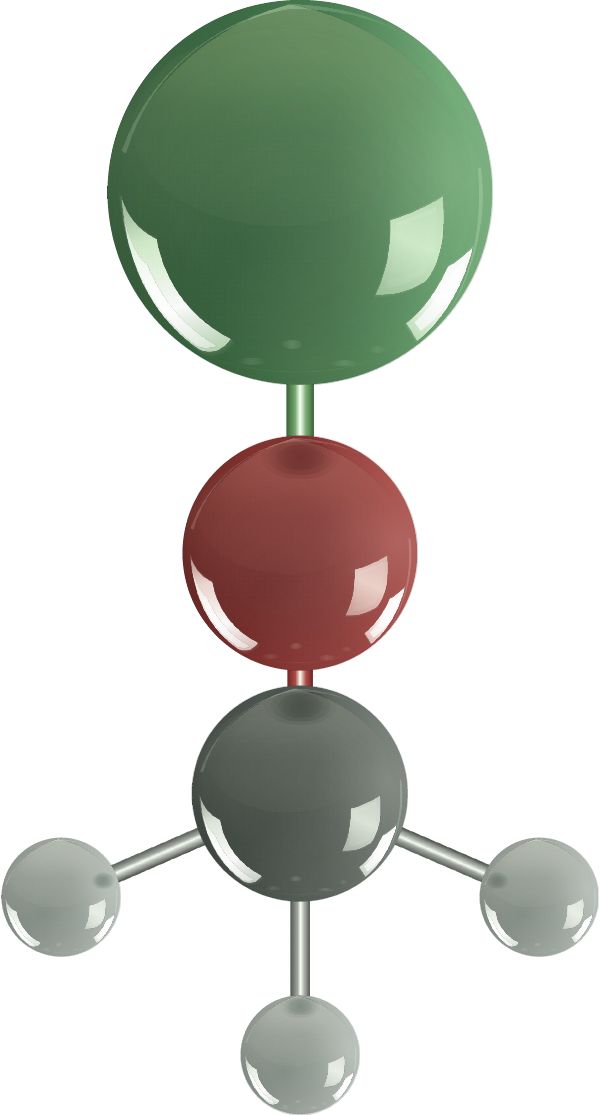}};
\coordinate (o) at (1,2.35);
\draw[black!70,dashed,line width=0.4pt] (1,-0.15)--(1,4.25);
\draw[ax] (1,4.25)--(1,4.6) node[above=0.5pt] {$z$};
\draw[ax] (o)--(0.28,1.92) node[left=0.3pt] {$x$};
\draw[ax] (o)--(2.25,2.35) node[right=0.3pt] {$y$};
\draw[line width=0.45pt,black!45] ({1+0.86*cos(180)},3.05) arc[start angle=180,end angle=360,x radius=0.86cm,y radius=0.21cm];
\draw[line width=0.45pt,black!45] ({1+0.86},3.05) arc[start angle=0,end angle=42,x radius=0.86cm,y radius=0.21cm];
\draw[line width=0.45pt,black!45] ({1-0.86},3.05) arc[start angle=180,end angle=138,x radius=0.86cm,y radius=0.21cm];
\draw[-{Stealth[length=1.3mm]},line width=0.8pt,draw=caoplus]  ({1+0.86*cos(285)},{3.05+0.21*sin(285)}) arc[start angle=285,end angle=345,x radius=0.86cm,y radius=0.21cm];
\draw[-{Stealth[length=1.3mm]},line width=0.8pt,draw=caominus] ({1+0.86*cos(255)},{3.05+0.21*sin(255)}) arc[start angle=255,end angle=195,x radius=0.86cm,y radius=0.21cm];
\node[text=caoplus,anchor=west]  at (1.86,2.9) {$n_+$};
\node[text=caominus,anchor=east] at (0.14,2.9) {$n_-$};
\draw[black!60,densely dotted,line width=0.5pt] (0.05,4.12)--(1,4.12);
\draw[ax,line width=0.7pt] (o)--(0.05,4.12) node[above left=-1pt] {$\mathbf N$};
\draw[line width=0.5pt] (0.93,4.12)--(1.07,4.12) node[right=0.5pt] {$k$};
\node[atm,text=black] at (1.8,3.4) {Ca};
\node[atm,text=black] at (1.5,2) {O};
\node[atm,text=black] at (1.5,1.1) {C};
\node[atm] at (0.22,0.4) {H};
\draw[black!70,dashed,line width=0.4pt] (1.55,-0.15)--(1.55,0.55);
\draw[ax] (1.55,-0.15)--(2.08,0.38) node[above right=-1pt] {lab $Z$};
\draw[line width=0.4pt] (1.55,0.2) arc[start angle=90,end angle=45,radius=0.35cm];
\node at (1.72,0.38) {$\beta$};
\end{tikzpicture}

%% file: ladder_tree_compact.tex
\definecolor{caoplus}{HTML}{0072B2}
\definecolor{caominus}{HTML}{D55E00}
\definecolor{caospec}{HTML}{8C8C8C}
\definecolor{caotarget}{HTML}{C64A17}
\begin{tikzpicture}[x=0.8cm,y=0.0099cm,font=\scriptsize,
  lev/.style={line width=1.1pt},
  lab/.style={above=0.3pt,inner sep=0.6pt,fill=white},
  salc/.style={text=black!55,inner sep=0.4pt},
  drv/.style={{Stealth[length=1.5mm]}-{Stealth[length=1.5mm]},line width=0.8pt},
  spec/.style={{Stealth[length=1.2mm]}-{Stealth[length=1.2mm]},draw=caospec,dashed,line width=0.5pt},
  pl/.style={inner sep=0.6pt,fill=white}]
\draw[line width=0.4pt] (-1.7,-10)--(-1.7,445);
\foreach \e in {0,100,200,300,400}{
  \draw[black!12,line width=0.3pt] (-1.7,\e)--(3.75,\e);
  \draw[line width=0.4pt] (-1.76,\e)--(-1.7,\e) node[left=0.5pt] {$\e$};}
\node[rotate=90] at (-2.72,215) {$(E-E_G)/hc\ (\mathrm{cm^{-1}})$};
\foreach \l in {-1,...,2}{\node[text=black!70] at (\l,-34) {$\l$};}
\node[text=black!70] at (3.03,-34) {$3$};
\node[text=black!70] at (3.72,-34) {$\ell$};
\draw[spec] (1.7,275.068)--(1.22,412.667);
\draw[drv,draw=caoplus]  (0.3,0)--(0.9,137.520)          node[pl,pos=.5,right=1pt] {P1};
\draw[drv,draw=caoplus]  (1.3,137.520)--(1.95,275.068)   node[pl,pos=.5,right=1pt] {P2};
\draw[drv,draw=caominus] (0.7,137.520)--(0.08,277.267)   node[pl,pos=.5,left=1pt] {P3};
\draw[drv,draw=caoplus]  (2.2,275.068)--(2.55,412.642)   node[pl,pos=.5,left=1pt] {P4};
\draw[drv,draw=caoplus]  (2.32,275.068)--(3.2,413.573)   node[pl,pos=.5,right=1pt] {P5};
\draw[drv,draw=caoplus]  (0.32,277.267)--(0.9,412.667)   node[pl,pos=.5,right=1pt] {P6};
\draw[drv,draw=caominus] (-0.3,277.267)--(-0.9,416.110)  node[pl,pos=.5,left=1pt] {P7};
\draw[lev] (-0.36,0)--(0.36,0) node[lab,midway] {$G$};
\draw[lev] (0.64,137.520)--(1.36,137.520) node[lab,midway] {$L_{01}$};
\draw[lev] (1.64,275.068)--(2.36,275.068) node[lab,midway] {$L_{02}$};
\draw[lev] (-0.36,277.267)--(0.36,277.267) node[lab,midway] {$L_{11}$};
\draw[lev,draw=caotarget] (2.42,412.642)--(2.98,412.642) node[lab,midway,text=caotarget] {$L_{03}$};
\draw[lev,draw=caotarget] (3.08,413.573)--(3.64,413.573) node[lab,midway,text=caotarget] {$L_{30}$};
\draw[lev,draw=caotarget] (0.64,412.667)--(1.36,412.667) node[lab,midway,text=caotarget] {$L_{12}$};
\draw[lev,draw=caotarget] (-1.36,416.110)--(-0.64,416.110) node[lab,midway,text=caotarget] {$L_{21}$};
\node[salc] at (2.70,470) {$\irr{a}_{1}$};
\node[salc] at (3.36,471) {$\irr{a}_{2}$};
\node[salc] at (1.00,469) {$\irr{e}$};
\node[salc] at (-1.00,473) {$\irr{e}$};
\end{tikzpicture}

%% file: appendix.tex
\section{Symmetry-adapted Fock encoding}

\subsection{Circular vibrational bases and experimental accessibility}
\label{app:circular_vibrations}

The circular basis is a unitary change of basis within a degenerate vibrational manifold and requires no modification of the molecule. Although the normal-coordinate angular momentum $L_Q=Q_1P_2-Q_2P_1$ generates rotations in the degenerate mode space, its relation to physical molecular angular momentum depends on the underlying atomic displacement patterns \cite{Herzberg1942,Amat1958}.

Circular vibrational excitation requires coherent driving of the two degenerate components with a relative phase of $\pm\pi/2$. For an infrared-active doublet, the interaction
\begin{equation}
H_{\mathrm{int}}
=-\sum_{j=1}^{2}
[\mathbf E(t)\cdot\mathbf d_j]Q_j,
\qquad
\mathbf d_j=
\left.\frac{\partial\boldsymbol{\mu}}{\partial Q_j}\right|_0,
\end{equation}
allows such excitation using resonant circularly polarized light when the transition-dipole derivatives have suitable orientations and strengths. Phase-controlled fields can accommodate more general couplings, while Raman excitation provides an alternative for Raman-active modes. Coherent vibrational excitation and time-resolved optical detection have been demonstrated experimentally \cite{Peschel2022,CoherentIR2024}. Circularly polarized molecular vibrations and their polarization-resolved pump--probe detection have been investigated theoretically \cite{Lin2025}.

\subsection{Proof of Corollary}~\ref{cor:shift}
\label{proof:shift}
Since \(\Gvib\) is one-dimensional, \(V_{\Grot}\otimes V_{\Gvib}\) is an irreducible carrier of species \(\Grot\otimes\Gvib\). Regrouping and exchanging tensor factors gives the natural \(\grp{G}\)-equivariant isomorphism
\[
V_{\Grot}\otimes V_{\Gspin}
\otimes V_{\Gvib}
\cong
V_{\Grot\otimes\Gvib}
\otimes V_{\Gspin}\otimes V_{\Gamma_0},
\]
with \(\Gamma_0\) the trivial irrep of \(\grp{G}\).
It therefore identifies the corresponding \(\Gmol\)-coupling spaces, proving equal multiplicities and simultaneous admissibility. Matching $V_{\Grot}\otimes V_{\Gvib}$ to the standard carrier of $\Grot\otimes\Gvib$ takes a fixed intertwiner, for example $\diag(1,-1)$ for $\irr{e}\otimes\irr{a}_{2}$ in $\grp{C_{3v}}$. After this choice the identification is fixed up to a phase. The nuclear-spin factor is unchanged, so its statistical weight is unchanged.

For Abelian \(\grp{G}\), all carriers are one-dimensional. Admissibility is therefore equivalent to \(
\Grot\otimes\Gspin\otimes\Gvib
=\Gmol,
\)
which has multiplicity one and yields \(\Grot
=\Gmol\otimes\Gspin^*
\otimes\Gvib^*\).
\hfill$\square$

\subsection{Vibrational states of calcium monomethoxide}
\label{app:caoch3}

$\mathrm{CaOCH_3}$ has six nuclei and 12 vibrational degrees of freedom, transforming as $4\irr{a}_{1}\oplus4\irr{e}$ under $\grp{C_{3v}}$, in the frame of Table~\ref{tab:molecular_examples}. With four doubly degenerate sets the vibrational Fock space is richer than that of phosphine, but a single rule assigns all of its carriers. On a product of circular occupation states of the $\irr{e}$ modes, $\bigotimes_s\occ{n_{s-}n_{s+}}$, the threefold rotation acts as $\rootunity{3}^{-\ell}$ with total vibrational angular momentum $\ell=\sum_s(n_{s+}-n_{s-})$, and $\elt{\sigma_{v}}(0)$ exchanges $n_{s+}\leftrightarrow n_{s-}$ in every mode simultaneously. Consequently a product state with $\ell\equiv2\pmod 3$ is a $\irket{\irr{e}}{1}$ carrier with partner $\irket{\irr{e}}{2}=\elt{\sigma_{v}}(0)\irket{\irr{e}}{1}$, a state with $\ell\equiv0$ and not invariant under $\elt{\sigma_{v}}(0)$ generates one $\irr{a}_{1}$ and one $\irr{a}_{2}$ state as $\tfrac{1}{\sqrt{2}}\,(1\pm\elt{\sigma_{v}}(0))$, and a $\elt{\sigma_{v}}(0)$-invariant state with $\ell=0$ is $\irr{a}_{1}$.

For the Ca--O--C bend $\nu_8$ this gives
\begin{align}
v_8=1:&\quad \irket{\irr{e}}{1}=\occ{10},\ \irket{\irr{e}}{2}=\occ{01};\nonumber\\
v_8=2:&\quad \ket{\irr{a}_{1}}=\occ{11},\ \irket{\irr{e}}{1}=\occ{02},\ \irket{\irr{e}}{2}=\occ{20};\nonumber\\
v_8=3:&\quad \ket{\irr{a}_{1}},\ket{\irr{a}_{2}}=\tfrac{1}{\sqrt{2}}\,(\occ{03}\pm\occ{30}),\ \irket{\irr{e}}{1}=\occ{21},\ \irket{\irr{e}}{2}=\occ{12},
\label{eq:caoch3_bend}
\end{align}
in the notation $\occ{n_{8-}n_{8+}}$. The combination band $\nu_7+\nu_8$ gives $\irr{e}\otimes\irr{e}=\irr{a}_{1}\oplus\irr{a}_{2}\oplus\irr{e}$ with $\ket{\irr{a}_{1}},\ket{\irr{a}_{2}}=\tfrac{1}{\sqrt{2}}\,(\ket{\ell_7{=}{+1},\ell_8{=}{-1}}\pm\ket{\ell_7{=}{-1},\ell_8{=}{+1}})$ and $\irket{\irr{e}}{1}=\ket{\ell_7{=}{+1},\ell_8{=}{+1}}$, $\irket{\irr{e}}{2}=\ket{\ell_7{=}{-1},\ell_8{=}{-1}}$. Every vibrational state of the molecule is thereby assigned a carrier, and the admissible species and coupling tensors are those of Table~\ref{tab:ammonia_species}. The $\irr{a}_{2}$ state at $v_8=3$, at $432\ \mathrm{cm^{-1}}$ in the harmonic approximation, is the lowest vibrational level that exchanges the $\irr{a}_{1}$ and $\irr{a}_{2}$ rotational partners of ortho spin (Corollary~\ref{cor:shift}).

\section{Molecular Fourier Transform}

\subsection{Loop monodromy and molecular symmetry}
\label{app:monodromy}

A closed loop in the molecular orientation space \(\Othree/\grp{G}\), when lifted from an orientation \(\orient\), may terminate at \(\orient\gel\), where \(\gel\in \grp{G}\). Because the lifted path remains within a single connected component of \(\Othree\), the corresponding group element must belong to the proper-rotation subgroup \(\grp{G}\cap\SOthree\). The associated monodromy is represented by \(D^{\Grot}(\gel)\). Consequently, non-Abelian loop monodromy requires a noncommutative image of \(\grp{G}\cap\SOthree\) under \(D^{\Grot}\). A multidimensional representation alone is insufficient. This discussion concerns symmetry-induced flat holonomy. Berry and dynamical connections, when present~\cite{Berry1984, Mead1992}, must be treated separately.

\subsection{Proof of Theorem~\ref{thm:fourier}} 
\label{proof:fourier}

Expand a sharp orientation in Wigner functions, $$\ket{\orient}=\sum_{J,\parity,m,k}\sqrt{\frac{2J+1}{16\pi^2}}\,\irrJp{J}{\parity}_{m,k}(\orient)^*\ket{{}^{J,\parity}_{\,m},k},$$ insert it into Eq.~\eqref{eq:coset_states} with $\orient=\crep\gel$, and pass to the $\grp{G}$-adapted basis $\cgket{\Gamma}{\nu}{{}^{J,\parity}_{\,m}\mult}$ (partner $\nu$, copy $\mult$) in which $\cev X_{\gel}$ acts as $D^{\Gamma}_{\nu',\nu}(\gel)\delta_{\mult\mult'}$. Splitting $\irrJp{J}{\parity}(\crep\gel)=\irrJp{J}{\parity}(\crep)\irrJp{J}{\parity}(\gel)$ and applying the Schur orthogonality relation $$\sum_{\gel} D^{\Grot}_{\lambda,\mu}(\gel)D^{\Gamma}_{\nu',\nu}(\gel)^{*}=\frac{|\grp{G}|}{d_{\Grot}}\delta_{\Gamma\Grot}\delta_{\lambda\nu'}\delta_{\mu\nu}$$ collapses the group sum to
\begin{equation}
\ket{\crep;\Grot^{\lambda\mu}}=\sum_{m^{(\mathrm r)}_{\Grot}(J,\parity)>0}\sum_{m,\mult}\sqrt{\frac{(2J+1)|\grp{G}|}{16\pi^2 d_{\Grot}}}\,\irrJp{J}{\parity}_{m;\lambda\mult}(\crep)^*\,\cgket{\Grot}{\mu}{{}^{J,\parity}_{\,m}\mult}.
\end{equation}
The prefactor is fixed by $\sqrt{\frac{2J+1}{16\pi^2}}\cdot\sqrt{\frac{d_{\Grot}}{|\grp{G}|}}\cdot\frac{|\grp{G}|}{d_{\Grot}}$, which is Eq.~\eqref{eq:induced-harmonic}. It reduces to $\sqrt{\frac{(2J+1)|\grp{G}|}{8\pi^2 d}}$ of Ref.~\cite{albert2025quantumtheorymolecularorientations} on replacing $\grp{O(3)}$ by $\grp{SO(3)}$. Contracting the index $\mu$ with the coupling tensor, which acts on neither $(\crep,\lambda)$ nor $(m,\mult)$, gives Eq.~\eqref{eq:ft_forward}. For orthogonality, $\sum_\lambda|\irrJp{J}{\parity}_{m;\lambda\mult}(\crep)|^2$ is invariant under $\crep\to{\crep}\gel$, so its integral over the quotient is $|\grp{G}|^{-1}$ times its integral over $\grp{O(3)}$, which by the Peter--Weyl orthogonality of $\grp{O(3)}$ is $d_{\Grot}\cdot16\pi^2/(2J+1)$. The product with the squared prefactor is one, and the off-diagonal terms vanish by the same argument. Completeness follows from completeness of the Wigner functions on $\grp{O(3)}$ restricted to the $\Grot$-covariant subspace. \hfill$\square$

\subsection{Symmetry operation on different state spaces}
\label{subsec:tour}

Before the full species tables, it is useful to watch what a single symmetry operation does to a concrete state and how the answer changes as sectors are added one at a time. We use phosphine and follow the two generator operations of $\grp{C_{3v}}$: the threefold rotation $\elt{C_{3}}=\rotax{z}(2\pi/3)$, which carries proton $j$ (at azimuth $2\pi(j-1)/3$) to proton $j+1$, and the reflection $\elt{\sigma_{v}}(0)=\elt{i}\,\rotax{y}(\pi)$ in the $xz$ plane, which fixes proton $1$ and exchanges protons $2$ and $3$. We take four state spaces in turn, all at the lowest rotational and vibrational excitation that shows the effect: (i) vibration alone; (ii) rotation and vibration, the rovibrational problem of conventional symmetric-top spectroscopy; (iii) rotation and nuclear spin, the two sectors of Ref.~\cite{albert2025quantumtheorymolecularorientations}; and (iv) all three. Table~\ref{tab:tour} summarizes the four steps. Throughout, $\rootunity{3}=\expe^{2\pi\iunit/3}$.

\emph{(i) Vibration alone.} One quantum in the degenerate bend $\nu_4$ spans a two-dimensional space. In the Cartesian basis, Eq.~\eqref{eq:creation_transform} gives the matrices of the normal-coordinate transformation itself, 
$U(\elt{C_{3}})
=
\frac12\left(\begin{smallmatrix}-1&-\sqrt3\\ \sqrt3&-1\end{smallmatrix}\right),\quad
U(\elt{\sigma_{v}}(0))=\left(\begin{smallmatrix}1&0\\0&-1\end{smallmatrix}\right).$
The rotation does not multiply the two components by a phase. Instead, it turns one displacement pattern into a superposition of both, so no single Cartesian component of $\nu_4$ has a definite symmetry. In the circular basis $\ket{\ell_4{=}{\mp1}}$ the same operations become
\begin{equation}
U(\elt{C_{3}})=\left(\begin{smallmatrix}\rootunity{3}&0\\0&\rootunity{3}^{2}\end{smallmatrix}\right),\qquad
U(\elt{\sigma_{v}}(0))=\left(\begin{smallmatrix}0&1\\1&0\end{smallmatrix}\right),
\label{eq:tour_vib_circ}
\end{equation}
so $\elt{C_{3}}$ multiplies a state of vibrational angular momentum $\ell$ by $\rootunity{3}^{-\ell}$ and $\elt{\sigma_{v}}(0)$ reverses the sense of circulation. The characters $(2,-1,0)$ on $\{\cls{E},\cls{2C_{3}},\cls{3\sigma_{v}}\}$ identify the carrier $\irr{e}$. 

Two quanta span $\{\ket{\ell_4{=}{+2}},\ket{\ell_4{=}0},\ket{\ell_4{=}{-2}}\}$, on which $U(\elt{C_{3}})=\diag(\rootunity{3},1,\rootunity{3}^{2})$ and $\elt{\sigma_{v}}(0)$ exchanges $\ell_4=\pm2$ and fixes $\ell_4=0$. The characters $(3,0,1)$ give $\operatorname{Sym}^{2}\irr{e}=\irr{a}_{1}\oplus \irr{e}$, the $\ell_4=0$ state being totally symmetric. For an experiment this is the familiar infrared selection rule: the dipole transforms as $\irr{a}_{1}\,(\mu_z)\oplus \irr{e}\,(\mu_x,\mu_y)$, so the $\nu_4$ fundamental is a perpendicular band, while $2\nu_4$ has a parallel ($\ell_4=0$) and a perpendicular ($\ell_4=\pm2$) component. What vibration alone cannot supply is a statistical weight: without nuclear spin there is nothing for the spin-statistics condition~\eqref{eq:gamma_mol} to act on.

\emph{(ii) Rotation and vibration.} We add the rotation at $J=1$, $\parity=+1$. By Eq.~\eqref{eq:rot_action} the molecule-frame operations act on the projection $k$ along the symmetry axis as
\begin{equation}
\elt{C_{3}}\ket{{}^{1,+}_{\,m},k}=\rootunity{3}^{k}\ket{{}^{1,+}_{\,m},k},\qquad
\elt{\sigma_{v}}(0)\ket{{}^{1,+}_{\,m},k}=(-1)^{1+k}\ket{{}^{1,+}_{\,m},-k},
\label{eq:tour_rot}
\end{equation}
while the laboratory projection $m$ is untouched. On a product state the two phases combine: $$\elt{C_{3}}\ket{k}\ket{\ell_4}=\rootunity{3}^{k-\ell_4}\ket{k}\ket{\ell_4},$$ and $\elt{\sigma_{v}}(0)$ sends $(k,\ell_4)\to(-k,-\ell_4)$, here with sign $+1$ for $|k|=1$. The four states with $|k|=|\ell_4|=1$ therefore split into two pairs. The pair $(k,\ell_4)=(1,1),(-1,-1)$ has $k-\ell_4=0$, is invariant under $\elt{C_{3}}$, and is exchanged by $\elt{\sigma_{v}}(0)$, so it gives
\begin{equation}
\begin{aligned}
    \tfrac{1}{\sqrt{2}}\,\bigl(\ket{1,{+1}}+\ket{{-1},{-1}}\bigr)\in \irr{a}_{1},\\
    \tfrac{1}{\sqrt{2}}\,\bigl(\ket{1,{+1}}-\ket{{-1},{-1}}\bigr)\in \irr{a}_{2},
\end{aligned}
\label{eq:tour_rovib}
\end{equation}
whereas the pair $(1,-1),(-1,1)$ acquires $\rootunity{3}^{2}$ and $\rootunity{3}$ and forms an $\irr{e}$ carrier. The $k=0$ state ($\irr{a}_{2}$) times $\nu_4$ ($\irr{e}$) gives a second $\irr{e}$. The six-dimensional space has characters $(6,0,0)$ and decomposes as $\irr{a}_{1}\oplus \irr{a}_{2}\oplus2\irr{e}$. This is the rule $k-\ell\equiv0\pmod 3$ for $\irr{a}$-type rovibrational levels, and Eq.~\eqref{eq:tour_rovib} shows where the $\ell$-type doubling of the $k\ell>0$ levels comes from: those levels form an $\irr{a}_{1},\irr{a}_{2}$ pair that a symmetry-allowed term can split, whereas the $k\ell<0$ levels form a degenerate $\irr{e}$ pair that no symmetric term can split. The perpendicular-band selection rule $\Delta k=\Delta\ell=\pm1$ with equal signs is the statement that the dipole preserves $k-\ell\bmod3$. Again, however, this space carries no spin, so it cannot tell which of these levels exist.

\emph{(iii) Rotation and nuclear spin.} Return to the vibrational ground state and add the three proton spins, the setting of Ref.~\cite{albert2025quantumtheorymolecularorientations}. Now each operation also permutes nuclei: $\elt{C_{3}}$ cycles the spins, $\ket{s_1s_2s_3}\to\ket{s_3s_1s_2}$, and $\elt{\sigma_{v}}(0)$ transposes the spins of protons $2$ and $3$. Because the protons are fermions, a physical state must be multiplied by $\chi_{\irr{a}_{2}}=(+1,+1,-1)$, i.e.\ it must be invariant under $\elt{C_{3}}$ and change sign under $\elt{\sigma_{v}}(0)$. Three trial states show how the rule operates at $J=1$, $\parity=+1$.
\begin{itemize}
\item $\ket{{}^{1,+}_{\,m},0}\otimes\ket{\uparrow\uparrow\uparrow}$: $\elt{C_{3}}$ gives $1\cdot1$ and $\elt{\sigma_{v}}(0)$ gives $(-1)\cdot(+1)=-1$. It is physical: an ortho state, one of four, since all four $I_{\mathrm{tot}}=\tfrac32$ spin states behave alike.
\item $\ket{{}^{1,+}_{\,m},0}\otimes\cgket{\irr{e}}{1}{\tfrac12\tfrac12}$: $\elt{C_{3}}$ gives $\rootunity{3}\neq1$. It is not physical, and no linear combination with other $k=0$ states can make it so.
\item $\tfrac{1}{\sqrt{2}}\,\bigl(\ket{{}^{1,+}_{\,m},1}\cgket{\irr{e}}{2}{\tfrac12M}-\ket{{}^{1,+}_{\,m},{-1}}\cgket{\irr{e}}{1}{\tfrac12M}\bigr)$: $\elt{C_{3}}$ gives $\rootunity{3}\cdot\rootunity{3}^{2}=1$, and $\elt{\sigma_{v}}(0)$ exchanges the two terms and so gives $-1$. It is physical: a para state, one for each $M=\pm\tfrac12$.
\end{itemize}
The $24$-dimensional space has characters $(24,0,-4)$ and contains $\irr{a}_{2}$ six times: four ortho states at $k=0$ and two para states at $|k|=1$, per laboratory projection. At $J=0$ the same count gives no physical state for $\parity=+1$ and four for $\parity=-1$. These are the observables of this step: nuclear-spin statistical weights ($4$ for $K=|k|=0,3,6,\ldots$, $2$ for the other $K$, before the $^{31}$P factor of $2$) and missing levels (only one parity component of each $K=0$ level exists).

\emph{(iv) Rotation, nuclear spin, and vibration.} Finally, excite one $\nu_4$ quantum with all three sectors present. The perrotation is the tensor product of the three actions above, $\cev X_{\gel}\otimes\mathrm{Perm}(\gel)\otimes U_{\mathrm{v}}(\gel)$, so on a product state the phases of all three factors multiply. The $48$-dimensional space at $J=1$, $\parity=+1$ has characters $(48,0,0)$ and contains $\irr{a}_{2}$ eight times. The eight physical states are:
\begin{itemize}
\item four \emph{ortho} states, $\tfrac{1}{\sqrt{2}}\,\bigl(\ket{1,{+1}}-\ket{{-1},{-1}}\bigr)\otimes\ket{\tfrac32,M_I}$: the $\irr{a}_{2}$ rovibrational combination of Eq.~\eqref{eq:tour_rovib}, species $(\irr{e},\irr{a}_{1},\irr{e})$;
\item two \emph{para} states of species $(\irr{a}_{2},\irr{e},\irr{e})$, $\ket{{}^{1,+}_{\,m},0}\otimes\tfrac{1}{\sqrt{2}}\,\bigl(\cgket{\irr{e}}{1}{\tfrac12M}\ket{\ell_4{=}{+1}}+\cgket{\irr{e}}{2}{\tfrac12M}\ket{\ell_4{=}{-1}}\bigr)$;
\item two \emph{para} states of species $(\irr{e},\irr{e},\irr{e})$, the GHZ states $\catT{1}$ of Eq.~\eqref{eq:nh3_ghz_momentum}.
\end{itemize}
Checking the last one directly: $\elt{C_{3}}$ multiplies $\ket{k{=}1}\irket{\irr{e}}{1}_{\mathrm{n}}\ket{\ell_4{=}{-1}}$ by $\rootunity{3}\cdot\rootunity{3}\cdot\rootunity{3}=1$, and $\elt{\sigma_{v}}(0)$ exchanges it with $\ket{k{=}{-1}}\irket{\irr{e}}{2}_{\mathrm{n}}\ket{\ell_4{=}{+1}}$, so the antisymmetric combination changes sign, as $\Gmol=\irr{a}_{2}$ requires. 

Step (iv) also changes what an experiment sees. In the vibrational ground state every $|k|=1$ level is para. With one quantum in $\nu_4$, the $|k|=1$ levels divide by the relative sign of $k$ and $\ell_4$: the $k\ell>0$ level becomes ortho, with only its $\irr{a}_{2}$ $\ell$-doubling component present (its $\irr{a}_{1}$ partner has no spin partner and is missing~\cite{Oka1973}), and the $k\ell<0$ level remains para. Counting only the proton factor, the statistical weights of the subbands of a perpendicular band therefore follow $k-\ell\bmod3$ rather than $k\bmod3$: weight $4$ when $k-\ell\equiv0$ and weight $2$ otherwise. This is the familiar strong--weak--weak alternation of $\Delta K=\pm1$ subbands~\cite{Hougen1962}, obtained here from the same projection that produces the entangled states. For a rigid $\grp{C_{3v}}$ top such as $\mathrm{PH_3}$ or $\mathrm{CaOCH_3}$ (Sec.~\ref{sec:experiment}) the counting is exactly as stated, while in ammonia each level with $k\neq0$ is in addition split into an inversion doublet.

\begin{table*}[t]
\caption{The four steps of Sec.~\ref{subsec:tour} for phosphine}, at $J=1$, $\parity=+1$ where rotation is present, per laboratory projection $m$. Characters are on $\{\cls{E},\cls{2C_{3}},\cls{3\sigma_{v}}\}$. ``Physical'' is the multiplicity of $\Gmol=\irr{a}_{2}$, which can be imposed only once nuclear spin is present.
\label{tab:tour}
\centering
\small
\begin{tabular}{lcccll}
\toprule
State space & dim & $\chi$ & decomposition & physical ($\irr{a}_{2}$) & what it predicts\\
\midrule
(i) $\nu_4$ & 2 & $(2,-1,0)$ & $\irr{e}$ & --- & perpendicular IR band\\
\phantom{(i)} $2\nu_4$ & 3 & $(3,0,1)$ & $\irr{a}_{1}\oplus \irr{e}$ & --- & parallel and perpendicular components\\
(ii) rotation $\otimes$ $\nu_4$ & 6 & $(6,0,0)$ & $\irr{a}_{1}\oplus \irr{a}_{2}\oplus2\irr{e}$ & --- & $k-\ell$ rule, $\ell$-type doubling\\
(iii) rotation $\otimes$ spin & 24 & $(24,0,-4)$ & $2\irr{a}_{1}\oplus6\irr{a}_{2}\oplus8\irr{e}$ & 6 (4 ortho, 2 para) & statistical weights, missing levels\\
(iv) rotation $\otimes$ spin $\otimes$ $\nu_4$ & 48 & $(48,0,0)$ & $8\irr{a}_{1}\oplus8\irr{a}_{2}\oplus16\irr{e}$ & 8 (4 ortho, 4 para) & weights set by $k-\ell$; GHZ state $\catT{1}$\\
\bottomrule
\end{tabular}
\end{table*}

\section{Experimental preparation of SALCs in calcium monomethoxide}
\label{app:caoch3_prep}

This section of appendix details the protocol in Sec.~\ref{sec:experiment} for preparing coherent superpositions associated with molecular symmetry-adapted linear combinations (SALCs) in calcium monomethoxide. Within the stated approximations, we derive the level energies and transition frequencies and the terms in the Hamiltonian (Appendix~\ref{app:caoch3_ham}), dipole matrix elements (Appendix~\ref{app:caoch3_dipoles}), and pulse sequence and experimental requirements (Appendix~\ref{app:caoch3_pulses}).

\subsection{Hamiltonian of calcium monomethoxide}
\label{app:caoch3_ham}

We calculate every level energy and transition wavenumber in this appendix from Eq.~\eqref{eq:caoch3_hamiltonian}~\cite{JohnstonDennison1935,Herzberg1945,Nielsen1951}, the same finite set of terms that the symmetry analysis in Sec.~\ref{sec:experiment} uses. Its constants combine experimental data and ab initio calculations, and the upper block of Table~\ref{tab:caoch3_constants} lists the values we use. The terms missing from Eq.~\eqref{eq:caoch3_hamiltonian} are not added to these energies. Instead, we estimate the size of each missing term and state what it requires of the pulses. The lower block of Table~\ref{tab:caoch3_constants} lists the constants used for these estimates.

\emph{Terms of Eq.~\eqref{eq:caoch3_hamiltonian}.} The $\nu_8$ fundamental and rotational constants $A$ and $B$ are measured~\cite{Jacox2003,Kozyryev2019,Namiki1998,Crozet2002}. We estimate the remaining parameters using PySCF~\cite{Sun2020PySCF}. Coupled-cluster normal modes give $\zeta_8=0.926$, while second-order vibrational perturbation theory gives $x_{8,8}=-0.53\ \mathrm{cm^{-1}}$ and the anharmonic contribution $g^{\mathrm{anh}}_{8,8}=0.63\ \mathrm{cm^{-1}}$~\cite{Nielsen1951,Watson1968}. The anharmonic estimates vary by up to $0.11\ \mathrm{cm^{-1}}$ with finite-difference step size and density functional. We include the bend-angular-momentum contribution through $g_{8,8}=A\zeta_8^2+g^{\mathrm{anh}}_{8,8}$~\cite{Watson1968} and choose $\tilde\omega_8=\nu_8-g_{8,8}$, so that Eq.~\eqref{eq:caoch3_hamiltonian} reproduces the measured fundamental. We use its central value, $\nu_8=144\ \mathrm{cm^{-1}}$. Its uncertainty of $\pm5\ \mathrm{cm^{-1}}$ shifts all seven transitions of the tree together and leaves their spacings unchanged.

\emph{Terms beyond Eq.~\eqref{eq:caoch3_hamiltonian}.} Two terms omitted in Sec.~\ref{sec:experiment} have measured ground-state constants. Neither enters the energies and wavenumbers quoted in this appendix, and we use each only as follows. Centrifugal distortion adds $-D_NN^2(N+1)^2-D_{NK}N(N+1)k^2-D_Kk^4$ to Eq.~\eqref{eq:caoch3_hamiltonian}~\cite{Namiki1998,Crozet2002}. Within the preparation tree, it would shift the energies relative to $G$ and the transition wavenumbers by at most $0.013\ \mathrm{cm^{-1}}$. This is about a tenth of the spread in the ab initio constants above, so adding it would not make the predicted lines more accurate, and we leave it out. Spin--rotation and proton hyperfine interactions~\cite{Namiki1998} split each rovibrational level into $J$ and $F$ components and shift the tree levels by at most $25$~MHz ($0.0008\ \mathrm{cm^{-1}}$). We omit these shifts from the energies as well. Each pulse must still address a single component, so we use this spin structure to set the Rabi frequency in Appendix~\ref{app:caoch3_pulses}.

The remaining terms have no measured constants for the bend levels, and we neglect them. We use ground-state rotational, distortion, and spin constants for all bend levels, neglecting their $v$ dependence. We also neglect $\ell$-type doubling ($\Delta k=\Delta\ell=\pm2$), whose $\nu_8$ constant is unmeasured~\cite{Nielsen1951}. In polar bend coordinates $(\rho,\vartheta)$, a cubic potential term $\propto\rho^3\cos3\vartheta$ mixes $\ell$ values differing by three. Its perturbative energy shifts, and those from cubic couplings to other modes, enter the constants of Eq.~\eqref{eq:caoch3_hamiltonian}, but we neglect the associated state mixing. All terms beyond Eq.~\eqref{eq:caoch3_hamiltonian} preserve molecular point group symmetry and can mix only levels of the same point group irreps~\cite{BunkerJensen2006}. Thus the species triples $(\Grot,\Gspin,\Gvib)$ in Table~\ref{tab:caoch3_levels} become approximate labels, while every tree level retains $\Gmol=\irr{a}_{2}$.

\begin{table}[h]
\centering
\small
\caption{Constants of the $\widetilde X\,{}^2A_1$ ground state, grouped by use. The upper block holds the constants of Eq.~\eqref{eq:caoch3_hamiltonian}; every level energy and transition wavenumber of the preparation tree is computed from these alone. The last two entries of this block are derived from the others and rounded here, and the calculations use the unrounded values. The lower block holds the constants of terms beyond Eq.~\eqref{eq:caoch3_hamiltonian}, which enter no tabulated energy: the distortion constants give only the size of the omitted shifts, and the spin constants set only the Rabi frequency $\Omega$ in Appendix~\ref{app:caoch3_pulses}. ``This work'' marks ab initio estimates; the other entries are literature values.}
\label{tab:caoch3_constants}
\begin{tabular}{lll}
\toprule
Quantity & Value & Source\\
\midrule
\multicolumn{3}{l}{\emph{Eq.~\eqref{eq:caoch3_hamiltonian}, used for all energies and wavenumbers}}\\
$\nu_8$ fundamental & $144\pm5\ \mathrm{cm^{-1}}$ & Refs.~\cite{Jacox2003,Kozyryev2019}\\
$A$ & $5.448303\ \mathrm{cm^{-1}}$ & Ref.~\cite{Crozet2002}\\
$B$ & $0.11626491\ \mathrm{cm^{-1}}$ ($3485.534(2)$~MHz) & Refs.~\cite{Namiki1998,Crozet2002}\\
$\zeta_8$ & $0.926$ & EA-EOM-CCSD, this work\\
$g_{8,8}=A\zeta_8^2+g^{\mathrm{anh}}_{8,8}$ & $4.672+0.63=5.30\ \mathrm{cm^{-1}}$ & $g^{\mathrm{anh}}_{8,8}$: B3LYP, this work\\
$x_{8,8}$ & $-0.53\ \mathrm{cm^{-1}}$ & B3LYP, this work\\
$\tilde\omega_8=\nu_8-g_{8,8}$ & $138.70\ \mathrm{cm^{-1}}$ & from $\nu_8$ and $g_{8,8}$\\
\midrule
\multicolumn{3}{l}{\emph{Terms beyond Eq.~\eqref{eq:caoch3_hamiltonian}, used only for estimates}}\\
$D_N$, $D_{NK}$ & $2.6885\times10^{-8}$, $2.3580\times10^{-6}\ \mathrm{cm^{-1}}$ & Refs.~\cite{Namiki1998,Crozet2002}\\
$D_K$ & $7.03\times10^{-5}\ \mathrm{cm^{-1}}$ & Ref.~\cite{Crozet2002}\\
$\varepsilon_{bc}$ (spin--rotation) & $12.451(11)$~MHz & Ref.~\cite{Namiki1998}\\
$a_F$, $T_{aa}$ (${}^1$H hyperfine) & $-0.421(27)$, $1.070(45)$~MHz & Ref.~\cite{Namiki1998}\\
\bottomrule
\end{tabular}
\end{table}
The preparation tree follows $k=\ell-3$, with $N=3$ or $4$ (Table~\ref{tab:caoch3_levels}). For a transition $(v,\ell,N,k)\to(v+1,\ell+q,N',k+q)$ with $q=\pm1$, Eq.~\eqref{eq:caoch3_hamiltonian} gives
\begin{equation}
\begin{aligned}
\tilde\nu={}&\tilde\omega_8+2x_{8,8}v+g_{8,8}(2q\ell+1)\\
&+B\bigl[N'(N'+1)-N(N+1)\bigr]+(A-B)(2qk+1)\\
&-2A\zeta_8\bigl[q(k+\ell)+1\bigr].
\end{aligned}
\label{eq:caoch3_lines}
\end{equation}
Table~\ref{tab:caoch3_pulses} lists the resulting wavenumbers of the seven transitions.

P1, P2, and P4 are consecutive $q=+1$ transitions with $\Delta N=0$. From one to the next, $v$, $\ell$, and $k$ of the lower level each increase by one, so Eq.~\eqref{eq:caoch3_lines} places these lines at equal spacings of $2x_{8,8}+2g^{\mathrm{anh}}_{8,8}+2A(1-\zeta_8)^2-2B=0.027\ \mathrm{cm^{-1}}$, or $0.81$~GHz. The anharmonic part, $+0.20\ \mathrm{cm^{-1}}$, and the rotational and Coriolis part, $-0.17\ \mathrm{cm^{-1}}$, nearly cancel. The spacing is therefore smaller than the spread in the ab initio constants and sensitive to the neglected $v$ dependence of the constants, so these lines require further spectroscopic measurements.

Competing transitions that start or end on a populated level lie at least $0.73$~GHz from every tree line, with one exception. In Eq.~\eqref{eq:caoch3_lines}, a line with $\Delta N=0$ does not depend on $N$. P4 therefore coincides with its $N=4$ counterpart, the transition between the level with $v=2$, $\ell=2$, $k=-1$, $N=4$ and $L_{30}$. Terms beyond Eq.~\eqref{eq:caoch3_hamiltonian} lift this coincidence: centrifugal distortion alone separates the two lines by $0.57$~MHz, and the unmeasured $v$ dependence of $B$ adds to it. Because this separation is small and uncertain, the protocol does not rely on it: P4 must finish before P5 populates $L_{30}$ (Appendix~\ref{app:caoch3_pulses}).

For transition strengths, let $\ket{k}$ denote the rotational state at the specified $N$, and $\ket{\irr{a}_{1}}_{\mathrm n}$ the ortho nuclear-spin state with $I=\tfrac32$. The circular bend occupations $n_\pm$ give $v=n_++n_-$ and $\ell=n_+-n_-$. Suppressing spin-coupling labels, each tree state combines two orthogonal reflection partners:
\begin{equation}
\ket{L_{n_-n_+}}
=\frac{1}{\sqrt2}
\Bigl(
\ket{\ell-3}\occ{n_-n_+}
+s_N\ket{3-\ell}\occ{n_+n_-}
\Bigr)\ket{\irr{a}_{1}}_{\mathrm n}.
\label{eq:caoch3_levels}
\end{equation}
The relative sign enforces $\Gmol=\irr{a}_{2}$:
$s_N=-(-1)^{N+k}\parity=(-1)^{N+1}$, using $k=\ell-3$, parity $\parity=(-1)^{v+1}$, and even $\ell+v$.
At $k=0$, $L_{03}$ ($N=3$) and $L_{30}$ ($N=4$) select bend species $\irr{a}_{1}$ and $\irr{a}_{2}$, respectively, realizing Corollary~\ref{cor:shift}. Their subscripts name levels: both use the first term $\ket{0}\occ{03}$ of Eq.~\eqref{eq:caoch3_levels}, with $N=3$ and $N=4$.

The amplitudes in Eq.~\eqref{eq:caoch3_target} describe coherence between complete molecular levels. After tracing out rotation and spin, the bend density operator remains invariant under molecular point group symmetry. Schur's lemma~\cite{hamermesh1989} therefore forbids coherence between inequivalent bend species and requires equal populations and no coherence between the two $\irr{e}$ rows. Only $L_{03}$ and $L_{30}$ individually leave the bend in a pure state.

\begin{table}[h]
\centering
\small
\caption{Preparation-tree levels, all with ortho nuclear spin ($I=\tfrac32$) and $M_F=0$. The $\ell$ and $k$ entries refer to the first component of Eq.~\eqref{eq:caoch3_levels}, except for $L_{03}$ and $L_{30}$, whose components have $k=0$ and $\ell=\pm3$. The state column uses the notation of Table~\ref{tab:ammonia_species}. Energies follow from Eq.~\eqref{eq:caoch3_hamiltonian} alone, with the constants in the upper block of Table~\ref{tab:caoch3_constants}; they include neither centrifugal distortion nor the spin--rotation and hyperfine shifts (Appendix~\ref{app:caoch3_ham}).}
\label{tab:caoch3_levels}
\begin{tabular}{lrrrrrrcccr}
\toprule
Level & $v$ & $\ell$ & $k$ & $N$ & $J$ & $F$ & $(\Grot,\Gspin,\Gvib)$ & state & role & $(E-E_G)/hc$ ($\mathrm{cm^{-1}}$)\\
\midrule
$G=L_{00}$ & 0 & 0 & $-3$ & 3 & $\tfrac72$ & 2 & $(\irr{a}_{2},\irr{a}_{1},\irr{a}_{1})$ & $\catP$ & start & 0.000\\
$L_{01}$ & 1 & 1 & $-2$ & 3 & $\tfrac72$ & 3 & $(\irr{e},\irr{a}_{1},\irr{e})$ & $\catB{1}$ (r:v) & tree & 137.520\\
$L_{02}$ & 2 & 2 & $-1$ & 3 & $\tfrac72$ & 4 & $(\irr{e},\irr{a}_{1},\irr{e})$ & $\catB{1}$ (r:v) & tree & 275.068\\
$L_{11}$ & 2 & 0 & $-3$ & 4 & $\tfrac92$ & 4 & $(\irr{a}_{2},\irr{a}_{1},\irr{a}_{1})$ & $\catP$ & tree & 277.267\\
$L_{03}$ & 3 & $\pm3$ & 0 & 3 & $\tfrac72$ & 5 & $(\irr{a}_{2},\irr{a}_{1},\irr{a}_{1})$ & $\catP$ & target & 412.642\\
$L_{30}$ & 3 & $\pm3$ & 0 & 4 & $\tfrac92$ & 5 & $(\irr{a}_{1},\irr{a}_{1},\irr{a}_{2})$ & $\catP$ & target & 413.573\\
$L_{12}$ & 3 & 1 & $-2$ & 4 & $\tfrac92$ & 5 & $(\irr{e},\irr{a}_{1},\irr{e})$ & $\catB{1}$ (r:v) & target & 412.667\\
$L_{21}$ & 3 & $-1$ & $-4$ & 4 & $\tfrac92$ & 5 & $(\irr{e},\irr{a}_{1},\irr{e})$ & $\catB{1}$ (r:v) & target & 416.110\\
\bottomrule
\end{tabular}
\end{table}

\subsection{Transition dipole matrix elements}
\label{app:caoch3_dipoles}

For an electric field polarized along the laboratory axis $Z$, the transition dipole matrix element $d=\bra{u}\mu_Z\ket{\mathrm{lo}}$ couples the lower and upper levels. We calculate $d$ in four steps: transform $\mu_Z$ to the molecular frame, evaluate the bend and rotational matrix elements, and recouple the spins~\cite{Edmonds1957,Zare1988,BrownCarrington2003}. We follow Ref.~\cite{Edmonds1957} for Wigner functions, angular-momentum coupling coefficients ($3j$ and $6j$ symbols), and reduced matrix elements.

\emph{Frame transformation.} The spherical dipole components are $\mu_{\pm1}=\mp\tfrac{1}{\sqrt2}\,(\mu_x\pm\iunit\mu_y)$ and $\mu_0=\mu_z$~\cite{Zare1988,BrownCarrington2003}. The molecular orientation $\orient$ is the rotation that carries the laboratory axes $(X,Y,Z)$ into the molecule-fixed axes $(x,y,z)$. It is specified by three Euler angles~\cite{Zare1988}: the polar coordinates (polar angle and azimuth) of the molecular $z$ axis in the laboratory frame, and the angle of rotation of the molecule about that axis. Rotating to the laboratory frame gives $\mu_Z=\sum_q [\wignerD{1}_{0q}(\orient)]^*\mu_q$, where $q=0,\pm1$ labels molecular-frame components and $\wignerD{1}_{0q}(\orient)=\bra{1,0}e^{-\iunit\alpha J_z}e^{-\iunit\beta J_y}e^{-\iunit\gamma J_z}\ket{1,q}=e^{-\iunit q\gamma}d^{1}_{0q}(\beta)$ is the Wigner $D$ function of the Euler angles $(\alpha,\beta,\gamma)$ of $\orient$~\cite{Zare1988}. The complex conjugate is needed because $(\alpha,\beta,\gamma)$ orient the molecular axes relative to the laboratory axes. Each coefficient $[\wignerD{1}_{0q}(\orient)]^*$ is then the direction cosine between the laboratory $Z$ axis and the molecular spherical component $q$. In Cartesian form, $\mu_Z=\cos\beta\,\mu_z-\sin\beta\cos\gamma\,\mu_x+\sin\beta\sin\gamma\,\mu_y$, and without the asterisk the $\mu_y$ term would change sign. Thus, a single laboratory polarization includes both $q=+1$ and $q=-1$ components.

\emph{Bend.} Symmetry places the permanent dipole along $z$ and confines its first-order response to an $\irr{e}$ displacement to the $x$--$y$ plane. The dipole derivatives for the $\nu_8$ coordinates $(Q_1,Q_2)$ of Appendix~\ref{app:circular_vibrations} are orthogonal and equal in magnitude~\cite{Herzberg1945,BunkerJensen2006}. Let $\mu_8=\bra{1}\mu_x\ket{0}$ be the fundamental transition dipole moment along one coordinate, and let $a_j$ and $a_j^\dagger$ annihilate and create Cartesian oscillator quanta. The linear dipole operators are then $\mu_x=\mu_8(a_1+a_1^\dagger)$ and $\mu_y=\mu_8(a_2+a_2^\dagger)$. Transforming to the circular operators of Appendix~\ref{app:circular_vibrations} gives $\mu_q=-q\mu_8(a_{-q}+a_q^\dagger)$ for $q=\pm1$.

The creation operator $a_q^\dagger$ adds one bend quantum with vibrational angular momentum $q\hbar$ about $z$, which circulates counterclockwise for $q=+1$ and clockwise for $q=-1$. The annihilation operator $a_{-q}$ removes one quantum circulating in the opposite direction. These processes change $v$ by $+1$ and $-1$, respectively, and, because $\ell=n_+-n_-$, both change $\ell$ by $q$~\cite{Nielsen1951}. Because $\mu_0$ has no first-order bend dependence, $\Delta\ell=0$ bend transitions are absent in this approximation. For an upward transition, the creation-operator matrix element gives
\begin{equation}
V_q=\bra{n_q+1}\mu_q\ket{n_q}=-q\mu_8\sqrt{n_q+1},\qquad n_{+1}=n_+,\quad n_{-1}=n_-.
\label{eq:caoch3_bend_factor}
\end{equation}

\emph{Rotation.} For laboratory projection $M_N$, the normalized symmetric-top wavefunction is $\langle\orient|NkM_N\rangle=[(2N+1)/8\pi^2]^{1/2}\,\wignerD{N}_{M_Nk}(\orient)^*$. Substituting into the matrix element of $\wignerD{1}_{0q}(\orient)^*$ gives an integral of three Wigner functions, yielding two $3j$ symbols~\cite{Edmonds1957,Zare1988}. Primes denote upper-level quantum numbers:
\begin{equation}
\bra{N'k'M_N'}\wignerD{1}_{0q}(\orient)^*\ket{NkM_N}=(-1)^{M_N'-k'}\sqrt{(2N'+1)(2N+1)}\begin{pmatrix}N'&1&N\\-M_N'&0&M_N\end{pmatrix}\begin{pmatrix}N'&1&N\\-k'&q&k\end{pmatrix}.
\label{eq:caoch3_rotation_me}
\end{equation}
Both symbols require $N'=N$ or $N\pm1$. The first gives the laboratory-projection dependence, and the second requires $k'=k+q$. Together with the bend selection rule, this gives the perpendicular-band rule $\Delta k=\Delta\ell=q$~\cite{Herzberg1945}, which conserves $k-\ell$ exactly. Molecular point group symmetry requires only conservation of $k-\ell\bmod3$, the residue determining compatibility with ortho or para spin (Appendix~\ref{subsec:tour}), because the dipole preserves nuclear spin.

\emph{Spins.} Neither $\mu_q$ nor $\wignerD{1}_{0q}$ acts on electron spin $S$ or proton spin $I$, so the ortho state $\ket{\irr{a}_{1}}_{\mathrm{n}}$ is conserved. Conservation of $k-\ell$ restricts the branch and mirror components of Eq.~\eqref{eq:caoch3_levels} to their respective upper-level partners.

In the coupled basis, an operator acting on rotation acquires one $6j$ symbol for each spin recoupling~\cite{Edmonds1957,BrownCarrington2003}. The two recouplings, $\mathbf J=\mathbf N+\mathbf S$ and $\mathbf F=\mathbf J+\mathbf I$, impose $\Delta J=0,\pm1$ and $\Delta F=0,\pm1$. The Wigner--Eckart theorem~\cite{Edmonds1957} supplies the dependence on the laboratory projection $M_F$ and requires $\Delta M_F=0$ for $Z$ polarization.

Combining these steps gives the branch-component matrix element $d_a$, before including the normalization of Eq.~\eqref{eq:caoch3_levels}:
\begin{align}
d_a={}&f_Mf_If_Sf_N\,V_q,\qquad
f_M=(-1)^{F'-M_F}\begin{pmatrix}F'&1&F\\-M_F&0&M_F\end{pmatrix},\nonumber\\
f_I={}&(-1)^{J'+I+F+1}\sqrt{(2F'+1)(2F+1)}\begin{Bmatrix}J'&F'&I\\F&J&1\end{Bmatrix},\qquad
f_S=(-1)^{N'+S+J+1}\sqrt{(2J'+1)(2J+1)}\begin{Bmatrix}N'&J'&S\\J&N&1\end{Bmatrix},\nonumber\\
f_N={}&(-1)^{N'-k'}\sqrt{(2N'+1)(2N+1)}\begin{pmatrix}N'&1&N\\-k'&q&k\end{pmatrix},
\label{eq:caoch3_dipole_factors}
\end{align}
where the curly (round) brackets denote Wigner $6j$ ($3j$) symbols~\cite{Edmonds1957}, and $f_N$ is the reduced rotational matrix element from Eq.~\eqref{eq:caoch3_rotation_me}. The factors $f_S$ and $f_I$ account for spin recoupling, and $f_M$ restores the laboratory-projection dependence.

\subsection{Pulse sequence and experimental requirements}
\label{app:caoch3_pulses}

Using the transition frequencies and dipole matrix elements, we construct selective pulses that distribute population through the tree and set the target amplitudes' relative phases.

\emph{Gate.} For pulse area $\Theta$ and phase $\phi$, the gate in Eq.~\eqref{eq:caoch3_gate} has the following matrix in the ordered basis $(\ket{\mathrm{lo}},\ket{u})$:
\begin{equation}
R_{u,\mathrm{lo}}(\Theta,\phi)=\begin{pmatrix}\cos\tfrac\Theta2&-\iunit\expe^{-\iunit\phi}\sin\tfrac\Theta2\\[2pt]-\iunit\expe^{\iunit\phi}\sin\tfrac\Theta2&\cos\tfrac\Theta2\end{pmatrix},
\label{eq:caoch3_gate_matrix}
\end{equation}
The gate leaves all other levels unchanged and transfers amplitude in both directions along the driven link. On the Bloch sphere, which represents two-level pure states, it rotates by $\Theta$ about an equatorial axis of azimuth $\phi$~\cite{AllenEberly1975,Nielsen_Chuang_2010}. A $\pi$ pulse exchanges the levels with phases $\ket{\mathrm{lo}}\to-\iunit\expe^{\iunit\phi}\ket{u}$ and $\ket{u}\to-\iunit\expe^{-\iunit\phi}\ket{\mathrm{lo}}$. A $\pi/2$ pulse creates an equal-weight superposition from either level. Since $R_{u,\mathrm{lo}}(\Theta,\phi)^{-1}=R_{u,\mathrm{lo}}(\Theta,\phi+\pi)$, reversing the pulse order and shifting each phase by $\pi$ inverts the sequence.

\emph{Physical realization.} A resonant field $\mathcal E(t)\cos(\omega t-\phi_L)$ implements this gate in the rotating-wave approximation~\cite{AllenEberly1975}, which neglects rapidly oscillating, nonresonant terms. Here $\mathcal E(t)$ is the field envelope, $\omega$ the angular frequency, and $\phi_L$ the field phase. The Rabi frequency $\Omega(t)=|d|\mathcal E(t)/\hbar$ sets the pulse area $\Theta=\int\Omega\,\diff t$, while $\phi=\phi_L+\arg(-d)$. Thus, $\phi_L=\phi$ for $d<0$ and $\phi_L=\phi+\pi$ for $d>0$. Only P5 has negative $d$ (Table~\ref{tab:caoch3_pulses}).

\emph{Requirements.} Realistic implementation of Eq.~\eqref{eq:caoch3_gate_matrix} requires five experimental conditions.

(i) Resonance: the tone must match the transition angular frequency $\omega_{u,\mathrm{lo}}$ with a detuning much smaller than $\Omega$.

(ii) Selectivity: every competing transition from a populated level must have angular-frequency detuning $\Delta\gg\Omega,\tau^{-1}$, where $\tau$ is the pulse duration. Off-resonant population errors are then of order $(\Omega/\Delta)^2$~\cite{AllenEberly1975}. Excluding the line that coincides with P4, which condition (iv) avoids, the nearest competing rovibrational line lies $\Delta/2\pi=0.73$~GHz from P4 (Appendix~\ref{app:caoch3_ham}). At the Rabi frequencies chosen below, all at most $44$~kHz, the estimated population error is below $4\times10^{-9}$.

Spin structure imposes a tighter constraint. It is the only term beyond Eq.~\eqref{eq:caoch3_hamiltonian} that sets a parameter of the protocol. We estimate it using the measured ground-state spin--rotation and proton hyperfine constants~\cite{Namiki1998} for all bend levels (lower block of Table~\ref{tab:caoch3_constants}). Each rovibrational line splits into components, including unwanted transitions from populated levels. The nearest lies $0.36$~MHz from P7 and has comparable strength. Requiring $\Omega/2\pi\le12$~kHz keeps its estimated population error below $10^{-3}$. The same estimate bounds the other six pulses less tightly. We therefore drive all seven links at one intensity, chosen so that P7 has $\Omega/2\pi=10$~kHz. Because $\Omega\propto|d|$, the other links then have $\Omega/2\pi$ between $9.8$ and $44$~kHz, each below its own bound, and every estimated population error stays below $7\times10^{-4}$.

(iii) Phase reference: all tones must share a clock defining the rotating frame of the target amplitudes.

(iv) Order: at each branch point, the partial pulse must precede the $\pi$ pulse, and each parent must be populated before driving its children. P4 must finish before P5 begins, so that $L_{30}$ stays empty throughout P4 and the line that coincides with P4 (Appendix~\ref{app:caoch3_ham}) has no population to drive. Gates on disjoint level pairs commute, allowing P6 and P7 to overlap with P4 and P5.

(v) Coherence: the sequence duration, at most $0.22$~ms here (see below), must be short compared with the bend-level lifetimes.

\emph{Intensity and duration.} Converting these Rabi frequencies into a laser intensity requires the absolute bend transition dipole $\mu_8$. The linear dipole operator $\mu_x=\mu_8(a_1+a_1^\dagger)$ of Appendix~\ref{app:caoch3_dipoles} is the first-order term $(\partial\mu_x/\partial Q_1)\,Q_1$, where the mass-weighted normal coordinate is $Q_1=(\hbar/2\omega)^{1/2}(a_1+a_1^\dagger)$ and $\omega$ is the harmonic angular frequency of $\nu_8$. Hence $\mu_8=(\hbar/2\omega)^{1/2}\,\partial\mu_x/\partial Q_1$~\cite{Herzberg1945,BunkerJensen2006}. We evaluate the derivative by central differences of the dipole moment along both $\nu_8$ components of the B3LYP force field of Appendix~\ref{app:caoch3_ham}, whose harmonic wavenumber is $150.7\ \mathrm{cm^{-1}}$, and obtain $\mu_8=0.150$~D. The same calculation gives a permanent dipole of $1.45$~D, $8\%$ below the measured $1.58(8)$~D~\cite{Namiki1998}. P7, with $|d|=0.124\,\mu_8$, then reaches $\Omega/2\pi=10$~kHz at the field amplitude $\mathcal E=107$~V/m, that is, at the intensity $I=c\varepsilon_0\mathcal E^2/2=1.5\ \mathrm{mW/cm^2}$~\cite{Foot2005}. We assume this intensity for all seven tones. A pulse of area $\Theta$ on link P$n$ then lasts $\Theta/\Omega_n$. The last column of Table~\ref{tab:caoch3_pulses} lists the $\pi$-pulse durations $\tau_\pi=\pi/\Omega_n$, from $11$ to $51~\mu\mathrm{s}$. With the pulses applied one after another, the single-SALC sequences (b) and (c) last $91$ to $112~\mu\mathrm{s}$. The equal superposition with every $c=\tfrac12$ takes $171~\mu\mathrm{s}$, or $145~\mu\mathrm{s}$ when P6 and P7 overlap P4 and P5 [condition (iv)]. No target in (a) takes longer than $217~\mu\mathrm{s}$, the sum of all seven $\tau_\pi$.

\emph{Solution.} We determine the pulse parameters in Eq.~\eqref{eq:caoch3_sequence} by working backward from the target amplitudes to $G$. At each branch, a partial pulse transfers amplitude from a parent $\mathrm{lo}$ to an initially empty child $u$. A subsequent $\pi$ pulse transfers the remainder to the other initially empty child $u'$. For initial parent amplitude $\beta$, Eq.~\eqref{eq:caoch3_gate_matrix} gives
\begin{equation}
R_{u',\mathrm{lo}}(\pi,\phi')\,R_{u,\mathrm{lo}}(\Theta,\phi)\,\beta\ket{\mathrm{lo}}=-\iunit\beta\bigl(\expe^{\iunit\phi}\sin\tfrac\Theta2\ket{u}+\expe^{\iunit\phi'}\cos\tfrac\Theta2\ket{u'}\bigr).
\label{eq:caoch3_split}
\end{equation}
For desired child amplitudes $\alpha$ and $\alpha'$, population conservation requires $|\beta|^2=|\alpha|^2+|\alpha'|^2$. Matching magnitudes gives $\tan\tfrac\Theta2=|\alpha|/|\alpha'|$, and matching phases gives $\phi=\arg\alpha-\arg\beta+\tfrac\pi2$ and $\phi'=\arg\alpha'-\arg\beta+\tfrac\pi2$. A parent with one child requires only a $\pi$ pulse. Both cases empty the parent, leaving no population in intermediate levels.

Write each target amplitude in Eq.~\eqref{eq:caoch3_target} as $c=|c|\,\expe^{\iunit\gamma}$, with $\gamma$ carrying the labels of $c$. The branch populations are $W_{\irr{a}}=|c^{\irr{a}_{1}}|^2+|c^{\irr{a}_{2}}|^2$ and $W_{\irr{e}}=|c^{\irr{e}}_{1}|^2+|c^{\irr{e}}_{2}|^2$. Applying Eq.~\eqref{eq:caoch3_split} gives the partial-pulse areas for P2, P4, and P6 in Eq.~\eqref{eq:caoch3_areas}, starting with $\tan\tfrac{\Theta_2}{2}=(W_{\irr{a}}/W_{\irr{e}})^{1/2}$. P1, P3, P5, and P7 are $\pi$ pulses. With $\phi_1=\phi_2=\phi_3=0$, successive applications give
\begin{align}
R_1\ket{G}&=-\iunit\ket{L_{01}},\nonumber\\
R_3R_2R_1\ket{G}&=-\bigl(\sqrt{W_{\irr{a}}}\ket{L_{02}}+\sqrt{W_{\irr{e}}}\ket{L_{11}}\bigr),\nonumber\\
R_7\cdots R_1\ket{G}&=\iunit\bigl(|c^{\irr{a}_{1}}|\,\expe^{\iunit\phi_4}\ket{L_{03}}+|c^{\irr{a}_{2}}|\,\expe^{\iunit\phi_5}\ket{L_{30}}+|c^{\irr{e}}_{1}|\,\expe^{\iunit\phi_7}\ket{L_{21}}+|c^{\irr{e}}_{2}|\,\expe^{\iunit\phi_6}\ket{L_{12}}\bigr).
\label{eq:caoch3_layers}
\end{align}
The final line equals $\iunit\ket{\Psi_T}$ when $\phi_4=\gamma^{\irr{a}_{1}}$, $\phi_5=\gamma^{\irr{a}_{2}}$, $\phi_6=\gamma^{\irr{e}}_{2}$, and $\phi_7=\gamma^{\irr{e}}_{1}$. Three adjustable areas and three relative phases supply the six real parameters of a normalized four-level target, excluding its global phase. The sequence therefore reaches every superposition in the target subspace. 

\begin{table}[h]
\centering
\small
\caption{Pulse sequences for Fig.~\ref{fig:caoch3_tree}, with gate area $\Theta_n$ and phase $\phi_n$ for $R_n$ [Eq.~\eqref{eq:caoch3_gate}]; a dash denotes an omitted pulse. (a) Arbitrary target, using the areas in Eq.~\eqref{eq:caoch3_areas} and target phases $\gamma$. (b) $L_{30}$, labeled by $\ket{\irr{a}_{2}}$; replacing P5 with P4 prepares $L_{03}$, labeled by $\ket{\irr{a}_{1}}$. (c) $L_{21}$, labeled by $\irket{\irr{e}}{1}$; replacing P7 with P6 prepares $L_{12}$, labeled by $\irket{\irr{e}}{2}$. Each sequence yields $\iunit\ket{\Psi_T}$. Couplings follow Eq.~\eqref{eq:caoch3_dipole_factors}; line positions follow Eq.~\eqref{eq:caoch3_lines} with the constants in the upper block of Table~\ref{tab:caoch3_constants}. They are rounded after the calculation, so they can differ by $0.001\ \mathrm{cm^{-1}}$ from differences of the rounded energies in Table~\ref{tab:caoch3_levels}. The last column gives the $\pi$-pulse duration $\tau_\pi=\pi/\Omega$ at the common intensity $I=1.5\ \mathrm{mW/cm^2}$ assumed in Appendix~\ref{app:caoch3_pulses}; a pulse of area $\Theta$ lasts $(\Theta/\pi)\,\tau_\pi$.}
\label{tab:caoch3_pulses}
\begin{tabular}{llrrrrcccr}
\toprule
Pulse & Link & $q$ & $\Delta N$ & $d/\mu_8$ & $\tilde\nu$ ($\mathrm{cm^{-1}}$) & (a) arbitrary & (b) $\ket{\irr{a}_{2}}$ & (c) $\irket{\irr{e}}{1}$ & $\tau_\pi$ ($\mu\mathrm{s}$)\\
\midrule
P1 & $G\to L_{01}$ & $+1$ & $0$ & $+0.121$ & 137.520 & $\pi,\ 0$ & $\pi,\ 0$ & $\pi,\ 0$ & 50.9\\
P2 & $L_{01}\to L_{02}$ & $+1$ & $0$ & $+0.220$ & 137.547 & $\Theta_2,\ 0$ & $\pi,\ 0$ & -- & 28.1\\
P3 & $L_{01}\to L_{11}$ & $-1$ & $+1$ & $+0.545$ & 139.746 & $\pi,\ 0$ & -- & $\pi,\ 0$ & 11.4\\
P4 & $L_{02}\to L_{03}$ & $+1$ & $0$ & $+0.231$ & 137.575 & $\Theta_4,\ \gamma^{\irr{a}_{1}}$ & -- & -- & 26.7\\
P5 & $L_{02}\to L_{30}$ & $+1$ & $+1$ & $-0.509$ & 138.505 & $\pi,\ \gamma^{\irr{a}_{2}}$ & $\pi,\ 0$ & -- & 12.2\\
P6 & $L_{11}\to L_{12}$ & $+1$ & $0$ & $+0.164$ & 135.400 & $\Theta_6,\ \gamma^{\irr{e}}_{2}$ & -- & -- & 37.8\\
P7 & $L_{11}\to L_{21}$ & $-1$ & $0$ & $+0.124$ & 138.843 & $\pi,\ \gamma^{\irr{e}}_{1}$ & -- & $\pi,\ 0$ & 50.0\\
\bottomrule
\end{tabular}
\end{table}

\emph{Readout.} Each SALC label identifies a distinct molecular target level. The weights $|c^{\irr{a}_{1}}|^2$, $|c^{\irr{a}_{2}}|^2$, $|c^{\irr{e}}_{1}|^2$, and $|c^{\irr{e}}_{2}|^2$ are therefore molecular populations, distinguishable by state-selective detection because the levels differ in $N$ or $k$.

Relative phases are measured by reversing part of the tree to interfere target amplitudes, as in Ramsey interferometry. Apply P5 as a $\pi$ pulse of phase $\phi_5$, then P4 as a $\pi/2$ pulse of phase $\phi$. The resulting population in $L_{02}$ is $P_{02}=\tfrac12|c^{\irr{a}_{1}}+\expe^{\iunit\varphi}c^{\irr{a}_{2}}|^2$, where $\varphi=\phi-\phi_5$ and phase information can be extracted.  These measurements probe molecular coherences, which disappear when rotation and spin are traced out and cannot be detected by measuring the bend alone.